\documentclass[11pt]{article}

\relax

\usepackage[letterpaper, margin=.85in]{geometry}

\def\keywordname{{\bfseries \emph{Keywords}}}%
\def\keywords#1{\par\addvspace\medskipamount{\rightskip=0pt plus1cm
\def\and{\ifhmode\unskip\nobreak\fi\ $\cdot$
}\noindent\keywordname\enspace\ignorespaces#1\par}}

\usepackage{graphicx}
\graphicspath{{converted_graphics/}}
\usepackage[utf8]{inputenc} 
\usepackage[T1]{fontenc}    

\usepackage{hyperref}       
\usepackage{url}            
\usepackage{doi}

\usepackage{titling}
\usepackage{authblk}
\usepackage[english]{babel}
\usepackage{xcolor}   
\usepackage{tikz}
\usetikzlibrary{positioning}
\usepackage{istgame}
\usepackage{amssymb}
\usepackage{amsmath}
\usepackage{amsfonts}
\usepackage{overpic}
\usepackage{natbib}
\usepackage[T1]{fontenc}
\usepackage{lineno}
\usepackage{setspace}
\usepackage{caption}
\usepackage{subcaption}
\usepackage{appendix}
\usepackage{amssymb}
\usepackage{pgfplots}
\pgfplotsset{compat=newest}

\title{Public policy for management of forest pests within an ownership mosaic}
\author[1*]{Andrew R. Tilman}
\author[1]{Robert G. Haight}
\date{December 12, 2024}
\affil[1]{USDA Forest Service, Northern Research Station, St. Paul, MN, USA}
\affil[*]{Corresponding Author, Email: \href{mailto:andrew.tilman@usda.gov}{andrew.tilman@usda.gov}}

\begin{document}
\maketitle

\begin{abstract}\noindent
Urban forests provide ecosystem services that are public goods with local (shade) to global (carbon sequestration) benefits and occur on both public and private lands. Thus, incentives for private tree owners to invest in tree care may fall short of those of a public forest manager aiming to optimize ecosystem service benefits for society. The management of a forest pest provides a salient focus area because pests threaten public goods provision and pest management generates feedback that mitigates future risks to forests. We use a game theoretic model to determine optimal pest treatment subsidies for a focal privately owned tree and use an optimization approach to guide targeted public treatment of a representative public tree. We find that optimal public subsidies for private tree treatment depend on assessed tree health and on the prevalence of the pest in the community, considerations absent from many existing programs. Next, by applying our pest treatment policies to a community-scale model of emerald ash borer forest pest dynamics, we predict ash mortality under a range of treatment scenarios over a 50-year time horizon. Our results highlight how designing policies that consider the public goods benefits of private actions can contribute to sustainable land management.
\end{abstract}

The UN’s 2030 Agenda for Sustainable Development defines a set of interconnected global goals, including transitioning toward cities that are resilient and green~\citep{Agenda2030}. Realizing this objectives will require major investments by governments, companies, and individuals~\citep{sachs2019six}. The crux of spurring sustainability transitions, especially in urban areas, is that the benefits of individual investments often accrue at broad scales and to many people. In other words, there are public goods problems embedded within many potential sustainability solutions and private incentives for action will often fall short~\citep{olson1965logic, bergstrom1986private}.  

This challenge arises in many contexts, including urban forest management.  
Urban forests, comprised of the public and private trees within urban areas, contain 5.5 billion trees in the US~\citep{nowak2018us}. These forests provide myriad ecosystem services, including carbon sequestration, erosion mitigation, and cooling, which provide benefits to the over 80\% of Americans who live in urban areas~\citep{dwyer1992assessing,roy2012systematic,livesley2016urban,pataki2021benefits}. However, in the US a minority of urban trees are publicly owned, with  these mainly found in parks and rights-of-way bordering streets~\citep{ArborDay2018}. As a result, urban forests are a complex mosaic of public and private lands managed by numerous individuals~\citep{dwyer2003sustaining,epanchin2010controlling}. Since the ecosystem service benefits of these forests spillover across ownerships, the investments needed to sustain urban forests are impure public goods. Thus, in the absence of appropriate policies, under-investment in stewardship on private land is expected~\citep{wichman2016incentives}. Public subsidy and cost-share programs~\citep{andreoni1996government, chizmar2021state} are tools that can boost private provisioning of public goods, yet it remains a challenge for policy makers to design cost effective programs \citep{ jacobsen2017public, chan2020optimal}. Moving toward cities that are sustainable, resilient, and green may benefit from carefully designed public policies that support individuals in taking actions that collectively contribute to sustainability.

The United States' Inflation Reduction Act of 2022 is making sustainability and climate mitigation investments~\citep{Bistline2023}, including through a USDA Forest Service grant program that supports tree planting and care on both public and private lands\citep{FSpress2023}.  For example, the Pittsburgh Canopy Alliance received \$8 million to plant and maintain trees along streets, in parks and public green spaces, and in residential and institutional property in neighborhoods where investments in greening will contribute to social justice \citep{UCFgrants2023}. The city of Minneapolis received an \$8 million grant for treatments to protect ash trees (\textit{Fraxinus spp.}) from emerald ash borer (\textit{Agrilus planipennis})(henceforth referred to as EAB) as well as removal of infested ash trees and replanting of diverse tree species~\citep{MPLSgrant2023}. Both approaches could contribute to more resilient future forests and more sustainable cities. We focus on the case of forest pest management, as it bears many of the hallmarks of the broader challenge of designing policies to promote sustainability on intermingled public and private lands. At the same time, given the unprecedented funding for urban forestry under the Inflation Reduction Act, it is especially salient to determine polices that could cost-effectively sustain urban forests in the face of threats from pests.  

The management of forest pests adds additional complexity to the general problem of spurring private contributions to public goods because each individual's control decisions affect not only the impact of the pest (or disease) on a focal tree but also it's spread across the landscape. For example, insecticide treatments, such as trunk injections with emamectin benzoate, provide effective EAB control to treated ash trees~\citep{mccullough2011evaluation} and also mitigate the risk of infestation of neighboring trees~\citep{mccullough2012evaluation}. This spillover effect results in a spatial-dynamic externality~\citep{wilen2007} that has been studied in the context of pest spread~\citep{epanchin2015individual}, including within management mosaics~\citep{sims2010dynamic,kovacs2014bioeconomic}. A spatially-explicit game theoretic analysis of the control of a mobile renewable resource with harmful local effects (such as a forest pest) has shown that decentralized management by individuals often diverges from socially optimal control policies, especially when the damages caused by the resource are intermediate~\citep{costello2017private}. In these cases, there may be a role for public policies that stimulate increased private investment in management.  
Subsidies and transfer payments are often proposed as mechanisms to align public and private incentives for pest management in urban forests. For example, \cite{Chen2022} use a principal-agent modeling framework to design programs in which a local government reimburses landowners for pest management costs based on numbers of infested or treated trees on the property. \cite{cobourn2019cooperative} show that transfer payments between municipalities can induce levels of pest control that make each city better off and approach a central planner’s goal of maximizing net social benefit.

We address the general challenge posed by forest pests to municipal forest managers in urban environments by considering three interrelated questions. First, how can public subsidy programs be designed to efficiently spur investments in forest pest management on private lands? Second, what are the optimal pest treatment decisions for publicly owned trees? Lastly, in a mosaic landscape of public and private lands, how effective are these policies at mitigating pest spread and maintaining forest health over time? These issues are made more complex because of numerous uncertainties facing both private and public decision makers, which we address. 

To address the first question, we construct a game theoretic model of subsidy policies for treatments to address forest pests on a focal privately owned tree, assuming that tree owners are myopic and consider only short-term costs and benefits over the duration of efficacy of treatment. To address the second question, we deploy an optimization approach to study treatment decisions for a focal public tree.  
Shifting from a static to dynamic perspective, we answer the third question via the development of an epidemiological model of the dynamics of pest spread and tree mortality over many decades. This model integrates the impacts of treatment of public and private trees in scenarios corresponding to the outputs of the initial models. We then explore the application of these three approaches in a case study of the management of EAB in an urban mosaic of public and private land. Our model is relevant to cities with EAB management plans (e.g.~\citep{BurnsvilleEAB}) that might benefit from refinements in how they strategically treat their publicly owned trees and how they encourage treatment of private trees. 
 
We find that optimal subsidy levels for private trees are dependent on tree health and how widespread a forest pest is within a community, considerations which are absent from many existing treatment programs.  
For EAB, comparing predicted long-term tree survival under scenarios where there is no public action to treat a pest and where optimal treatment policies are deployed across public and private ownership shows how these lands interact to shape collective risks to forest health. Our approach provides a road map for the design of public policies that aim to stimulate collaborative pest management in an ownership mosaic.

\section*{Treatment subsidies for private trees}

 We construct a game theoretic model of the treatment of a single tree growing on private land that is susceptible to a forest pest that can be managed with treatments administered by a tree care specialist. The model considers strategic interactions between two stakeholders: a municipal forester and a tree owner. The municipal forester offers pest treatment subsidies for the care of private trees with an objective of optimizing forest ecosystem service benefits subject to the cost of the subsidy policy. The private tree owner makes a decision whether or not to treat their tree with the objective to maximize the benefits they receive from their tree, taking into account treatment costs.  Henceforth, we call these stakeholders the municipality and the tree owner. A game theoretic approach is necessary because the objectives of each stakeholder are intertwined. This approach allows us to consider uncertainty that the municipality may have about the tree owner and incorporate this into the design of the subsidy program.  
 
\begin{figure}[ht]
    \centering
    \includegraphics[width=.6\textwidth]{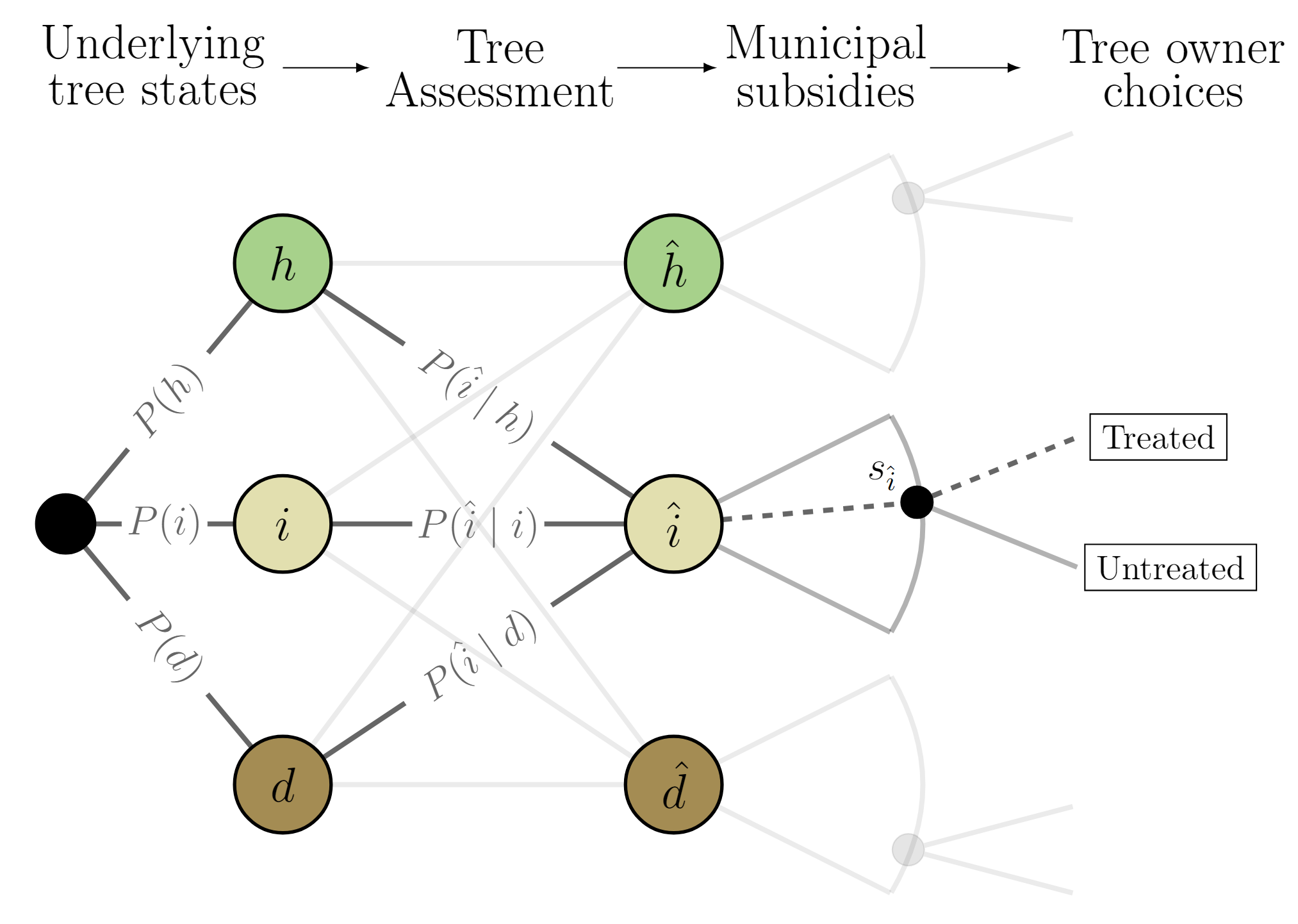}
    \caption{The structure of the game theoretic model, from assessment of a tree's health to treatment outcomes. A focal tree can fall into one of three states (healthy, $h$; infested, $i$; dead / dying, $d$) with likelihoods given by community level forest pest prevalence. Next, the health status of the tree is assessed (assessed as healthy, $\hat h$; infested, $\hat i$; dead / dying, $\hat d$), with assessment errors possible. This splits the game into three branches, one for each assessed tree state. The game that follows for each assessed health state can be analyzed independently. We highlight the case of a tree that has been assessed as being in an infested state. The municipality sets a subsidy level that reduces the cost of treatment for a tree owner and the tree owner responds to this in their decision to treat their tree or not. The arc in the figure represents the continuous range of subsidy levels that the municipality can set. Here we show a case where the subsidy is high enough that the tree owner decides to have their tree treated.}
    \label{fig:gameStructure}
\end{figure}

 The structure of the model is shown in Figure~\ref{fig:gameStructure}. Beginning with a focal tree that falls into one of three pest infestation states (healthy, infested, or dying / dead), the state of the tree is (imperfectly) assessed based on observable characteristics. For example, in the context of EAB management, arborists can assess ash tree health by observing the tree’s crown condition, which is correlated with  infestation\citep{flower2013relationship,knight2014monitoring}. 
 However, an ash tree with no visible crown dieback may have either no infestation or a low level of infestation that has yet to cause visible damage. Conversely, crown dieback observed in a ash tree could be due to a different cause. 
 Similar assessment inaccuracies are likely to arise in the context of other forest pests and diseases. We define assessment accuracy probabilities that allow the stakeholders to estimate the likelihood of each underlying tree health state using Bayes' theorem.  

For each possible assessed tree health state, we analyze municipally subsidized treatment of a privately owned ash tree as a dynamic game of incomplete information. This allows us to consider subsidy policies that vary for different tree health states. Figure~\ref{fig:gameStructure} highlights the structure of the game for a tree that is assessed as infested. In this case, the municipality selects a non-negative subsidy level that partially offsets the costs of treatment for a tree owner. In response to the subsidy level available and the price of treatment, the tree owner decides whether to treat their tree or not. The tree owner has a unique `type' which we define as the value that they assign to avoiding tree mortality over the period of efficacy of the treatment. While the tree owner knows this value, the municipality only knows the distribution from which this value is drawn. This introduces incomplete information for the municipality. Therefore, we assume that the municipality aims to maximize their expected utility given their beliefs about the distribution of tree owner values. These assumptions allow us to solve for the Perfect Bayesian Equilibrium (a refinement of the Nash equilibrium) of the game. Most analyses of Perfect Bayesian Equilibria concern how actions reveal information about a players type. Our analysis is greatly simplified because the tree owner (the player about whom there is incomplete information) makes the final decision in our model and therefore any information revelation cannot impact the municipality's strategy.

We proceed to introduce key aspects of tree health assessment, treatment effectiveness, and treatment dependent tree mortality risk perception that underlie the (expected) payoffs of the tree owner and the municipality. Then we will derive the benefits of treatment from the perspectives of the tree owner and municipality. Using these benefits, we construct the objective functions of each player in the game, and describe how this game can be analyzed. Finally, we define the optimal subsidy policy and resulting treatment outcomes for a representative privately owned tree.

\paragraph{Tree health assessment}
We assume an arborist assesses the health status of a tree as healthy, $\hat h$, infested, $\hat i$, or dead / dying, $\hat d$, with an arbitrary assessed state denoted  
$\hat \phi\in\{\hat h,\hat i, \hat d\}$.
A tree's health status is only partially observable, with the true underlying state 
$\phi\in\{h,i,d\}$
of the tree possibly diverging from the assessed state. We assume that surveillance data are available at the municipality scale, with publicly known frequencies of each underlying tree state given by $P(h)$, $P(i)$, and $P(d)$. 
Since assessment techniques are imperfect predictors of the true underlying health of a tree, we define the accuracy of assessment with probabilities. For example,
$P(\hat h \mid i)$
is the probability that a tree is assessed as healthy given that it actually is infested. There are nine of these probabilities, which can, in general, be written as 
$P(\hat \phi \mid \phi)$
for some assessed state, $\hat \phi$, and some true underlying state, $\phi$. 

When a tree owner is deciding whether to treat their tree and when the municipality is setting subsidy policies for trees, they use the information in the assessment of the tree's health to update their beliefs about the likelihood of the tree having each of the three possible underlying states using Bayes' theorem. These probabilities can be written as 
$P(\phi \mid \hat \phi)$
for some true underlying state, $\phi$, given some assessed state, $\hat\phi$, and calculating these probabilities relies on the publicly-known community-level prevalence of healthy, infested and dying trees as well as the accuracy of assessment.
 
\paragraph{Effectiveness of treatment}
Two pathways of treatment effects are considered. First, a healthy tree's risk infestation given exposure to a forest pest may be reduced by treatment. This is akin to vaccination of the tree. We define $\epsilon_h$ as the effectiveness of treatment at preventing establishment of infestation with a pest (in other words, treatment effectiveness for healthy trees, hence the $h$ subscript). At one extreme, if $\epsilon_h=1$ the treatment is 100\% effective and there will be no chance of infestation over the time horizon of effectiveness of treatment, conversely if $\epsilon_h=0$ then treatment will have no impact on the infestation risk of a healthy tree. 
The second pathway of treatment impact that we consider is the potential for treatment of an infested tree to reverse the infestation and return the tree to a healthy state. We define $\epsilon_i$ as the effectiveness of treatment for an infested tree and it represents the likelihood of tree recovery given treatment (in other words, treatment effectiveness for infested trees, hence the $i$ subscript). 

\paragraph{Perceived tree mortality risks}
We assume that the municipality and the tree owner form risk perceptions about the likelihood of tree mortality over a fixed time horizon and employ these risk assessments in decision making.   
The decisions of the municipality and the tree owner influence the direct risk of mortality (or conversely, likelihood of survival) for a focal tree. Their decisions can also have spillover effects on the spread of a pest within a community and thus influence mortality risks for non-focal trees. We assume that a private tree owner consider only the direct impacts of their actions, whereas the municipality considers both direct and spillover effects because they are aiming to optimize forest ecosystem service benefits at a broad scale. We develop estimates for the perceived direct and spillover impacts of treatment and non-treatment by drawing on epidemiological models.

We assume risk perception is formed via forecasting a tree's mortality probability given current infestation levels remain constant over the time horizon of treatment efficacy. This forecast employs a simplified and tractable epidemiological model which results in equations for the risk of mortality over time under treatment or non-treatment (see Supplementary Material Section~\ref{SI-focalRisk} for a complete description of the risk perception model). We define $\mu_{t\phi}$ as the perceived mortality probability for a treated tree (throughout the manuscript, a `$t$' subscript corresponds to `treated')  in health state $\phi$ over a time horizon $\tau^*$ which corresponds the to duration of efficacy of treatment. Similarly, we define $\mu_{u\phi}$ as the perceived mortality probability for an untreated tree (throughout the manuscript, a `$u$' subscript corresponds to `untreated') in health state $\phi$ over the same time horizon. 

In addition to the direct impact of treatment on the survival of a focal tree, by eliminating a source of pest spread the treatment can also affect other trees in the community, resulting in a dynamic externality~\citep{wilen2007}. We model these potential spillover effects of treatment by drawing on the basic pest reproduction number ($R_0$) in epidemiology which defines the expected number of infections caused by a single infection in a fully susceptible population. In the context of forest pest management, this represents the expected number of newly infested trees attributable to a focal tree, which can depend of treatment. For example, healthy trees that are treated are less likely to become infested and therefore less likely to cause mortality of other trees. Similarly, an infested tree that is treated may recover and reduce the expected spillover mortality risk. Using this approach, we estimate spillover tree mortality risk under treatment and non-treatment (see Supplementary Material Section~\ref{SI-SpilloverRisk} for a complete description of the spillover risk perception model).  We define $\lambda_{t\phi}$ as the perceived spillover mortality risk (expected number of trees in the community that will experience mortality due to the focal tree) for a treated tree in health state $\phi$. We similarly define $\lambda_{u\phi}$ as the perceived spillover mortality risk for an untreated tree. In our model, private tree owners ignore spillover mortality risk when making treatment decisions because this spillover is an externality from the perspective of a private tree owner. However, from the perspective of the municipality, this spillover risk is central because it estimates how both public and private treatment influences community tree survival.

\paragraph{Value of treatment}\label{PriVal}
In our model a tree owner makes decisions by comparing the costs and benefits of two alternative outcomes: tree survival versus tree mortality. We assume that the value of a surviving tree to its owner is $V_o$ and that the cost associated with tree mortality, such as removal costs, is $W_o$ for the owner. Throughout the manuscript, `$o$' subscripts denote terms that correspond to the owner of a private tree.  We assume $V_o$ and $W_o$ are assessed over a time horizon that corresponds to the length of effectiveness of treatment. The likelihoods of survival versus mortality and the assessed state of the tree define the payoff to a tree owner (ignoring treatment costs) of having a treated or untreated tree. These payoffs are 
\begin{align}
\pi_{t\hat \phi} &= 
    P( h \mid \hat \phi)\left(\left(1-\mu_{th}\right)V_o-\mu_{th}W_o\right)
    +P( i \mid \hat \phi)\left(\left(1-\mu_{ti}\right)V_o-\mu_{ti}W_o\right)
    -P( d \mid \hat \phi)W_o\\
\pi_{u\hat \phi} &=
     P( h \mid \hat \phi)\left(\left(1-\mu_{uh}\right)V_o-\mu_{uh}W_o\right)
    +P( i \mid \hat \phi)\left(\left(1-\mu_{ui}\right)V_o-\mu_{ui}W_o\right)
    -P(d \mid \hat \phi)W_o
\end{align}
for a treated, $t$, and untreated $u$ tree in some assessed health state $\hat\phi$. See Table~\ref{tab:variables} for definitions of all variables and parameters used in all models.
The benefit of treatment from the owners perspective is the difference in their payoff between treatment and non-treatment for a tree with a particular assessed state, given by 
\begin{equation}
\pi_{t\hat \phi} - \pi_{u\hat \phi}=\Delta_o k_{\hat \phi}
\end{equation}
where $\Delta_o=V_o+W_o$ is the benefit of avoiding mortality of a focal tree to its owner and 
\begin{equation}
k_{\hat \phi}=P(h\mid\hat \phi)\left(\mu_{uh}  - \mu_{th}\right) + P(i\mid\hat \phi)\left(\mu_{ui}  - \mu_{ti}\right)   
\end{equation}
is the change in the likelihood of tree survival that results from treatment of a tree in assessed state $\hat\phi$ (for a more detailed derivation, see Supplementary Material Section~\ref{suppPriVal}). While the value of avoiding tree mortality does not depend on its assessed health state, the likelihood that treatment averts mortality does.

A tree provides benefits not just to its owner but also to the broader community because many of the ecosystem service benefits of urban trees are public goods. Therefore a tree owner's valuation of a surviving tree, $V_o$, may not align with the municipal value of a surviving tree, $V_m$, which we assume is also assessed over the time horizon of treatment efficacy. Similarly, $W_m$ is the social cost of tree mortality, which we assume does not include removal costs. From the perspective of the municipality, the benefit of avoiding mortality of a single tree is thus 
$\Delta_m=V_m+W_m$. While the tree owner considers only direct risks to the survival of their own tree, the municipality also considers the spillover impact of treatment or non-treatment on changes in the likelihood of survival of other trees. This introduces a second source of divergence between the payoffs of the municipality and the private tree owner.  
These values and costs, together with perceived direct and spillover mortality risks, assessment accuracy, and community level infestation rates can be used to construct the payoff from the perspective of the municipality as 
\begin{align}
    \Pi_{t\hat \phi} &=
     P( h \mid \hat \phi)\left[V_m-\Delta_m\left(\mu_{th}+\lambda_{th}\right)\right]
    +P( i \mid \hat \phi)\left[V_m-\Delta_m\left(\mu_{ti}+\lambda_{ti}\right)\right]
    -P(d \mid \hat \phi)W_m\\
    \Pi_{u\hat \phi} &=
     P( h \mid \hat \phi)\left[V_m-\Delta_m\left(\mu_{uh}+\lambda_{uh}\right)\right]
    +P( i \mid \hat \phi)\left[V_m-\Delta_m\left(\mu_{ui}+\lambda_{ui}\right)\right]
    -P(d \mid \hat \phi)W_m
\end{align}
for a treated and untreated focal tree that is assessed to be in some health state $\hat\phi$ (derivations of these payoff equations are presented in Supplementary Material Section~\ref{suppSocVal}). The benefit of treatment from the socially-oriented perspective of the municipality can be written as
\begin{equation}
\Pi_{t\hat \phi} - \Pi_{u\hat \phi} = \Delta_m \left(k_{\hat \phi}+l_{\hat\phi}\right)
\end{equation}
where 
\begin{align}
l_{\hat\phi} = P(h\mid\hat \phi)\left(\lambda_{uh}  - \lambda_{th}\right) + P(i\mid\hat \phi)\left(\lambda_{ui}  - \lambda_{ti}\right)  
\end{align}
is the expected change in community tree survival given treatment of a focal tree in assessed health state $\hat\phi$. These equations show that private and municipal benefits of treatment across the scenarios share $k_{\hat \phi}$, which describes the effectiveness of treatment for the focal tree given the information that is available. However, whenever $\Delta_m\neq\Delta_o$ or $l_{\hat\phi}>0$, the private and social incentives for treatment will not align.

\paragraph{Cost of treatment} 
Treatments for forest pests often rely on licensed tree care firms and professionals for their application because of the use of specialized equipment and insecticides. We analyze a model which implicitly considers the role of these firms by implementing a subsidy-dependent treatment cost function that emerges from a model of Bertrand competition among multiple symmetric firms. In Supplementary Material Section~\ref{suppAnalysis}, two key simplifying results about subsidy policies and resulting treatment costs are obtained. First, we show that it is always optimal for the municipality to offer the same subsidy level to all firms. Second, we show that given equal subsidies and symmetric firms, the subsidy will be completely passed on to the tree owner in the form of lower prices. As a consequence of these results, we assume that the price a tree owner pays for treatment is
\begin{equation}
c - s_{\hat\phi}
\end{equation}
where $c$ is the marginal cost of treatment for the firms and $s_{\hat\phi}$ is the subsidy level offered by the municipality for treatment of a tree with assessed health state $\hat\phi$.

\paragraph{Game structure}\label{Game}
Once a tree has had its health status assessed, the municipality chooses subsidy levels, $s_{\hat\phi}$, that are available to firms for the treatment of a tree assessed in state $\hat\phi$, which results in a treatment cost for the tree owner of $c - s_{\hat\phi}$. The tree owner then decides whether to treat their tree or not. While the tree owner knows how much they value avoiding tree mortality, $\Delta_o$, the municipality only know that $\Delta_o$ is drawn from a uniform distribution between $a$ and $b$, denoted $\Delta_o\sim U[a,b]$. A strategy for the tree owner defines their likelihood of choosing treatment across the full range of values from $a$ to $b$.
Formal definitions of the strategy spaces of the players can be found in Supplementary Material Section~\ref{suppStrat}.

\paragraph{Objective functions}\label{Util}
Given the value of treated and untreated trees and the cost of subsidies, we construct the utility function for the municipality. We assume that subsidies, $s_{\hat \phi}\geq0$ are costly to the municipality. This leads to a utility function for the municipality given by 
\begin{equation}
U_{m\hat\phi} = \begin{cases}
    \Pi_{t\hat \phi}-s_{\hat \phi}, & \text{if tree assessed as $\hat \phi$ is treated}\\
    \Pi_{u\hat \phi} & \text{if tree assessed as $\hat \phi$ is untreated}
\end{cases}
\end{equation}
and we assume that the municipality aims to maximize their expected utility when they choose the subsidy levels to offer for each assessed tree health state, $\hat\phi$. 
Finally, we consider the utility of a private tree owner, which will depend on their valuation of the tree. These valuations can be used to construct the utility function of the tree owner as
\begin{equation}
U_{o\hat\phi} = \begin{cases}
    \pi_{t\hat \phi}- c + s_{\hat \phi}, & \text{if the tree is treated}\\
    \pi_{u\hat \phi}, & \text{if the tree is untreated}
\end{cases}
\end{equation}
where the owners benefit from their private valuation of a treated or untreated tree, and if the tree is treated, they bear the cost of treatment. We assume that the tree owner makes a treatment decision that maximizes their utility. We use these utility functions to derive the Perfect Bayesian Equilibrium for the game.

\paragraph*{Game equilibrium}
The analysis of the game is conducted via backward induction, starting with the tree owner's optimal strategy that maximizes their utility given an arbitrary subsidy level. Given this, we compute the the optimal subsidy policy of the municipality that maximizes their expected utility, given their beliefs about how the tree owner will respond to the subsidy level offered. In total, this will yield strategies for the municipality and the tree owner that satisfy the conditions of a Perfect Bayesian Equilibrium, assuring that no player will have an incentive to unilaterally deviate from their strategy. Details of the analysis can be found in Supplementary Material Section~\ref{suppAnalysis}. 

\begin{figure}[t]
    \centering
    \includegraphics[width=.95\textwidth]{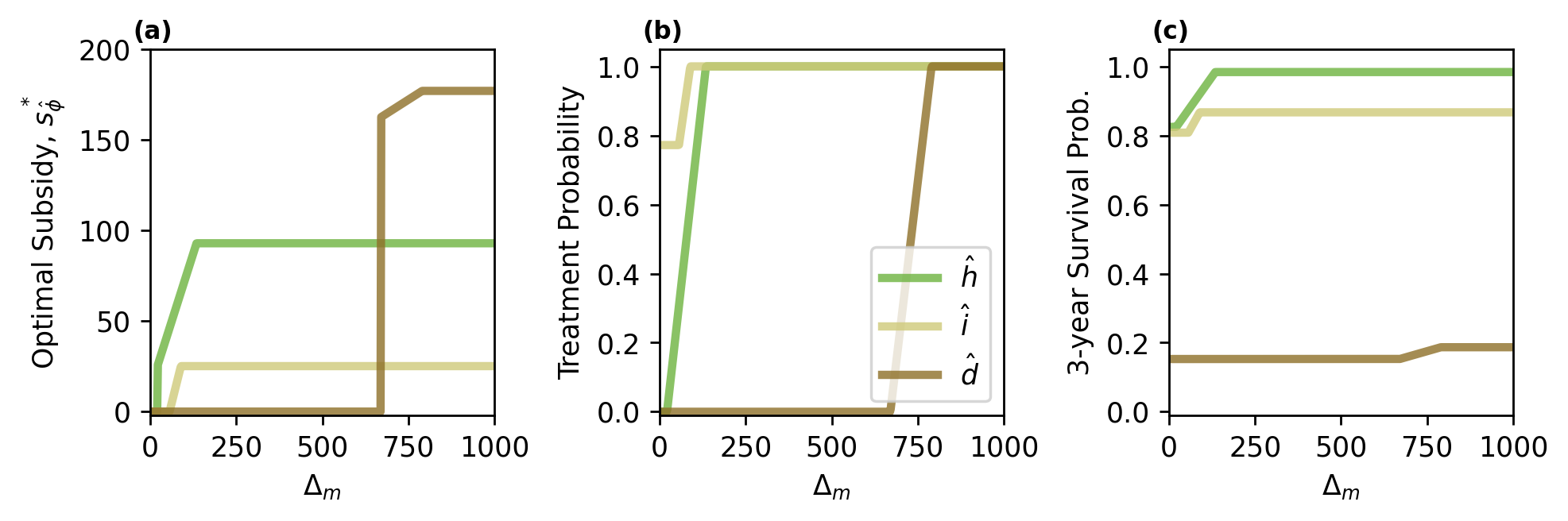}
    \caption{Optimal subsidy levels, treatment probabilities, and expected 3-year tree survival for privately owned trees. These values vary across assessed tree health states and depend on the social benefit of avoiding tree mortality, $\Delta_m$. These figures show a case where the community infestation state is $(P(h),P(i),P(d))=(0.8,0.15,0.05)$, implying that a pest is widespread but has yet to cause significant tree mortality. Remaining parameters that generated this illustrative example can be found in Table~\ref{param_table}}
    \label{fig:optSub1}
\end{figure}

We focus on the optimal subsidy policy of the municipality and how this optimal policy impacts treatment rates of a privately owned tree. The equilibrium subsidy policy, defined by the magnitude of subsidy available given different parameter conditions are met, is 
\begin{equation}
s^*_{\hat\phi}=
\begin{cases}
    0 
        & c\leq ak_{\hat\phi}\\
    c-ak_{\hat\phi} 
        &  c> ak_{\hat\phi} \;\text{ and }\;c+bk_{\hat\phi}-2ak_{\hat\phi}\leq \Delta_m\left(k_{\hat\phi} + l_{\hat\phi}\right) \\
    \frac{1}{2}\left(\Delta_m\left(k_{\hat\phi} + l_{\hat\phi}\right) +c-bk_{\hat\phi}\right) 
        & c> ak_{\hat\phi} \;\text{ and }\;\mid c-bk_{\hat\phi}\mid < \Delta_m\left(k_{\hat\phi} + l_{\hat\phi}\right) < c+bk_{\hat\phi}-2ak_{\hat\phi}\\
    0 
        &  c> ak_{\hat\phi} \;\text{ and }\; \Delta_m\left(k_{\hat\phi} + l_{\hat\phi}\right) \leq \mid c-bk_{\hat\phi}\mid
\end{cases}
\label{optSub}
\end{equation}
where the first case arises when the cost of treatment is low enough that a tree owner will opt for treatment regardless of their type, even in the absence of a subsidy. The second case corresponds to the case where the social value of treatment is high enough to justify a subsidy level that induces even the tree owners with the lowest value for treatment of their tree to choose treatment. The third case corresponds to an intermediate equilibrium where the optimal subsidy level results in a some types of tree owners choosing treatment and others opting not to treat. The final case arises when the social benefit of treatment is too low to justify subsidizing treatment.

When the optimal subsidy levels are positive, they result in increased treatment probabilities for a privately owned tree. The probability that the owner of a focal tree assessed in health state $\hat\phi$ will opt for treatment given that equilibrium strategies are followed is
\begin{equation}
P_{t\hat\phi o}^*=
\begin{cases}
    1 
        &  c \leq ak_{\hat\phi} \;\text{ or }\;c+bk_{\hat\phi}-2ak_{\hat\phi}\leq \Delta_m \left(k_{\hat\phi} + l_{\hat\phi}\right) \\
    \frac{\Delta_m \left(k_{\hat\phi} + l_{\hat\phi}\right) -c+bk_{\hat\phi}}{2k_{\hat\phi}(b-a)} 
        & c > a \;\text{ and }\;\mid c-bk_{\hat\phi}\mid < \Delta_m \left(k_{\hat\phi} + l_{\hat\phi}\right) < c+bk_{\hat\phi}-2ak_{\hat\phi}\vspace{3.5pt}\\
    \frac{bk_{\hat\phi}-c}{k_{\hat\phi}(b-a)} 
        & bk_{\hat\phi}>c> ak_{\hat\phi} \;\text{ and }\; \Delta_m \leq  bk_{\hat\phi}-c\\
     0 
        & c\geq bk_{\hat\phi}\;\text{ and }\; \Delta_m \left(k_{\hat\phi} + l_{\hat\phi}\right) \leq  c-bk_{\hat\phi}
\end{cases}
\label{eq:privTreat}
\end{equation}
where the first case arises when treatment costs are low relative to private benefits or when the social value of treatment is high. The second case results under intermediate subsidy levels, and the final two cases result when no subsidies are applied and either some types of tree owners still treat their tree (case 3) or none choose to (case 4). Figure~\ref{fig:optSub1} shows a representative example of optimal subsidies and treatment rates across assessed health classes of trees as a function of the social value of avoiding tree mortality, $\Delta_m$. In this case, trees assessed as infested have a positive treatment probability even in the absence of subsidies. Only for high levels of $\Delta_m$ do trees assessed as dying receive subsidies and become treated. Despite treatment probabilities being lower for healthy trees at low values of $\Delta_m$, trees assessed as healthy have higher 3-year survival probabilities than infested trees (based on mortality risk perceptions, $\mu_{t\phi}$ and $\mu_{u\phi}$). This results because mortality requires infestation as a precondition and this delays mortality for healthy trees.

\begin{table}[htbp]
\begin{center}
    \begin{tabular}{c|l } 
     Term & Definition \\ 
     \hline
     $\phi\in\{h,i,d\}$ & Underlying tree state: healthy, infested, dead/dying  \\
     $\hat \phi\in\{\hat h,\hat i,\hat d\}$ & Assessed tree state: healthy, infested, dead/dying \\
     $P(\phi)$ & (Prior) probability of a tree being in state $\phi$\\
     $P(\hat\phi\mid\phi)$ & Probability of a tree in some state $\phi$ being assessed as $\hat\phi$ (health assessment accuracy) \\
     $P(\phi\mid\hat\phi)$ & (Posterior) probability of a tree being in state $\phi$ given that it was assessed as $\hat\phi$\\
     $\tau^*$ &  Planning time horizon \\
     $V_o$ & Value of a surviving tree to its owner over $\tau^*$ (`$o$' subscript indicates the tree owner perspective)\\
     $V_m$ & Value of a surviving tree to society over $\tau^*$ (`$m$' subscript corresponds to a municipal perspective)\\
     $W_o$ & Cost of tree mortality to its owner, including removal costs\\
     $W_m$ & Cost of tree mortality to society \\
     $W'_m$ & Cost of tree mortality to society, including removal costs\\
     $\Delta_o$ & Private value of avoiding tree mortality, $V_o+W_o$ (drawn from a uniform distribution, $U(a,b)$) \\ 
     $\Delta_m$ & Social value of avoiding tree mortality, $V_m+W_m$  \\
     $\Delta'_m$ & Social value of avoiding tree mortality including removal costs, $V_m+W'_m$  \\
     $\mu_{t\phi}$& Perceived mortality risk for a treated (subscript `$t$') tree in health state $\phi$ \\
     $\mu_{u\phi}$& Perceived mortality risk for an untreated (subscript `$u$')tree in health state $\phi$ \\
     $\lambda_{t\phi}$& Perceived spillover mortality risk for a treated tree in health state $\phi$ \\
     $\lambda_{u\phi}$& Perceived spillover mortality risk for an untreated tree in health state $\phi$ \\
     $k_{\hat\phi}$ & Treatment-based change in a tree's survival probability (tree assessed in state $\hat\phi$)\\
     $l_{\hat\phi}$ & Treatment-based change in survival of other trees (a focal tree assessed in state $\hat\phi$)\\
     $\pi_{t\hat\phi}$ & Payoff to a tree owner of treating a tree assessed as $\hat\phi$ \\
     $\pi_{u\hat\phi}$ & Payoff to a tree owner of not treating a tree assessed as $\hat\phi$ \\
     $\Pi_{t\hat\phi}$ & Payoff to society of treating a tree assessed as $\hat\phi$ \\
     $\Pi_{u\hat\phi}$ & Payoff to society of not treating a tree assessed as $\hat\phi$ \\
     
     $c$ & Unsubsidized cost of treatment\\
     $s_{\hat\phi}$ & Treatment subsidy offered for a tree in assessed health state $\hat\phi$\\
     $U_o$ & Utility of tree owner \\
     $U_m$ & Utility of municipality \\
     $\alpha$ & Recovery rate of infested trees that are treated\\
     $\beta$ & Pest spread rate\\
     $\gamma$ & Pest-induced mortality rate\\
     $\epsilon_h$ & Effectiveness of treatment for a healthy tree\\
     $\epsilon_i$ & Effectiveness of treatment for an infested tree\\
    \end{tabular}
\end{center}
     \caption {Descriptions and definitions of variables and parameters in the model.}
     \label{tab:variables}
\end{table}

\section*{Treatment of public trees}
Now let us consider an optimization approach applied to municipally owned trees. Much of the setup and notation remains from the prior game theoretic model. We assume that public trees can be treated at a cost, $c$, by the municipality. The problem for a forest manager is to identify the subset of trees for which the social benefit of treatment is greater than the public cost. While the social value of a surviving tree, $V_m$, remains unchanged, the social cost of mortality of a municipally owned tree may be higher than the social cost of a private tree's mortality. In the case of a municipally owned tree, we assume this cost is inclusive of the removal cost of the tree. Thus we define $W'_m$ as the social cost of tree mortality, including removal costs. The objective of the municipality for the treatment of public trees is to choose treatment probabilities to maximize
\begin{equation}
U_{m\hat\phi} = \begin{cases}
    \Pi'_{t\hat \phi}-c, & \text{if tree assessed as $\hat \phi$ is treated}\\
    \Pi'_{u\hat \phi} & \text{if tree assessed as $\hat \phi$ is untreated}
\end{cases}
\end{equation}
Where the expected benefit of treatment is 
\begin{equation}
\Pi'_{t\hat \phi} - \Pi'_{u\hat \phi} = \Delta'_m \left(k_{\hat \phi}+l_{\hat\phi}\right)
\end{equation}
and the net social benefit of avoiding tree mortality is $\Delta'_m=V_m+W'_m$. As before, we denote the change in the survival probability in response to treatment for a focal tree assessed in state $\hat\phi$ as $k_{\hat\phi}$ and the spillover effect of treatment on changes in expected community-level tree survival as $l_{\hat\phi}$.
This can be used to define the cases where a municipal tree should be treated. The optimal probability of treating a public tree is given by 
\begin{equation}
P^*_{t\hat\phi m}=
\begin{cases}
    1        &  c\leq \Delta'_m\left(k_{\hat\phi}+l_{\hat\phi}\right)\\
    0        &  c > \Delta'_m\left(k_{\hat\phi}+l_{\hat\phi}\right).
\end{cases}
\label{eq:pubTreat}
\end{equation}
This equation provides a strikingly simple proscription for treatment of public trees. However, considerable nuance is encoded within the $k_{\hat\phi}+l_{\hat\phi}$ because these terms depend on treatment effectiveness, assessment accuracy, the state of EAB infestation within the community and the perceived direct and spillover mortality risks to trees. 

\section*{Forest pest dynamics}
The models presented to examine the treatment of a focal public or private tree are static and based on risk perceptions and treatment choices at a single point in time. These simplifying assumptions result in an analytically tractable approach for treatment and subsidy programs, but beg the question of how well such programs might perform over time.  We now construct a complimentary model that describes the temporal dynamics of tree infestation and mortality for a population of public and private trees. We build a compartmental epidemiological model~\citep{kermack1927contribution} of forest pest spread across a population of public and private trees so that the impact of subsidy policies and treatment decisions on the temporal dynamics of tree infestation and mortality can be evaluated. 
  
We categorize the population of trees into classes with distinct properties that influence the spread of the pathogen, subdividing the population into healthy, infested, and dying classes for municipally owned and privately owned trees. Rates of transfer between compartments describe pest spread, tree mortality, and tree recovery following treatment. We denote municipal trees by the frequency of healthy trees, $H_m$, infested trees, $I_m$, and dead or dying trees, $D_m$ (using capital letters to distinguish the population of trees from individual tree states considered earlier). The municipally owned trees interact with three classes of privately owned trees, $H_o$, $I_o$, and $D_o$ ($o$ subscripts denote that the ownership of these classes of trees is private, as before). We measure the frequency of trees in each category so that we have $H_m+I_m+D_m+H_o+I_o+D_o=1$. 

We denote the rate of change of the fraction of trees as $\frac{dH_m}{d\tau}=\dot H_m$ for healthy municipally owned trees and follow similar notation for the rates of change of other classes of trees. The dynamics of pest spread and tree infestation are governed by
\begin{align}
    &\dot H_m = -\beta\left(1-\epsilon_h P_{thm}\right) H_m(I_m+I_o) + \alpha\epsilon_iP_{tim} I_m \\
    &\dot H_o = -\beta\left(1-\epsilon_h P_{tho}\right) H_o(I_m+I_o) + \alpha\epsilon_iP_{tio} I_o \\
    &\dot I_m = \beta\left(1-\epsilon_h P_{thm}\right) H_m(I_m+I_o) - \gamma\left(1-\epsilon_i P_{tim}\right) I_m - \alpha\epsilon_i P_{tim} I_m\\
    &\dot I_o = \beta\left(1-\epsilon_h P_{tho}\right) H_o(I_m+I_o) - \gamma\left(1-\epsilon_i P_{tio}\right) I_o - \alpha\epsilon_i P_{tio} I_o\\
      &\dot D_m = \gamma\left(1-\epsilon_i P_{tim}\right) I_m \\
      &\dot D_o = \gamma\left(1-\epsilon_i P_{tio}\right) I_o
\end{align}
where $P_{tio}$, $P_{tho}$, $P_{tim}$, and $P_{thm}$ correspond to treatment probabilities of trees that are infested or healthy and privately or municipally owned. These probabilities can be defined to create various treatment scenarios that depend on the state of the system. For instance, treatment probabilities may increase when the community level of infestation ($I_o+I_m$) is higher. We can also construct treatment scenarios to reflect the results of the game theoretic and optimization models of treatment described by equation~\ref{eq:privTreat} and equation~\ref{eq:pubTreat}. However, these treatment probabilities apply to trees based on their assessed state, whereas the dynamics described in this section depend on treatment probabilities based on a tree's actual health state. These transformations can be calculated with Bayes' theorem, described in Supplementary Material Sections~\ref{epiMuniTreat} and \ref{epiPrivTreat}.

In contrast to the game theoretic and optimization models, where we relied on mortality risk perception, for example $\mu_{th}$, over a fixed time horizon, this epidemiological model depends only on treatment effectiveness parameters for healthy and infested trees to describe the dynamics of pest spread. The treatment effectiveness parameter for healthy trees, $\epsilon_h\in[0,1]$, describes how effective treatment is at preventing infestation of healthy trees. The treatment effectiveness parameter for infested trees, $\epsilon_i\in[0,1]$, describes the likelihood that treatment will result in tree recovery. The propensity for the pest to spread in the absence of treatment is given by $\beta$, which integrates the contact rate of pest to new hosts and the propensity for infestations to become established. The recovery rate for infested trees that are treated effectively is given by $\alpha$. The mortality rate of infested trees that are untreated or for which treatment is ineffective is $\gamma$. A complete derivation of the model structure can be found in Supplementary Material Section~\ref{suppCompartmentalModel}.

\begin{figure}[htbp]
    \centering
    \includegraphics[width=.95\textwidth]{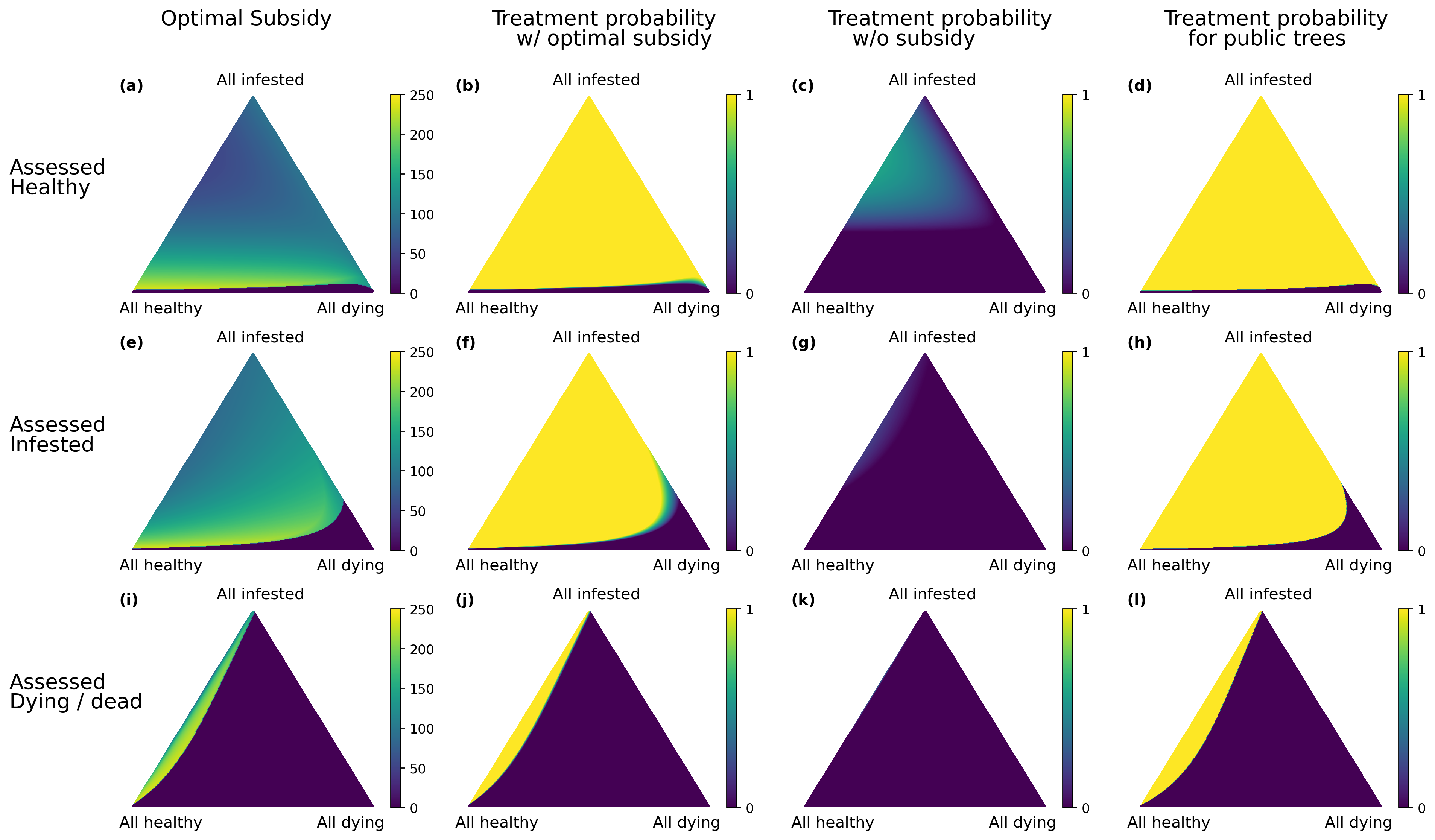}
    \caption{For a social value of avoiding tree mortality, $\Delta_m=1150$, we show simplex plots that illustrate equilibrium policies and outcomes across all possible infestation states ($P(h)+P(i)+P(d)=1$), where the corners of the simplex plots correspond to a state where all trees are in one health class. Each row of panels displays results for a different assessed tree health state. The first column shows how optimal subsidy levels depend on the community infestation state of the pest outbreak. The second column illustrates treatment probabilities for a focal tree with optimal subsidies. The third column illustrates the likelihood of treatment in the absence of subsidies and the last column of plots shows the socially optimal treatment choices for publicly owned trees. For these parameter values, private trees are rarely treated in the absence of subsidies, showing the importance for municipal programs for managing pests on private land.}
    \label{fig:simplex1}
\end{figure}

\section*{Emerald ash borer case study}
In this section, we integrate the our static game theoretic and optimization modeling approaches with our dynamical modeling of pest spread to explore the impact of a range of subsidy and treatment scenarios on the dynamics of Emerald ash borer (EAB). EAB is a non-native insect that attacks and kills North American ash~\citep{herms2014emerald}, an important part of urban forests throughout the United States, planted because of their tolerance to drought and soil compaction~\citep{macfarlane2005characteristics}. Ash trees provide ecological benefits such as wildlife habitat and stormwater absorption, social benefits such as pleasing environments for people, and human health benefits such as reduced stress and cooling~\citep{dwyer1992assessing,pataki2021benefits}. One highly vulnerable ash species, Black ash (\textit{Fraxinus nigra}), is a cultural keystone species for many Native Peoples in North America with spiritual, cultural, and utilitarian significance~\citep{costanza2017precarious,siegert2023biological}. Since EAB was discovered near Detroit, Michigan, and Windsor, Ontario, in 2002, the insect has spread to 36 states in the U.S. and five Canadian provinces ~\citep{EABinfo}, killed millions of ash trees~\citep{herms2014emerald}, and cost homeowners and cities hundreds of millions of dollars in losses of property value and damage mitigation~\citep{kovacs2010cost,aukema2011economic}. In response, cities have developed EAB management plans for public trees, including surveillance of tree health, application of systemic insecticides for tree protection, and preemptive removal of infested or unhealthy trees~\citep{herms2014emerald}. Studies have found that insecticide treatments are effective~\citep{mccullough2011evaluation} at preventing infestation. Simulation and optimization studies of urban ash populations show that emphasizing treatment rather than tree removal can result in greater net social benefits~\citep{mccullough2012evaluation,kovacs2014bioeconomic,kibics2021multistage, sadof2023urban}. However, given that investing in treatment for ash trees is a public good, we expect private tree owners to under-invest in treatment in the absence of public policies that support treatment. We apply our models of optimal subsidies and optimal treatment choices to the case of EAB and assess the extent to which these policies sustain ash health at the community scale. 

We create three treatment scenarios for privately owned trees where either no treatment occurs, no public subsidies are provided for treatment, or where the optimal subsidy policy is in place for privately owned trees. Leveraging our optimization approach for municipally owned trees, we consider two cases: a non-treatment scenario and an optimal treatment scenario. This creates six scenarios (described in Supplementary Material Section~\ref{TreatScenarios}) that we use to illustrate the impact of treatment and subsidy policies on the dynamics of spread of EAB within a landscape of interacting public and private trees. We examine the relative outcomes of a focus on public land, private land, or both in designing treatment programs and examine the impact of intervention timing on maintaining ash populations. To tailor our case study to the realities of EAB, we draw on the literature for estimates of parameter values, summarized in Table~\ref{tab:EABcaseStudy} and described in Supplementary Material Section~\ref{suppEABpars}).

\begin{table}
\small
    \centering
    \begin{tabular}{c|ccccccccc}
    Parameter  & $P(\hat h \mid h)$&$P(\hat i \mid h)$&$P(\hat d \mid h)$&$P(\hat h \mid i)$&$P(\hat i \mid i)$&$P(\hat d \mid i)$&$P(\hat h \mid d)$& $P(\hat i \mid d)$ & $P(\hat d \mid d)$ \\
        Value & 0.89 & 0.1 & 0.01  & 0.49 & 0.5 & 0.01 & 0.01 & 0.19 & 0.8\\\hline  
        Parameter & $\Delta_o$ & $\Delta_m$ & $\Delta'_m$ & $\beta$ & $\gamma$ & $\alpha$ & $\tau^*$ & $\epsilon_h$ & $\epsilon_i$ \\
        Value & $U(675,1100)$ & 1150 & 1850 & 1 & 0.3 & 1 & 3 & 0.97 & 0.5 \\
        
    \end{tabular}
    \caption{EAB case study parameters}
    \label{tab:EABcaseStudy}
\end{table}

To illustrate how these parameter values translate into treatment scenarios, in Figure~\ref{fig:simplex1} we present the optimal subsidy levels and resulting treatment probabilities for private trees in each assessed tree state and for any state of EAB prevalence (by plotting the subsidy levels on a two dimensional simplex displaying the frequencies of healthy, infested and dying trees). This is compared with the treatment probabilities in the absence of subsidies and the optimal treatment choices for public trees. Each row of panels corresponds to a different assessed tree health state. The first column (panels (a), (e), and (i)) shows the optimal subsidy policy for private trees of each assessed health state and the second shows the resulting treatment probabilities, which are used to construct our `optimal private subsidies' treatment scenario for private trees (panels (b), (f), and (j)) . The third column in Figure~\ref{fig:simplex1} (panels (c), (g), and (k))  shows private tree treatment probabilities in the absence of subsidies, highlighting the vast gulf between treatment choices in the presence and absence of optimal subsidies. We use these treatment probabilities to define our `no private subsidies' treatment scenario for private trees. The final column of Figure~\ref{fig:simplex1} (panels (d), (h), and (l)) shows the optimal treatment choices for public trees, which is used to construct our `optimal public treatment' scenario for municipally owned trees. For these parameters, optimal subsidy polices nearly align treatment of public and private trees. This need not be the case, however. There exist cases where privately owned trees are more likely to be treated because the municipality can leverage the willingness to pay of private tree owners and get these trees treated at a lower cost than public trees due to cost sharing between the parties, there also exist cases where public trees are treated at higher rates than private trees, even in the presence of optimal subsidies.

\begin{figure}
     \centering
    \includegraphics[width=.9\textwidth]{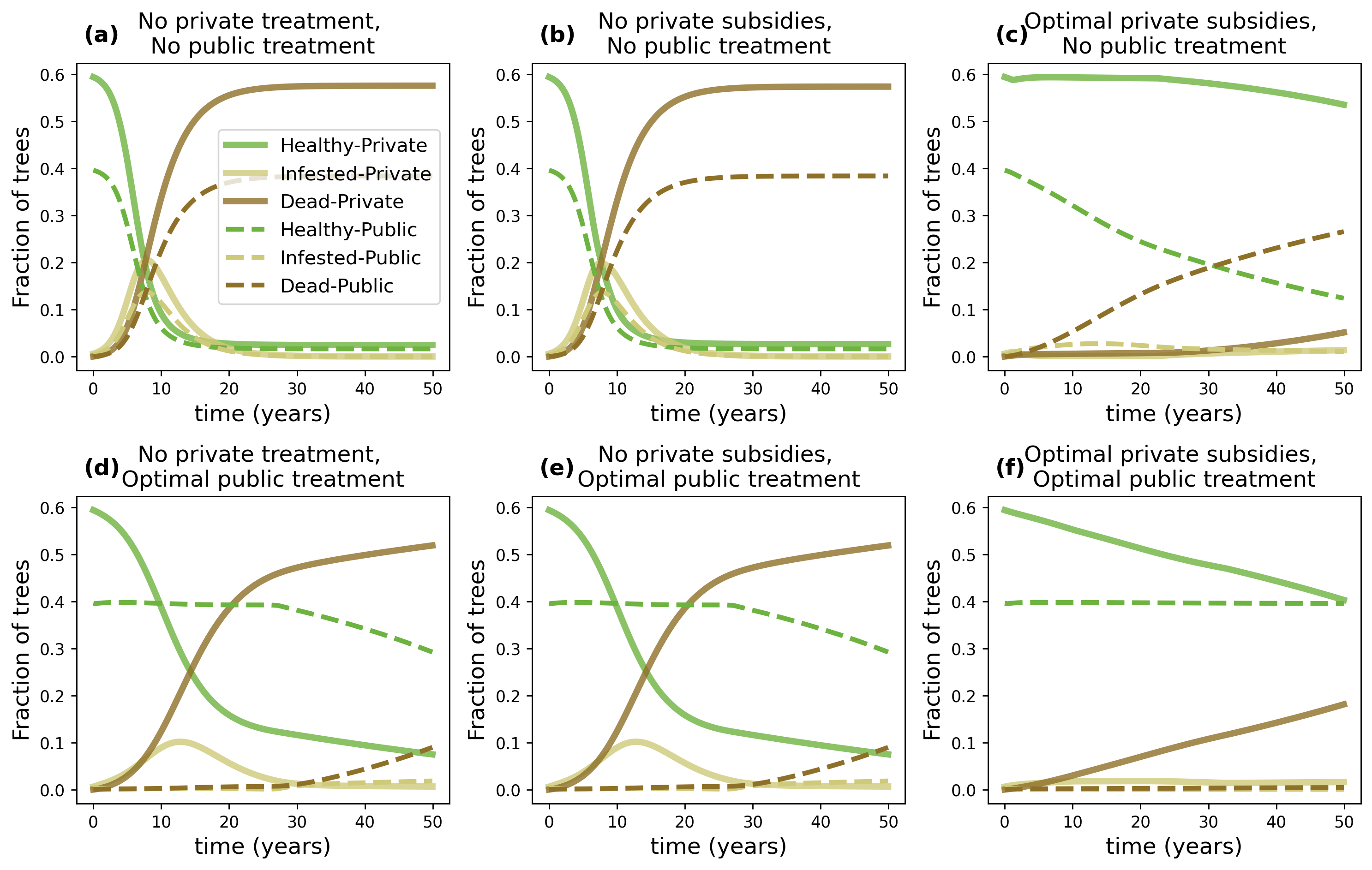}
     \caption{The predicted 50-year dynamics of EAB spread in an urban forest based on the epidemiological model. In each panel the vertical axis is the fraction of trees in a given health and ownership state. The first row (panels (a), (b), and (c)) shows cases where public trees are untreated whereas the second row (panels (d), (e), and (f)) show cases with optimal treatment of public trees.  The first column (panels (a) and (d)) shows simulations where privately owned trees are untreated, the second column (panels (b) and (e)) shows the dynamics of EAB spread when there are no subsidies for the treatment of privately owned trees, and the final column (panels (c) and (f)) shows cases where optimal subsidy levels are set for the treatment of private trees. For these figures, 40\% of ash trees are public and 60\% are privately owned. Panel (a) corresponds to the `no treatment' scenario, panel (b) corresponds to the `no public action' scenario and panel (f) corresponds to the `optimal policies' scenario. Supplementary Material Figure~\ref{supp:fig:dynamics} shows the joint dynamics averaged across public and private trees.}
    \label{fig:dynamics}
     \end{figure}

We have considered optimal policies for treatment of privately owned trees and seen that these policies could increase the rate of treatment of private trees and can result in more similar management of public and private trees. However, our static modeling approaches do not allow for an assessment of the impact of these policies on EAB spread or ash mortality over time. We use the epidemiological model of forest pest dynamics developed to project the dynamics of EAB spread within a community 60\% of ash trees that are privately owned and 40\% are public. We initialize the simulations to reflect the recent arrival of EAB to a community when nearly all trees are EAB-free (99\%) but a few have EAB infestations (1\%). In Figure~\ref{fig:dynamics}, we show six scenarios of public and private treatment decisions that were constructed based on the outcomes of the static game theoretic and optimization models presented in Figure~\ref{fig:simplex1}. When subsidies are not implemented and public trees are untreated (panel (b)), ash mortality is indistinguishable from the no treatment scenario (panel (a)) because treatment rates for private trees are low throughout the simulation. The optimal subsidy policy for private trees combined with optimal treatment choices for public trees (panel (f)) still results in EAB spread, almost entirely among privately owned trees. However, 50-year ash survival at the community level is approximately $80\%$, so the impacts of EAB on aggregate urban forest health in this case will be far more limited that if there was no treatment and over $80\%$ of ash trees died during the first 20 years of infestation. 

\begin{figure}[ht]
     \centering
    \includegraphics[width=.95\textwidth]{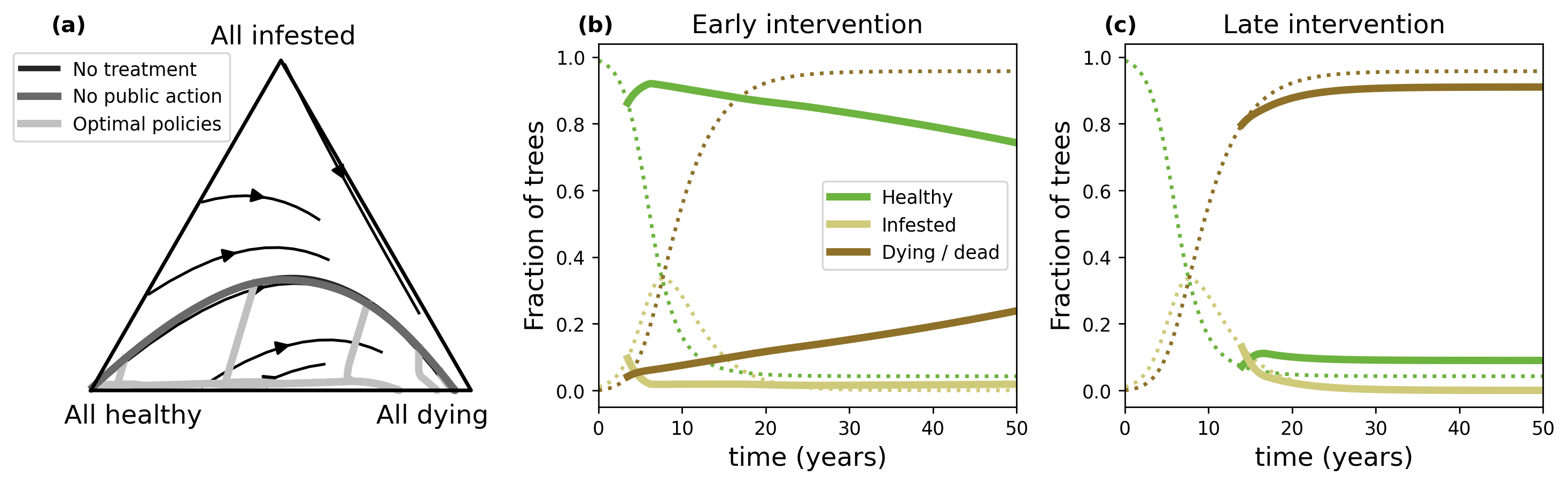}
     \caption{Intervention timing impacts the success of EAB treatment policies.  The panel (a) illustrates the dynamics of EAB under a case where no trees are treated with black arrows that show the direction of dynamics. Rather than using the no treatment case as a baseline, we consider a `no public action' baseline scenario where there are no subsidies for the treatment of private trees and no publicly owned trees are treated (shown in dark grey). Under this `no public action' baseline, private tree owners rarely treat their trees. From this no public action baseline, we simulate the introduction of `optimal policies' for EAB management. Under our `optimal policy' scenario, we assume optimal treatment of public trees (described in Supplementary Material Section~\ref{epiMuniTreat}) and optimal subsidy policies for privately owned trees(described in Supplementary Material Section~\ref{epiPrivTreat}). The impact of intervention timing is illustrated by simulating the introduction of optimal policies at 3.5 year intervals (shown in light grey). The impact of intervention timing on 50-year ash tree survival is shown in panels (b) and (c), where outcomes under cases where public policies are enacted 3.5 years and 14 years after EAB arrival, respectively. In these two panels, dashed lines correspond to the `no public action' baseline while solid lines correspond to the `optimal policy' scenario.}
    \label{fig:simplexDynam}
     \end{figure}

Population dynamics under these scenarios are represented in a simplex plot in panel (a) of Figure~\ref{fig:simplexDynam}. This helps illustrate the relationship between pest dynamics and treatment policies because as an outbreak unfolds the subsidy levels and treatment outcomes change dynamically. The dynamics in the `no treatment' case are shown with arrows and black arcs that wends toward the state where nearly all ash trees die.  For the scenario where there is `no public action' (public trees untreated, private trees unsubsidized), there is little treatment by private tree owners, resulting in an indistinguishable trajectory of pest spread. In this plot, we also show the dynamics that would result after the introduction of `optimal policies' that guide treatment of public trees and subsidies for privately owned trees. Several cases are shown with different intervention timings. Once the municipality implements an optimal subsidy policy and treatment plan, the EAB infestation levels plummet followed by a period of much slower spread (as shown by the light grey lines heading toward the bottom of the simplex). Panels (b) and (c) in Figure~\ref{fig:simplexDynam} show that earlier implementation of the optimal policy results in greatly increased long-run healthy ash populations.

In sum, the integration of game theoretic, optimization, and epidemiological modeling applied to the case of EAB shows how treatment polices designed via analytically tractable static models can enhance the stewardship of ash trees across private and public lands and slow the spread of EAB at community scales. Given the high prevalence of privately owned trees in urban areas, achieving these outcomes necessitates coordinated action on both public and private land. While the implementation of treatment and subsidy programs is costly, our approach assures that subsidies and treatment are only used when the social benefits (via increased ecosystem service provision and reduced spillover mortality risk) exceed the costs of the programs. We show that early implementation of treatment on public land and support for subsidy programs for private tree owners increases ash survival over a 50-year time horizon, extending the life of ash trees long enough that the community level impact of pest-induced mortality will be less noticeable when paired with replanting. This has the potential to greatly reduce the level of disruption that EAB can otherwise cause in urban communities .

\section*{Discussion}
Urban forests and trees provide a wide range of ecosystem services with benefits that accrue from local to global scales. This can be seen in the relationship between tree cover and home values. Trees on a property increase a home's value, but so does having higher tree cover within the neighborhood~\citep{kovacs2022tree}. This illustrates that some ecosystem service benefits of trees are local public goods. Carbon sequestration by trees provides an even starker example because the benefit of reduced atmospheric carbon dioxide is a global public good~\citep{buchholz2021global}. Therefore, sustaining the ecosystem service benefits of privately owned urban trees resembles an impure public goods problem. This has been central to the challenge of achieving sustainability because public goods result in a conflict between the collective good and our individual self-interest~\citep{levin2010crossing,levin2014public}. In the case of urban forest management, this implies that private tree owners may under-invest in management interventions that sustain the ecosystem service benefits of their trees. These issues are especially acute in the context of forest pest management because investments in control have spatiotemporal spillover benefits that reduce risks to other nearby trees. In urban forests, these phenomena arise within a mosaic of public and private lands with many uses and distinct management aims. The confluence of these factors creates a unique challenge for urban landscape-scale management of forest pests. 

While total eradication of a non-native pest species may not be a realistic target~\citep{green2021functional}, slowing the spread of an invasive pest, such as EAB, may still be beneficial~\citep{mccullough2012evaluation}.For example, if the ash trees in a community are decimated over a 5-15 year period, this  significantly disrupts the ecosystem service benefits of trees in areas where ash predominate and imposes major tree removal costs to the public and private tree owners. On the other hand, if these same losses are extended across a 50-year period, then the rate of tree mortality would be closer to the baseline rate within urban areas and the impact of EAB on the provision of ecosystem services would be far less pronounced. This suggests that the both the reduction and delay in mortality seen in our results would be beneficial to communities.

A limitation of our static models is that we extrapolate from an analysis of a representative public or private tree to a community where there are many trees by assuming that the outcomes derived from focusing on one tree can be applied to the whole community. This may not be the case if tree care firms or the municipality face increasing marginal costs of treatment (due to a constrained supply of arborists who can administer treatments, for example). While this could be an important feature for managing emergent pests, it is outside the scope of this work.  Relatedly, our dynamic model is not spatially explicit and instead assumes that all the trees within the community interact. This assumption would be unrealistic if the model were applied at large spatial scales, but may be reasonable at the scale of a community or municipality. A benefit of this simplicity is that it yields intuition into what determines optimal subsidy policies and testing their impact with the dynamical model is not computationally intensive. As such, these models could form the basis of a tool for considering policies for pest treatment in urban areas. 

Recent investments in urban forestry have emphasized environmental justice~\citep{FSpress2023}. While our model doesn't explicitly weigh the distributional effects of treatment subsidies, we can nonetheless consider how the management approach studied in this paper might hinder (or advance) these objectives. The optimal subsidy levels derived in our game theoretic analysis fall into three categories, two of which should not impact inequality. When no subsidy is optimal, then there will be no impact on inequality because no one receives a subsidy. Similarly, when the optimal subsidy is high enough to ensure that all individuals treat their trees we also need not worry about the impact on inequality of the policy because it impacts all ash tree owners the same way. In this case, assuring that the program is being administered equitably could advance environmental justice. 

However, there are cases that could generate feedback effects between inequality and the environment~\citep{Hamann2018}. When an intermediate subsidy level is optimal, individuals with higher valuations of tree avoiding tree mortality opt for treatment and those with lower valuations decline treatment. Research has found that the relationship between socioeconomic factors and urban forests is complex, context dependent, and shaped by legacy effects~\citep{roman2018human,grove2020legacy,koo2023environmental}. However, positive relationships between income and demand for urban forests have been documented~\citep{zhu2008demand,le2017atlanta}. If we assume a positive correlation between socioeconomic indicators and willingness to pay for treatment of ash trees, intermediate subsidy levels could exacerbate inequality because the subsidy will flow to those with higher socioeconomic status. In addition to the direct effect of the subsidy on inequality, there is also the potential for diverging treatment decisions to impact EAB spread and ash decline. For example, in communities with high levels of residential segregation~\citep{massey1988dimensions,logan2017national}, this could manifest as spatial EAB hotspots that disproportionately impact historically marginalized communities and reinforce inequitable patterns of urban tree canopy cover~\citep{Schwarz2015,locke2021residential}. In these cases, it may be worth considering alternative objective functions for the municipality that explicitly weigh distributional effects and equity so that harmful positive feedback effects between inequality and the environment can be avoided.  

The management of forest pests in an urban mosaic of public and private land poses unique public goods problems that can be addressed with carefully designed management interventions. We integrate game theoretic, optimization, and dynamical systems approaches, and uncertainty, to study how forests comprised of interacting public and private ownerships can be cost-effectively sustained in the face of threats from pests. Our case study on EAB illustrates how designing public programs to bolster private incentives for investment in public goods could be employed across a range of management challenges and contribute to collective action for sustainability.

\bibliographystyle{apalike}
\bibliography{biblio}

\section*{Acknowledgments}
The findings and conclusions in this publication are those of the authors and should not be construed to represent any official USDA or U.S. Government determination or policy. Code that generates figures is available at $\href{https://github.com/atilman/EAB-treatment_code}{github.com/atilman/EAB-treatment\_code}$

\appendix
\newpage

\title{\textit{Supplementary Material for:}\\Public policy for management of forest pests within an ownership mosaic}
\maketitle
\renewcommand{\thefigure}{S.M.~\arabic{figure}}
\setcounter{figure}{0}
\renewcommand{\thetable}{S.M.~\arabic{table}}
\setcounter{table}{0}
\spacing{0.5}
\tableofcontents
\singlespacing

\section{Introduction}
Forests pests, such as emerald ash borer (EAB), have had devastating impacts on trees in North American forests~\citep{kovacs2010cost}. In the years since the arrival of EAB, treatments have been developed which dramatically reduce the risk of infestation and death of treated ash trees~\citep{mccullough2011evaluation}. These treatments may also slow the spread of EAB at landscape scales if deployed strategically~\citep{mccullough2012evaluation}. However, achieving coordinated deployment of these treatments is challenging due to the mosaic of forest ownership types in the region, the costs of treatment, and the public goods nature of many of the ecosystem services that mature ash trees provide. 
Due to these factors, in the absence of active intervention on the part of municipalities or land management agencies, private market forces are likely to lead to sub-optimal outcomes from the perspective of society. Similar management challenges can arise for other forest pests and studying the policies for EAB control may shed light on approaches that could improve responses to emerging pest threats.

We develop and analyze a game theoretic model to understand how municipal subsidy programs for treatments to address forest pests on private land can be optimized to achieve cost-effective stewardship of tree populations. We also analyze optimal treatment decisions for public trees and show how policies can be designed to prioritize management across public and private lands. The game theoretic model that we develop to study subsidy policies is based on perceived risks to trees from a forest pest at a single point in time. In order to examine the dynamics of tree infestation and mortality we develop an epidemiological model of public and private trees that are susceptible to a forest pest. We synthesize these two approaches by applying a range of treatment scenarios that result from the game theoretic model to the dynamic epidemiological model of pest spread. As the pest spreads, this alters the subsidy levels and treatment choices that result from the game theoretic model, allowing for the study of the complex feedback effects between treatment, pest spread, and tree mortality.  

\section{Game theoretic model for privately owned trees}~\label{suppModel}
We consider the municipally subsidized treatment of a privately owned tree susceptible to a forest pest by a tree care firm as a dynamic game of incomplete information in three stages. While we focus on a single  representative tree, our results are later applied to a population of trees. The health status of the tree is independently assessed and this status is know by the municipality, the tree care firms and the owner of the tree in question. Next, the municipality chooses subsidies that are available to the tree care firms, with the potential for these subsidy levels to vary across firms and tree health status. The firms then simultaneously submit bids to the tree owner defining the price they will charge for treating the tree. Finally, the tree owner decides whether to accept the bid of either firm, or opt not to treat their tree. We assume that the value of the tree to its owner is drawn from a distribution and that the while the tree owner knows their personal valuation (referred to as their `type'), the municipality and tree care firms only know the distribution from which this valuation is drawn. This introduces incomplete information. However, since the tree owner is the last player to take an action in the game, we do not need to update other players beliefs about the type of the tree owner. Instead, we can simply analyze this dynamic game via backward induction with players at each stage choosing a strategy to maximize their expected utility given the distribution of types of tree owners. This makes our analysis simpler than is often the case for a Perfect Bayesian Equilibrium. First, we introduce key aspects of tree health, assessment, and treatment that underlie the payoffs of the tree owner and the municipality. Then we will derive the expected benefits of treatment from the tree owner and municipal perspective.

\subsection{Tree health assessment}\label{suppAssessment}
We assume that an arborist affiliated with or independent from the tree care firms or the municipality assess the health of a tree and shares this information with the tree care firms, tree owner, and municipality. In the case of EAB, arborists are able to assess infestation with reasonable accuracy by visually observing crown density. For other forest pests, alternative assessment techniques may be used.
In cases where tree assessment is undertaken by the tree care firms, our results will still apply as long as we assume that all firms use the same assessment methodology and thus when any firm that assesses a tree's health, they will reach the same conclusion about its health state state. 

We assume an arborist will assess the health status of a tree in one of three states, 
\begin{equation}
\hat \phi\in\{\hat h,\hat i, \hat d\},
\end{equation}
with the three possible assessed states being: healthy, $\hat h$, infested, $\hat i$, or dead / dying, $\hat d$. 
We denote a tree's true underlying state 
\begin{equation}
\phi\in\{h,i,d\}
\end{equation}
where the true state of the tree can ether be healthy, $h$, infested, $i$, or dead / dying, $d$. 

We assume that surveillance data are available at the municipality scale, with publicly known frequencies of each underlying tree state given by $P(h)$, $P(i)$, and $P(d)$. 

We allow for assessment techniques to be imperfect predictors of the true underlying health status of a tree. We define the accuracy of assessment with probabilities. For example,
\begin{equation}
P(\hat h \mid d)
\end{equation}
is the probability that an dying tree is determined to be healthy by a tree care specialist ($\hat h$ corresponds to the assessed state, whereas $d$ without a hat corresponds to the true underlying state of the tree). There are several of these probabilities, which can be written as 
\begin{equation}
P(\hat \phi \mid \phi)
\end{equation}
for some assessed state, $\hat \phi$ and some true underlying state $\phi$. These assessment accuracy probabilities have been estimated by researchers for EAB~\citep{flower2013relationship}, but our analysis can be applied for arbitrary accuracy values.

\subsection{Inferring underlying tree health}\label{suppBayes}
When a tree owner is deciding whether to treat their tree or when a municipality is setting subsidy policies for trees assessed in state $\hat i$, for example, they need to know how likely it is that the true health of the tree matches or diverges from this assessment. These probabilities can be computed with Bayes' Theorem. For example, 
\begin{equation}
P( h \mid \hat h) = \frac{P(\hat h \mid h)P(h)}{P(\hat h \mid h)P(h)+P(\hat h \mid i)P(i)+P(\hat h \mid d)P(d)}
\end{equation}
is the probability that a tree is healthy given that it was assessed as such by an arborist. There is also a chance that a tree given a healthy assessment is actually infested, with this probability given by
\begin{equation}
P( i \mid \hat h) = \frac{P(\hat h \mid i)P(i)}{P(\hat h \mid i)P(i)+P(\hat h \mid h)P(h)+P(\hat h \mid d)P(d)}.
\end{equation}
The likelihoods of all true underlying health states of a tree given its assessed health can be similarly computed and will be pivotal in defining the expected utility functions of the municipality, firms and tree owner.  

\subsection{Effectiveness of treatment}\label{suppTreatment}
There are two pathways of impact that treatments for a forest pest could have. First, healthy trees could be treated to reduce their risk of becoming infested by a pest. This is akin to vaccination of the tree. We define $\epsilon_h\in[0,1]$ as the effectiveness of treatment at preventing infestation with a pest. If $\epsilon_h=1$ the treatment is 100\% effective and there will be no chance of infestation, and if $\epsilon_h=0$ then treatment will have no impact of the infestation risk of a healthy tree. 
When treatments are applied to an infested tree, they can reverse an infestation and return a tree to a healthy state if they are effective. We define $\epsilon_i\in[0,1]$ as the effectiveness of treatment for an infested tree and it represents the likelihood of tree recovery given treatment. 
These effectiveness values of treatment can be estimated from the literature, with ~\cite{mccullough2011evaluation} finding that the effectiveness of treatment at preventing infestation of a healthy tree is very high, with $\epsilon_h\in[.95,1]$. On the other hand, the effectiveness of treatment at returning a tree to a healthy state is lower and more uncertain, with values ranging from $\epsilon_i\in[.35,.9]$. These values inform the perceived risk of tree mortality that tree owners and municipalities form when they make decisions. 

\subsection{Perceived direct mortality risk}\label{SI-focalRisk}
The effectiveness of treatment can reduce the risk of infestation given exposure to a pest and increase the likelihood of tree recovery for infested trees. When a tree owner or municipality make decisions, we assume they do so based on their expectations of the likelihood of tree survival versus mortality (a binary outcome) over some fixed time horizon. We call these expectations their risk perceptions for a tree and we model the formation of these expectation / perceptions with an approach that is analytically tractable and consistent with epidemiological models of pest spread.

We break down the impact of treatment or non-treatment on tree owner and municipality perceptions of tree mortality risk given that a tree has a healthy, infested, or dying underlying state. For a tree with a healthy or infested state, the perceptions of mortality risk (and conversely survival likelihood) depend on the treatment choice because treatment effects the rates of infestation and tree death. We assume that risk perceptions are formed by forecasting mortality of a tree over a fixed time horizon assuming that the community level of infestation that drives pest spread is fixed at its current level. These assumptions can be used to write a system of equations that describe the dynamics of mortality probability for a tree in a community with $I_0$ equal to the fraction of trees in the community that are currently infested, with $\beta$ the transmission rate of the pest, and with $\gamma$ defined as the death rate from infestation. The system of equations for an untreated focal tree is   
\begin{align}
        &\dot h = -\beta I_0 h  \\
    &\dot i = \beta I_0 h - \gamma i \\
      &\dot d = \gamma i
\end{align}
where $h$, $i$ and $d$ are the probabilities that the focal tree is in the corresponding health state. Because we assume that mortality risk perception is undertaken given the simplifying assumption that $I_0$ is fixed, the system of equations presented above is mathematically equivalent to a radioactive decay chain with two steps. This system of differential equations is among those for which an analytical solution exists, described by the Bateman equation~\citep{bateman1910solution}. We use this solution to derive the expected mortality risk for each health class of untreated trees over a fixed planning horizon. The probability of the focal tree being in a dead/dying health state $\tau$ years into the future given the probabilities of being in each health state at present are $h_0$, $i_0$, and $d_0$ can be written as
\begin{equation}
    d[t] = d_0 + i_0 \left(1 - e^{-\gamma \tau}\right) + h_0 \frac{\beta I_0 \left(1 - e^{-\gamma \tau}\right) - \gamma \left(1 - e^{-\beta I_0 \tau}\right)}{\beta I_0 -
    \gamma}
\end{equation}
for an untreated focal tree. This model reflects the structure of the epidemiological model studied later in this manuscript, but is simplified to apply only to a focal tree and to ignore the potential for community infestation levels to change over time. However, the benefit of this approach is that it is tractable and allows for the estimation of mortality risk by extrapolating dynamics based on present conditions. By setting the initial probability of each health state to 1, we can compute the estimated mortality probability over any time horizon for a tree that has any initial health state. For our analyses, we let the time horizon equal the duration of effectiveness of treatment so that individuals and the municipality make treatment decisions based on the impact of treatment on mortality over the period of efficacy of treatment. We denote the mortality risk for an untreated healthy tree as 
\begin{equation}
    \mu_{uh} = \frac{\beta I_0 \left(1 - e^{-\gamma \tau^*}\right) - \gamma \left(1 - e^{-\beta I_0 \tau^*}\right)}{\beta I_0 -
    \gamma}
\end{equation}
which is equal to $d[\tau]$ when $h_0=1$ and $\tau=\tau^*$ is the duration of effectiveness of treatment. For an infested tree, we have $i_0=1$ and the mortality risk is 
\begin{equation}
    \mu_{ui} = 1 - e^{-\gamma \tau^*}.
\end{equation}
For a tree that is already dying, $d_0=1$, mortality risk is trivially
\begin{equation}
    \mu_{ud} = 1.
\end{equation}

For a treated tree the system of equations that governs risk perceptions of the tree owner and the municipality is
\begin{align}
        &\dot h = -\beta I_0 \left(1-\epsilon_h\right) h  \\
    &\dot i = \beta I_0\left(1-\epsilon_h\right) h - \gamma \left(1-\epsilon_i\right)i \\
      &\dot d = \gamma i
\end{align}
so that as the effectiveness of treatment at protecting healthy trees, $\epsilon_h\in[0,1]$, increases, the risk of a tree becoming infested declines. The rate of mortality is also diminished as treatment effectiveness, $\epsilon_i\in[0,1]$, for infested trees increases. Following the same approach as above, the mortality probability for a focal tree at time $\tau$ years into the future is 
\begin{equation}
    d[t] = d_0 + i_0 \left(1 - e^{-\gamma \left(1-\epsilon_i\right)\tau}\right) + h_0 \frac{\beta I_0 \left(1-\epsilon_h\right) \left(1 - e^{-\gamma \left(1-\epsilon_i\right) \tau}\right) - \gamma \left(1-\epsilon_i\right) \left(1 - e^{-\beta I_0 \left(1-\epsilon_h\right)\tau}\right)}{\beta I_0\left(1-\epsilon_h\right) -
    \gamma\left(1-\epsilon_i\right)}
\end{equation}
which shows how treatment can delay both infestation and mortality for initially healthy trees. 
We denote the mortality risk for a treated healthy tree as 
\begin{equation}
    \mu_{th} = \frac{\beta I_0 \left(1-\epsilon_h\right) \left(1 - e^{-\gamma \left(1-\epsilon_i\right) \tau^*}\right) - \gamma \left(1-\epsilon_i\right) \left(1 - e^{-\beta I_0 \left(1-\epsilon_h\right)\tau^*}\right)}{\beta I_0\left(1-\epsilon_h\right) -
    \gamma\left(1-\epsilon_i\right)}
\end{equation}
which is equal to $d[\tau]$ when $h_0=1$ and $\tau=\tau^*$ is the duration of effectiveness of treatment. For an infested tree, we have $i_0=1$ and the mortality risk is 
\begin{equation}
    \mu_{ti} = 1 - e^{-\gamma \left(1-\epsilon_i\right)\tau^*}.
\end{equation}
For a tree that is already dying, $d_0=1$, mortality risk is trivially
\begin{equation}
    \mu_{td} = 1.
\end{equation}

Studies on the use of insecticides for the treatment of ash trees to prevent and treat EAB infestation find that trunk injection with emamectin benzoate can be effective for 2-4 years~\citep{sadof2022factors}. For main text case study on EAB we use a 3-year time horizon, but this approach of estimating mortality risk can be applied over any horizon.

\subsection{Perceived spillover mortality risk}\label{SI-SpilloverRisk}
A key feature of managing for invasive pests is considering the risk of spread of the pest across the landscape. Management actions taken for a focal tree can influence the risk to other trees in the community and treatment at the community scale may be able to slow or arrest the spread of a forest pest. This potential spillover benefit of treatment can also be estimated using approaches from epidemiology. Whereas the benefit of treatment for a focal tree can have private as well as public benefits, any spillover benefit of treatment on risk reduction for other trees in the community will be a public good and therefore we assume that the municipality but not the tree owner will consider spillover effects of treatment in their decision making. 

We estimate the impact of a focal tree on community tree mortality based on the basic reproduction number of the pest ($R_0$ in epidemiology) in terms of the number of new host trees infested by a single infested focal tree given all other trees are healthy. Based on a simple compartmental epidemiological model of a community's trees, the basic reproduction number of a pest is $R_0=\beta/\gamma$. However, as the fraction of healthy trees declines, so does the effective reproduction number and the expected number of new infestations attributable to a single infested untreated tree becomes $R_{\text{eff}}=\beta H/\gamma$ where $H$ is the fraction of community trees that are healthy.
Because we have assumed that infestation invariably results in mortality, we can define the magnitude of spillover mortality risk for an untreated infested tree as

\begin{equation}
    \lambda_{ui} = \frac{\beta H_0}{\gamma}
\end{equation}
given that the current fraction of healthy trees is $H_0$. While the estimated risk to the focal tree was a probability of mortality, the spillover risk is an expectation of the number of tree mortality events that will be attributable to a focal tree. From the perspective of the municipality, the sum of these risks will be what determines their decision-making about subsidy policies for private trees and treatment choices for public trees. For untreated trees that are healthy, we assume that the municipality estimates the spillover mortality risk by first estimating the likelihood of infestation within the time horizon and then multiplying this by the expected spillover mortality risk given infestation, as shown above. 

Following the methods outlined in section~\ref{SI-focalRisk}, we can compute the expected infestation probability over a time horizon of $\tau^*$ for healthy untreated tree and a current community infestation level of $I_0$ as

\begin{equation}
    i[\tau^*] = 1 - e^{-\beta I_0 \tau^*}
\end{equation}
which can be combined with the spillover risk given infestation, resulting in a spillover mortality risk for a healthy untreated tree of

\begin{equation}
    \lambda_{uh} = \left( 1 - e^{-\beta I_0 \tau^*}\right)\frac{\beta H_0}{\gamma}.
\end{equation}
The epidemiological model we study assumes that dead / dying trees don't contribute to transmission and therefore the spillover mortality attributable to such a tree independent of treatment and is trivially $\lambda_{ud}=\lambda_{td}=0$.

Spillover risk is altered by treatment for healthy and infested trees. We assume that treatment can reduce the infestation probability of a focal tree as well as result in tree recovery. Given that a treated tree is infested, we assume that treatment has no direct impact on the `infectiousness' of the tree. However, treatment can have an indirect effect by altering the expected duration of infestation (e.g. if the recovery rate, $\alpha$, is faster than the mortality rate, $\gamma$). Given these assumptions, the spillover mortality risk for a treated infested tree is 
\begin{equation}
    \lambda_{ti} = \frac{\beta H_0}{\gamma(1-\epsilon_i)+\alpha\epsilon_i}
\end{equation}

Treatment can reduce the expected infestation probability over a time horizon of $\tau^*$ for a healthy treated tree which, given a current community infestation level of $I_0$, is

\begin{equation}
    i[\tau^*] = 1 - e^{-\beta (1-\epsilon_h) I_0 \tau^*}
\end{equation}
which can be combined with the spillover risk given infestation, resulting in a spillover mortality risk for a healthy treated tree of

\begin{equation}
    \lambda_{th} = \left( 1 - e^{-\beta I_0 (1-\epsilon_h)\tau^*}\right)\frac{\beta H_0}{\gamma(1-\epsilon_i)+\alpha\epsilon_i}.
\end{equation}
These estimated spillover mortality risks will be used by the municipality but not the tree owner when estimating the benefits of treatment. 

\subsection{Private value of treatment}\label{suppPriVal}
Given the assessed health of the tree by an arborist, the tree owner can now break their payoff down into six cases: treatment or not, given an assessment that a tree is healthy, infested or dying. The value of the tree to its owner under these four cases is 

\vspace{-20pt}
\begin{align}
\pi_{t\hat h} &= 
    P( h \mid \hat h)\left(\left(1-\mu_{th}\right)V_o-\mu_{th}W_o\right)
    +P( i \mid \hat h)\left(\left(1-\mu_{ti}\right)V_o-\mu_{ti}W_o\right)
    -P( d \mid \hat h)W_o\\
\pi_{t\hat i} &=
    P( h \mid \hat i)\left(\left(1-\mu_{th}\right)V_o-\mu_{th}W_o\right)
     +P( i \mid \hat i)\left(\left(1-\mu_{ti}\right)V_o-\mu_{ti}W_o\right)
    -P( d \mid \hat i)W_o\\
\pi_{t\hat d} &=
    P( h \mid \hat d)\left(\left(1-\mu_{th}\right)V_o-\mu_{th}W_o\right)
     +P( i \mid \hat d)\left(\left(1-\mu_{ti}\right)V_o-\mu_{ti}W_o\right) 
    -P( d \mid \hat d)W_o\\
\pi_{u\hat h} &=
     P( h \mid \hat h)\left(\left(1-\mu_{uh}\right)V_o-\mu_{uh}W_o\right)
    +P( i \mid \hat h)\left(\left(1-\mu_{ui}\right)V_o-\mu_{ui}W_o\right)
    -P(d \mid \hat h)W_o\\
\pi_{u\hat i} &=
     P( h \mid \hat i)\left(\left(1-\mu_{uh}\right)V_o-\mu_{uh}W_o\right)
    +P( i \mid \hat i)\left(\left(1-\mu_{ui}\right)V_o-\mu_{ui}W_o\right)
    -P(d\mid\hat i)W_o\\
\pi_{u\hat d} &=
     P( h \mid \hat d)\left(\left(1-\mu_{uh}\right)V_o-\mu_{uh}W_o\right)
    +P( i \mid \hat d)\left(\left(1-\mu_{ui}\right)V_o-\mu_{ui}W_o\right)
    -P(d\mid\hat d)W_o
\end{align}
where $V_o$ is the value of a surviving tree and $W_o$ is the cost of  tree mortality to a tree owner. Since a tree provide benefits to communities and society as a whole, whereas the costs of care are primarily shouldered by the tree owner, it is likely that a tree owners valuation will not align with the value of the tree in the eyes of a socially-minded municipality.  We find that the private benefit of avoiding the mortality a single tree, 
$$\Delta_o=V_o+W_o,$$
shapes the benefit of treatment from the perspective of a tree owner. We define the value of treatment from the owners perspective as the difference in payoff between treatment and non-treatment for a tree with a particular health assessment. If the tree's assessment indicates infestation, then the benefit of treatment is
\begin{align*}
\pi_{t\hat i} - \pi_{u\hat i} 
            &= (V_o+W_o)\left(
              P(i\mid\hat i)(\mu_{ui}-\mu_{ti})
            +  P(h\mid\hat i)(\mu_{uh}-\mu_{th}) \right)\\
            &= \Delta_o k_{\hat i}
\end{align*}
where 
\begin{equation}
k_{\hat i}=P(i\mid\hat i)(\mu_{ui}-\mu_{ti}) +  P(h\mid\hat i)(\mu_{uh}-\mu_{th})
\end{equation}
is the change in the survival probability of the focal tree when it is treated and its assessed health state is infested and $\Delta_o$ is the value to the tree owner of avoiding tree mortality. Similarly, if  the tree owner is told their tree is healthy, this benefit function is 
\begin{align}
\pi_{t\hat h} - \pi_{u\hat h} = \Delta_o\left(
              P(i\mid\hat h)(\mu_{ui}-\mu_{ti})
            +  P(h\mid\hat h)(\mu_{uh}-\mu_{th}) \right) = \Delta_o k_{\hat h},
\end{align}
where 
\begin{equation}
k_{\hat h} = P(i\mid\hat h)(\mu_{ui}-\mu_{ti}) +  P(h\mid\hat h)(\mu_{uh}-\mu_{th})
\end{equation}
is the change in the survival probability of the focal tree in response to treatment given it has a healthy assessed state over the tree owner's planning time horizon $\tau^*$. 
Finally, we can write benefit of treating a tree reported to be dead or dying as
\begin{align}
\pi_{t\hat d} - \pi_{u\hat d} = \Delta_o\left(P(i\mid\hat d)(\mu_{ui}-\mu_{ti})
            + P(h\mid\hat d)(\mu_{uh}-\mu_{th}) \right) = \Delta_o k_{\hat d},
\end{align}
where 
\begin{equation}
k_{\hat d} = P(i\mid\hat d)(\mu_{ui}-\mu_{ti}) + P(h\mid\hat d)(\mu_{uh}-\mu_{th})
\end{equation}
which shows that the only benefit from treating a tree assessed as dying is if this assessment is incorrect. 

\subsection{Social value of treatment}\label{suppSocVal}
We define the social value of a surviving tree as $V_m$ and the social cost of tree mortality as $W_m$. From the perspective of the municipality, this implies that the value of avoiding the mortality of a single tree is $$\Delta_m=V_m+W_m.$$ 
These values and costs, together with the treatment effectiveness, assessment accuracy, and community level infestation rates can be used to construct the social payoffs of treatment or non-treatment for each assessed tree state, with  
\begin{align*}
    \Pi_{u\hat h} =\hspace{6pt}
     &P( h \mid \hat h)\left[\left(1-\mu_{uh}-\lambda_{uh}\right)V_m-\left(\mu_{uh}+\lambda_{uh}\right)W_m\right]\\
    +&P( i \mid \hat h)\left[\left(1-\mu_{ui}-\lambda_{ui}\right)V_m-\left(\mu_{ui}+\lambda_{ui}\right)W_m\right]\\
    -&P(d \mid \hat h)W_m
\end{align*}
defining the social value of an untreated tree that is assessed as healthy, which accounts for the risk of mortality of the focal tree ($\mu_{u\phi}$)  and the spillover mortality risk for other trees in the community ($\lambda_{u\phi}$). The municipality's payoffs from the other assessed health and treatment combinations can be similarly written, with only subscripts varying. The benefit of treatment from the socially oriented perspective of the municipality, given the assessed  health of the tree can be written as 
\begin{align*}
\Pi_{t\hat h} - \Pi_{u\hat h} &= \Delta_m
              \left(P(i\mid\hat h)(\mu_{ui}-\mu_{ti} + \lambda_{ui}-\lambda_{ti})+ P(h\mid\hat h)(\mu_{uh}-\mu_{th} + \lambda_{uh}-\lambda_{th})  \right) = \Delta_m \left(k_{\hat h} + l_{\hat h}\right)\\
\Pi_{t\hat i} - \Pi_{u\hat i} &= \Delta_m
              \left(P(i\mid\hat i)(\mu_{ui}-\mu_{ti}+ \lambda_{ui}-\lambda_{ti})+ P(h\mid\hat i)(\mu_{uh}-\mu_{th}+ \lambda_{uh}-\lambda_{th}) \right)= \Delta_m \left(k_{\hat i} + l_{\hat i}\right)\\
\Pi_{t\hat d} - \Pi_{u\hat d} &= \Delta_m
              \left(P(i\mid\hat d)(\mu_{ui}-\mu_{ti}+ \lambda_{ui}-\lambda_{ti})+ P(h\mid\hat d)(\mu_{uh}-\mu_{th}+ \lambda_{uh}-\lambda_{th})  \right)= \Delta_m \left(k_{\hat d} + l_{\hat d}\right)        
\end{align*}
for the three possible health assessment outcomes where we define 
\begin{equation}
l_{\hat \phi} = P(i\mid\hat\phi)(\lambda_{ui}-\lambda_{ti}) +  P(h\mid\hat  \phi)(\lambda_{uh}-\lambda_{th})
\end{equation}
as the change in the expected number of surviving trees (other than the focal tree) that results from treatment of a focal tree in some assessed health state $\hat\phi$.

\begin{figure}[ht]
    \centering
    \includegraphics[width=\textwidth]{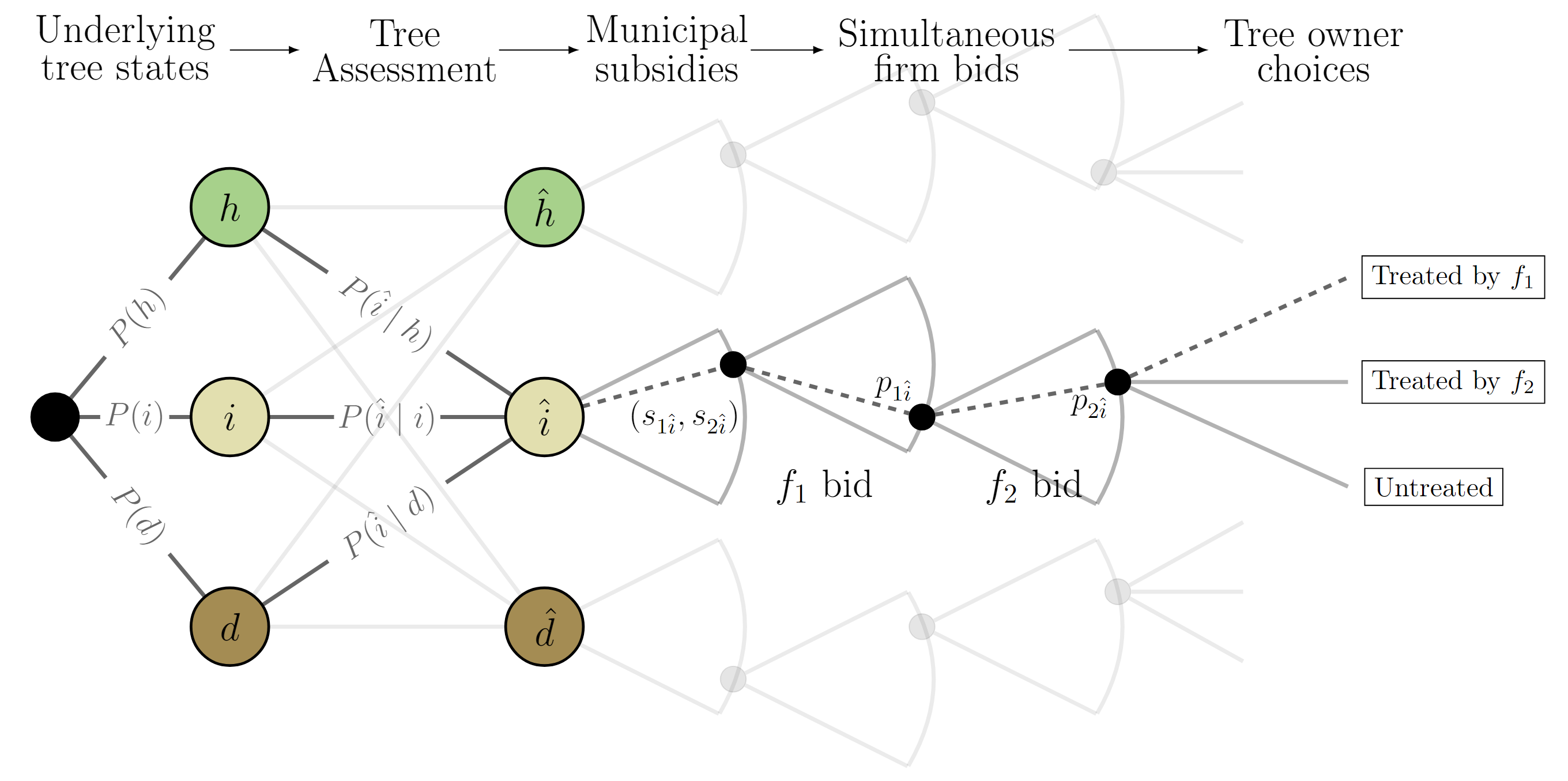}
    \caption{The structure of the game theoretic model, from assessment of a tree's health to treatment outcomes. A focal tree can fall into one of three states with likelihoods given by community level forest pest prevalence. Next, the health status of the tree is assessed, with assessment errors possible. This splits the game into three branches, one for each assessed tree state. The game that follows for each assessed health state can be analyzed independently. We highlight the case of a tree that has been assessed as being in an infested state. First, the municipality sets subsidy levels that each firm will receive if they treat a tree with this assessed state. The arc in the figure at this stage represents the continuous range of subsidy pairs that the municipality can set for the firms. Next, knowing the assigned subsidy levels, the firms simultaneously submit bids for treatment of a focal tree, \`a la Bertrand competition. Finally, the tree owner will decide whether to treat their tree and which firm's bid to select. Here we show a case where firm 1 offers a lower bid, and the tree owner decides to have their tree treated by firm 1.}
    \label{SI-fig:conceptual}
\end{figure}

\subsection{Game structure}\label{suppGame}
Once a tree has had its health status assessed, we subsequently consider a game in three stages, as illustrated in Figure~\ref{SI-fig:conceptual}. First, the municipality chooses subsidy levels, $s_{1\hat\phi}$ and $s_{2\hat\phi}$ that are available to firm 1, ($f_1$) and firm 2, ($f_2$) for the treatment of a tree assessed in as being in state $\hat\phi$. Next, the tree firms simultaneously submit bids to the tree owner for the price they will charge for treating the tree, with the bids of firm 1 and 2 being $p_{1\hat\phi}$ and $p_{2\hat\phi}$, respectively. Finally, the tree owner decides whether to accept the bid of either firm, or opt not to treat their tree. While the tree owner knows that their value of treatment given the assessed status of the tree is $\Delta_o k_{\hat\phi}$, the firms and municipality only know that $\Delta_o$ is uniformly distributed from $a$ to $b$, i.e.
\begin{equation}
\Delta_o\sim U[a,b].
\end{equation}

\subsection{Strategy spaces}\label{suppStrat}
The strategy space of the municipality is defined as the set of non-negative subsidy pairs that they could offer to the firms for treatment of a tree with an assessed health state $\hat\phi$,
\begin{equation}
    \sigma_{m\hat\phi} = \{(s_{1\hat \phi},s_{2\hat \phi}) \mid s_{1\hat\phi}\geq0,s_{2\hat\phi}\geq0\}.
\end{equation}
The strategy space of firm 1 is the set of non-negative prices that they could offer for treatment of the tree,
\begin{equation}
    \sigma_{f_1\hat\phi} = \{p_{1\hat \phi} \mid p_{1\hat\phi}\geq0\},
\end{equation}
with a similar strategy space for firm 2, given by
\begin{equation}
    \sigma_{f_2\hat\phi} = \{p_{2\hat \phi}\mid p_{2\hat\phi}\geq0\}.
\end{equation}
Finally, the strategy of the tree owner is given by a mapping between their type, $\Delta_o$, which is drawn from a uniform distribution and the probability that they choose any of their actions
\begin{equation}
    A_{o\hat\phi} = \{\text{treat w/}f_1, \text{treat w/}f_2, \text{untreated}\}.
\end{equation}
We denote a mapping that forms their strategy as 
\begin{equation}
    g_{o\hat\phi}:[a,b]\to P(A_{o\hat\phi})
\end{equation}
so that a strategy $g_{o\hat\phi}$ will assign a probability distribution over the set $A_{o\hat\phi}$ for every possible value of $\Delta_o\in[a,b]$. Therefore the strategy space of the tree owner can be written as the set of possible strategies
\begin{equation}
    \sigma_{o\hat\phi}=\left\{g_{o\hat\phi}:[a,b]\to P(A_{o\hat\phi})\right\}.
\end{equation}

\subsection{Utility functions}\label{suppUtil}
Given the value treated and untreated trees, we construct the utility functions for the municipality. We assume that subsidies are paid at a cost to the municipality. This leads to a utility function for the municipality given by 
\begin{equation}
U_{m\hat\phi} = \begin{cases}
    \Pi_{t\hat \phi}-s_{1\hat \phi}, & \text{if tree assessed as $\hat \phi$ is treated by firm 1}\\
    \Pi_{t\hat \phi}-s_{2\hat \phi}, & \text{if tree assessed as $\hat \phi$ is treated by firm 2}\\
    \Pi_{u\hat \phi} & \text{if tree assessed as $\hat \phi$ is untreated.}
\end{cases}
\end{equation}

The utility function for firm 1 depends on the the cost of administering the treatment, $c$, the price they bid, $p_{1\hat\phi}$ and the level of the subsidy that they receive, $s_{1\hat\phi}$. This leads to a utility function given by 
\begin{equation}
U_{f_1\hat\phi} = \begin{cases}
    p_{1\hat\phi}+s_{1\hat\phi}-c, & \text{if firm 1 treats tree}\\
    0, & \text{if firm 1 does not}
\end{cases}
\end{equation}
for firm 1. A similar function holds for firm 2, since we assume that the costs for each firm are the same. This leads to a utility function of
\begin{equation}
U_{f_2\hat\phi} = \begin{cases}
    p_{2\hat\phi}+s_{2\hat\phi}-c, & \text{if firm 2 treats tree}\\
    0, & \text{if firm 2 does not.}
\end{cases}
\end{equation}

Finally, we consider the utility of a private tree owner, which will depend on their valuation of a treated or untreated tree and the price they pay for treatment. These valuations can be used to construct the utility function of the tree owner as
\begin{equation}
U_{o\hat\phi} = \begin{cases}
    \pi_{t\hat \phi}-p_{1\hat\phi}, & \text{if firm 1 treats the tree}\\
    \pi_{t\hat \phi}-p_{2\hat\phi}, & \text{if firm 2 treats the tree}\\
    \pi_{u\hat \phi}, & \text{if the tree is untreated}.
\end{cases}
\end{equation}

\section{Analysis}\label{suppAnalysis}
The analysis of the game in this case will proceed via backward induction, starting from determining the optimal strategy of a tree owner in response to an arbitrary pair of bids and working back to the subsidy policy that maximizes the expected utility of the municipality. We assume that the tree owner chooses their strategy to maximize their utility, that the firms offer treatment prices to maximize their expected utility, given when they know about the distribution of $\Delta_o$, and that the municipality maximizes their expected utility given the bidding strategies they expect firms to employ and the distribution of $\Delta_o$. 

\subsection{Tree owner behavior}\label{suppOwner}
We assume that the tree owner knows the assessed state of their tree is some $\hat\phi\in\{\hat h, \hat i, \hat d\}$. They simultaneously receive a pair of bids from the tree care firms for treatment of the tree, $(p_{1\hat\phi},p_{2\hat\phi})$. Any type of tree owner with type $\Delta_o\in[a,b]$ will maximize their utility by following the actions defined by the strategy mapping
\begin{equation}
g_{o\hat\phi}^*\left[\Delta_o\right] = \begin{cases}
     \text{treat w/}f_1 & \text{if } p_{1\hat\phi}<p_{2\hat\phi} \text{  and  } p_{1\hat\phi}\leq \Delta_o k_{\hat\phi}\\
    \text{treat w/}f_2 &\text{if } p_{2\hat\phi}<p_{1\hat\phi} \text{  and  } p_{2\hat\phi}\leq \Delta_o k_{\hat\phi}\\
    \text{flip coin} &\text{if } p_{1\hat\phi}=p_{2\hat\phi} \text{  and  }  p_{2\hat\phi}\leq \Delta_o k_{\hat\phi}\\
    \text{don't treat} &\text{if } \Delta_o k_{\hat\phi}<\text{min}\{p_{1\hat\phi},p_{2\hat\phi}\}
\end{cases}
\end{equation}
where the `flip coin' action corresponds to the case where the tree owner treats their tree but randomly selects one of the firms. As desired, this strategy mapping assumes that the tree owner knows their own valuation of avoiding tree mortality, $\Delta_o$, and also knows the impact that treatment will have, given the tree health assessment that they receive, $k_{\hat\phi}$. Together, these define the benefit of treatment of a tree, $\Delta_o k_{\hat\phi}$, and only when this benefit is at least as great as the price offered by the lowest bidder will the owner choose to have their tree treated.

\subsection{Firm behavior}\label{suppFirms}
We assume that the firms utility functions mirror each other and therefore we can compute the best response for firm 1 and apply this to firm 2 with subscripts switched. We assume that each firm seeks to maximize their expected utility given the bid of the opposing firm, their beliefs about the value of treatment to the tree owner, and the assessed state of the tree. This optimization assumes an arbitrary subsidy level is offered to their firm for treatment.

First, we will construct the expected utility function of firm 1. The firms utility depends on how profitable treatment is if they win the bid and how likely it is that they win the bid. This leads to
\begin{equation}
E[U_{f_1}] = \left(p_{1\hat\phi}+s_{1\hat\phi}-c\right)P(f_1 \text{ wins bid})
\end{equation}
for the expected utility of firm 1 which can be further broken down by defining the probability function of firm 1 winning the bid. The probability of winning the bid depends on the bid of the other firm and the beliefs of the firm about the benefit of treatment to the tree owner. 

There are several cases for this probability function. First, if $p_{1\hat\phi}>p_{2\hat\phi}$ then firm 1 will never win the bid. If $p_{1\hat\phi}<p_{2\hat\phi}$ then firm 1 will win the bid if $p_{1\hat\phi}<\Delta_o k_{\hat\phi}$. If the two bids are equal, we assume for simplicity that the odds that firm 1 wins the bid are half of what they would be if they had the lowest bid. However, the results of the analysis still hold for any case where the tree owner has a non-zero probability of selecting each of the equal bids. Putting this together, we have 
\begin{equation}
P(f_1 \text{ wins bid}) = \begin{cases}
     P\left(p_{1\hat\phi}\leq \Delta_o k_{\hat\phi}\right) & \text{if } p_{1\hat\phi}<p_{2\hat\phi} \\
    1/2\;P\left(p_{1\hat\phi}\leq \Delta_o k_{\hat\phi}\right) & \text{if } p_{1\hat\phi}=p_{2\hat\phi} \\
    0 &\text{if } p_{1\hat\phi}>p_{2\hat\phi} 
\end{cases}
\end{equation}
which can be further broken down by calculating the likelihood that a bid would be acceptable to the tree owner. First, recalling that it is common knowledge that the value of avoiding tree mortality to its owner, $\Delta_o\sim U[a,b]$, follows a uniform distribution, then the firm's expectations about the distribution of the benefit of treatment to the tree owner, given the assessed health of tree is 
\begin{equation}
\Delta_o k_{\hat\phi}\sim U[ak_{\hat\phi},bk_{\hat\phi}],
\end{equation}
which can be used to calculate the probability that the firms bid would be low enough to have a chance of being accepted. This probability function is
\begin{equation}
P\left(p_{1\hat\phi}\leq \Delta_o k_{\hat\phi}\right)
=\begin{cases}
     1 &  p_{1\hat\phi}\leq ak_{\hat\phi} \\
    \frac{bk_{\hat\phi}-p_{1\hat\phi}}{k_{\hat\phi}(b-a)} &  ak_{\hat\phi}<p_{1\hat\phi}<bk_{\hat\phi} \\
    0 & bk_{\hat\phi}\leq p_{1\hat\phi} .
\end{cases}
\end{equation}

This can be used to construct the expected utility function of firm 1, which is 
\begin{equation}
E[U_{f_1}] = \begin{cases}
     p_{1\hat\phi}+s_{1\hat\phi}-c &  p_{1\hat\phi}\leq ak_{\hat\phi} \text{ and } p_{1\hat\phi} < p_{2\hat\phi} \\
    \left(p_{1\hat\phi}+s_{1\hat\phi}-c\right)\frac{bk_{\hat\phi}-p_{1\hat\phi}}{k_{\hat\phi}(b-a)} &  ak_{\hat\phi}<p_{1\hat\phi}<bk_{\hat\phi} \text{ and } p_{1\hat\phi} < p_{2\hat\phi}\\
    1/2\left(p_{1\hat\phi}+s_{1\hat\phi}-c \right) &  p_{1\hat\phi}\leq ak_{\hat\phi} \text{ and } p_{1\hat\phi} = p_{2\hat\phi}  \\
    1/2\left(p_{1\hat\phi}+s_{1\hat\phi}-c\right)\frac{bk_{\hat\phi}-p_{1\hat\phi}}{k_{\hat\phi}(b-a)} &  ak_{\hat\phi}<p_{1\hat\phi}<bk_{\hat\phi} \text{ and } p_{1\hat\phi} = p_{2\hat\phi}  \\
    0 & bk_{\hat\phi}\leq p_{1\hat\phi} \text{ or } p_{1\hat\phi} > p_{2\hat\phi}
\end{cases}
\end{equation}
We will first restrict our consideration to those cases where the second firm's bid is higher, and derive firm 1's optimal behavior given this assumption.
The optimal bid can be computed by taking the partial derivative of the expected utility with respect to $p_{1\hat\phi}$, which yields 
\begin{equation}
\frac{\partial E[U_{f_1}]}{\partial p_{1\hat\phi}} = \begin{cases}
     1 &  p_{1\hat\phi}\leq ak_{\hat\phi} \\
    \frac{bk_{\hat\phi}-s_{1\hat\phi}+c-2p_{1\hat\phi}}{k_{\hat\phi}(b-a)} &  ak_{\hat\phi}<p_{1\hat\phi}<bk_{\hat\phi} \\
    0 & bk_{\hat\phi}\leq p_{1\hat\phi} 
\end{cases}
\end{equation}
and shows that when firm 1 knows they have the lowest bid they will never set $p_{1\hat\phi} < ak_{\hat\phi}$ since this only reduces their profitability without increasing their chances of treating a tree. Also, if there is to be an optimal bid that falls in the middle case above, it would be 
\begin{equation}
p_{1\hat\phi}=1/2\left(bk_{\hat\phi}+c-s_{1\hat\phi}\right)
\end{equation}
assuming that firm 1 knows they are the lowest bidder, because this is the bid where $\frac{\partial E[U_{f_1}]}{\partial p_{1\hat\phi}} = 0$ given that $ak_{\hat\phi}<p_{1\hat\phi}<bk_{\hat\phi}$. However, care must be taken to assure that this bid does indeed fall within the relevant range of the state space. 

In sum, an uncontested firm 1, which doesn't have to worry about competing with firm 2 has an optimal bid policy of 
\begin{equation}
 p_{1\hat\phi}^* = \begin{cases}
 ak_{\hat\phi} 
    & c+bk_{\hat\phi}-2ak_{\hat\phi}<s_{1\hat\phi}\\
    
     1/2\left(bk_{\hat\phi}+c-s_{1\hat\phi}\right) 
     & c-bk_{\hat\phi}\leq s_{1\hat\phi}\leq c+bk_{\hat\phi}-2ak_{\hat\phi} \\
     p_{1\hat\phi}\geq c-s_{1\hat\phi} 
     &  s_{1\hat\phi}< c-bk_{\hat\phi} \\
\end{cases}
\end{equation}
This optimal bid policy will be useful later when considering the case where only one firm receives a subsidy. It also will contribute to the analysis of how firm 1 should respond to an arbitrary bid placed by firm 2. 

In order to construct the best response relation for firm 1, first consider the case where the subsidy to firm 1 is so low that they can not profitably win the bid, regardless of what the other firm does. This is the case where $s_{1\hat\phi}< c-bk_{\hat\phi}$ and in this case we will assume that the firm chooses some bid $p_{1\hat\phi}\geq c-s_{1\hat\phi}$. While any bid greater than that of the second firm will lead to the same payoff of zero, we assume that the first firm will limit their bids to those that would not lead to negative utility should they win the bid. 

Now we will turn our attention to the cases where subsidies, costs of treatment and the benefits of treatment to tree owners are such that it is feasible for a firm to win a profitable bid. These are the cases where $s_{1\hat\phi} \geq c-bk_{\hat\phi}$. 
If firm 2's bid is high, then firm 1 won't need to competitively bid. Instead, firm 1 will follow their uncontested optimal bidding policy defined as $p_{1\hat\phi}^*$. Because we are in the case where subsidies are sufficient to induce potentially winning bids, we know that one of the first two conditions on $p_{1\hat\phi}^*$ will result. Therefore, for $s_{1\hat\phi} \geq c-bk_{\hat\phi}$ and $p_{1\hat\phi}^*< p_{2\hat\phi}$, the best response of firm 1 is $p_{1\hat\phi}^*$. 

If the bid of firm 2 is so low that firm 1 can not bid less than firm 2 and have a chance of a positive payoff, then any bid greater than that of firm 2 will lead to the same outcome of 0 payoff, whereas a bid lower than that of firm 2 will lead to a negative expected payoff. We further restrict the bids of firm 1 in these cases to bids that would not lead to negative payoffs should they for some reason end up winning the bid. Therefore, if  $p_{2\hat\phi}<c-s_{1\hat\phi}$, the best response of firm 1 will be any bid $p_{1\hat\phi}\geq c-s_{1\hat\phi}$.

For intermediate bids of firm 2, that fall between $c-s_{1\hat\phi}$ and $p_{1\hat\phi}^*$, there is the potential for competition among the firms. If firm 1 bids higher than firm 2 in this case, they will receive a payoff of zero. On the other hand, any bid greater than $c-s_{1\hat\phi}$ but strictly less than $p_{2\hat\phi}$ will lead to a positive expected payoff. These are bids that are less than that which firm 1 would bid were there no competition from firm 2, and therefore firm 1 will want to choose the highest price is that falls within this range. However, if they match the bid of firm 2, they will win the bid with half the probability that they would if they just marginally decreased their bid. Therefore, in this intermediate case, firm 1's best response is to slightly undercut the bid of firm 2 by setting their price at $p_{2\hat\phi}-\epsilon$. In this setting there always exists some very small $\epsilon>0$ that leads to higher expected utility than $\epsilon=0$. 

These three cases can be aggregated and used to construct the best response of firm 1 to an arbitrary bid by firm 2. We will define 
\begin{equation}
    p_{1\hat\phi}'=\max\left\{\frac{1}{2}\left(bk_{\hat\phi}+c-s_{1\hat\phi}\right),ak_{\hat\phi} \right\}
\end{equation}
as the uncontested bid of firm 1, that they would choose if they were able to act as a monopolist. We use this to write the best response as 
\begin{equation}
\hat p_{1\hat\phi}\left(p_{2\hat\phi}\right) = \begin{cases}
p_{1\hat\phi}\geq c-s_{1\hat\phi} 
     &  s_{1\hat\phi}< c-bk_{\hat\phi} \text{    or    } s_{1\hat\phi}<c-p_{2\hat\phi}\\  
 p_{2\hat\phi}-\epsilon 
    &s_{1\hat\phi} \geq c-bk_{\hat\phi} \text{  and  }  c-s_{1\hat\phi}\leq p_{2\hat\phi} \leq p_{1\hat\phi}'\\
     p_{1\hat\phi}'
     & s_{1\hat\phi} \geq c-bk_{\hat\phi} \text{  and  } p_{1\hat\phi}'<p_{2\hat\phi} 
\end{cases}
\end{equation}
which shows that if the subsidy is so low that either firm 1 cannot profitably win the bid or profitably undercut firm 2, then any large enough bid is a best response. On the other hand, if the bid of firm 2 is not competitive, then firm 1 will act like a monopolist and select their bid, $p_{1\hat\phi}'$ which aligns with the bidding policy we calculated as $p_{1\hat\phi}^*$. For the intermediate cases, firm 1 will undercut the bid of firm 2 to guarantee that they win the bid because winning the bid will lead to positive profit, whereas losing the bid leads to nothing and bidding equal to the other firm leads to half as much profit. In these cases, expected profit decreases as their bid price decreases because the loss of profit of treating a tree at a lower price is not offset by the increased likelihood that the tree owner will accept the bid. Therefore the firm will seek to minimally undercut its competitor. 

We focused on the best response of firm 1 to an arbitrary bid of firm 2. However, the firm's have symmetric utility functions and therefore the best response of firm 2 will exactly mirror that of firm 1 with subscripts switched, given by
\begin{equation}
\hat p_{2\hat\phi}\left(p_{1\hat\phi}\right) = \begin{cases}
p_{2\hat\phi}\geq c-s_{2\hat\phi} 
     &  s_{2\hat\phi}< c-bk_{\hat\phi} \text{    or    } s_{2\hat\phi}<c-p_{1\hat\phi}\\  
 p_{1\hat\phi}-\epsilon 
    &s_{2\hat\phi} \geq c-bk_{\hat\phi} \text{  and  }  c-s_{2\hat\phi}\leq p_{1\hat\phi} \leq p_{2\hat\phi}'\\
     p_{2\hat\phi}'
     & s_{2\hat\phi} \geq c-bk_{\hat\phi} \text{  and  } p_{2\hat\phi}'<p_{1\hat\phi} \\
\end{cases}
\end{equation}
where we define 
\begin{equation}
    p_{2\hat\phi}'=\max\left\{\frac{1}{2}\left(bk_{\hat\phi}+c-s_{2\hat\phi}\right),ak_{\hat\phi} \right\}
\end{equation}
as the price the firm 2 would bid if they have a chance to treat the tree profitably and expect the other firm not to submit a competitive bid. 

\begin{figure}
    \centering
    \includegraphics[width=.45\textwidth]{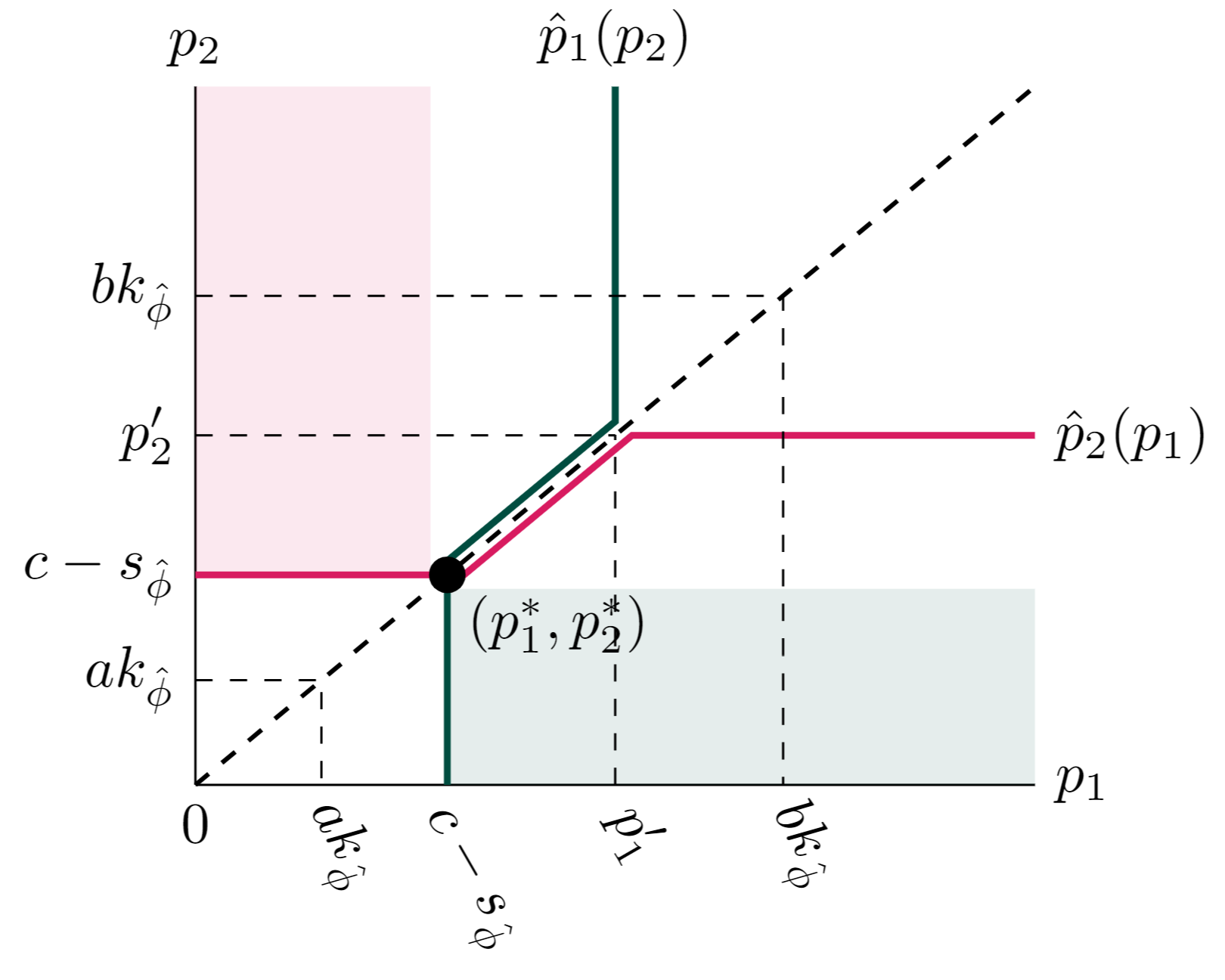}
    \caption{The best response of firm 1, to an arbitrary price set by firm 2, $\hat{p}_1(p_2)$, is plotted along with the best response of firm 2 to an arbitrary price set by firm 1, $\hat{p}_2(p_1)$. Nash equilibria occur at intersections of the best response relations. The willingness to pay for treatment of a tree by its owners is uniformly distributed between $ak_{\hat\phi}$ and $bk_{\hat\phi}$ and if the low bid by one of the firms is less than the realized willingness to pay then the tree owner will accept that bid. Each firm receives a subsidy, $s_{\hat\phi}$ when they treat a tree assessed in health state $\hat\phi$. In the absence of competition, each firm would act as a monopolist and charge a price $p'$ for treatment. Also, neither firm will bid less than $c-s_{\hat\phi}$ because $c$ is the marginal cost of treatment for the firm. If they bid less than their subsidized marginal cost they would be worse off treating the tree. The shaded regions are those regions where a firm is indifferent among all bids in the region. In this case the shaded regions do not intersect and therefore bids in these regions will not be part of a Nash equilibrium. For intermediate bids of their competitor, a firm has an incentive to undercut the other bid. Putting this together allows one to construct the best response relations for each firm and show that, in this case, there is a unique Nash equilibrium: these best response relations only intersect at only one point, marked on this plot. ($ak_{\hat\phi}=1.5$, $bk_{\hat\phi}=7$ $c=10$, $s_{1\hat\phi}=s_{2\hat\phi}=s_{\hat\phi}=7$)}
    \label{fig:NashEQ}
\end{figure}

\begin{figure}[htbp]
     \centering
     \begin{subfigure}[ht]{0.3\textwidth}
         \centering
         \captionsetup{width=.9\linewidth}
         \includegraphics[width=\textwidth]{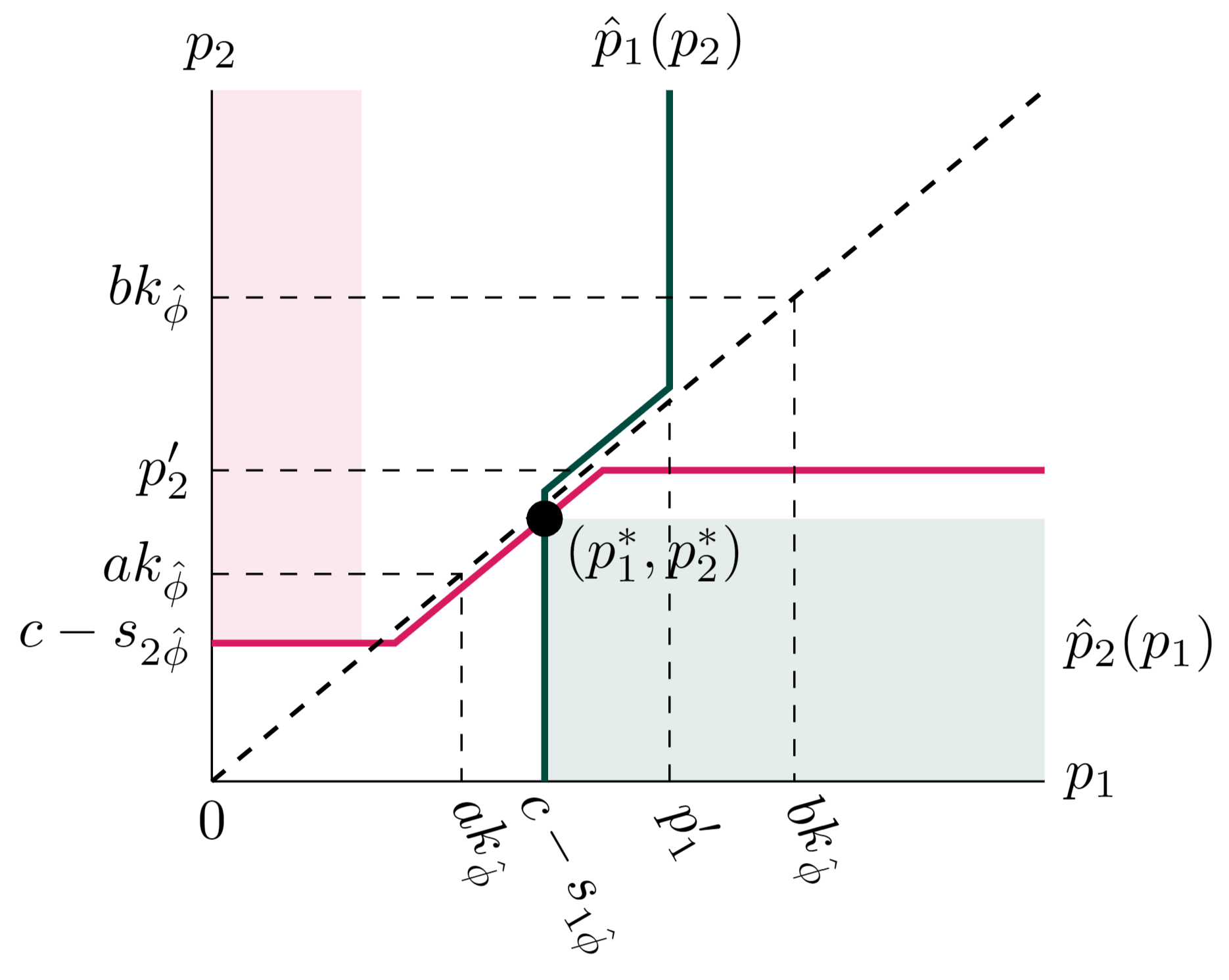}
         \caption{Firm 2's bid undercuts firm 1 ($ak_{\hat\phi}=3$, $bk_{\hat\phi}=7$ $c=10$, $s_{1\hat\phi}=6$, $s_{2\hat\phi}=8$)}
         \label{fig:c_low}
     \end{subfigure}
     \begin{subfigure}[ht]{0.32\textwidth}
         \centering\captionsetup{width=.9\linewidth}
         \includegraphics[width=\textwidth]{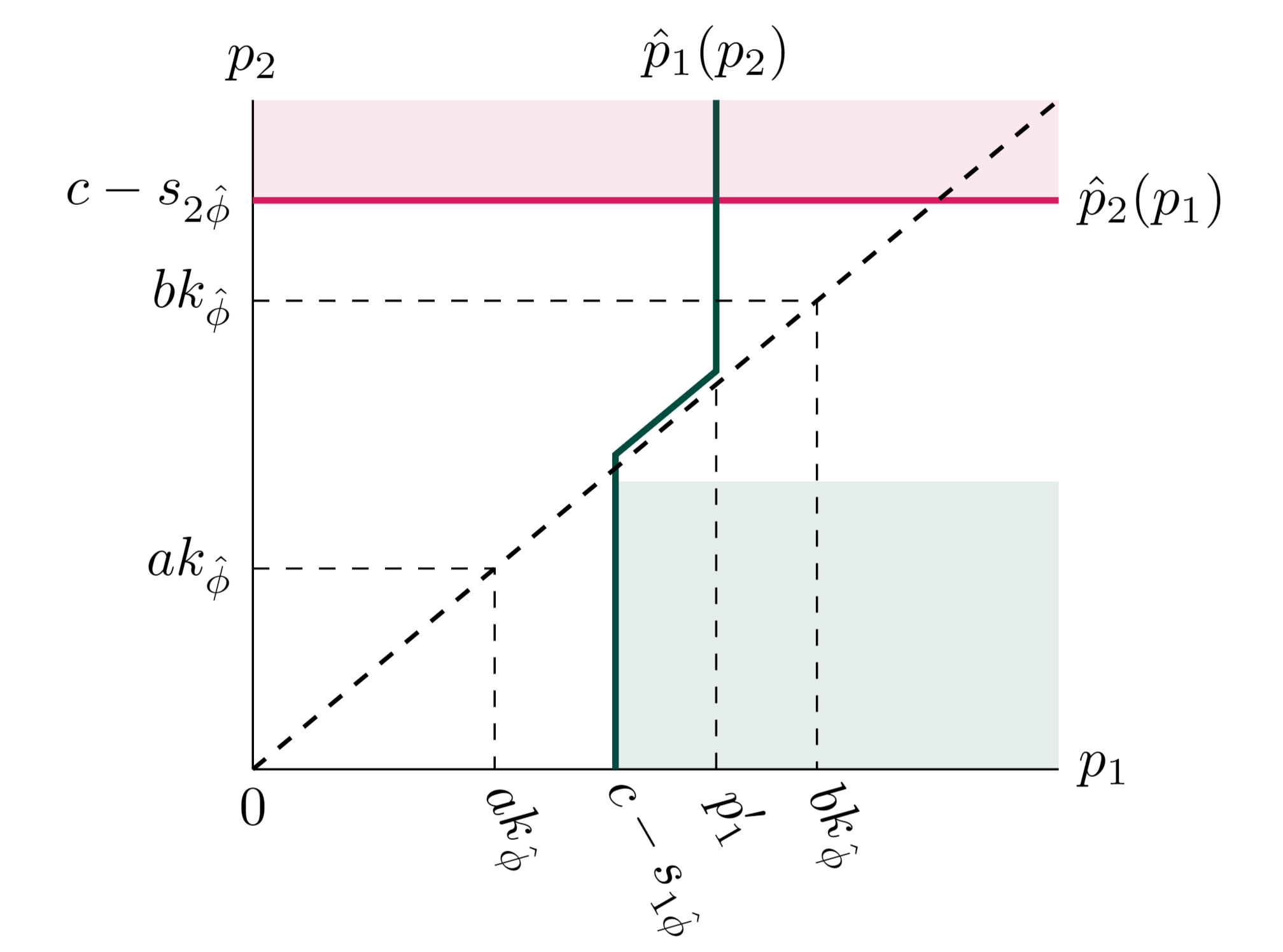}
         \caption{Firm 1 acts as a monopolist ($ak_{\hat\phi}=3$, $bk_{\hat\phi}=7$ $c=10$, $s_{1\hat\phi}=5.5$, $s_{2\hat\phi}=1.5$)}
         \label{fig:2c}
     \end{subfigure}
          \begin{subfigure}[ht]{0.3\textwidth}
         \centering\captionsetup{width=.9\linewidth}
         \includegraphics[width=\textwidth]{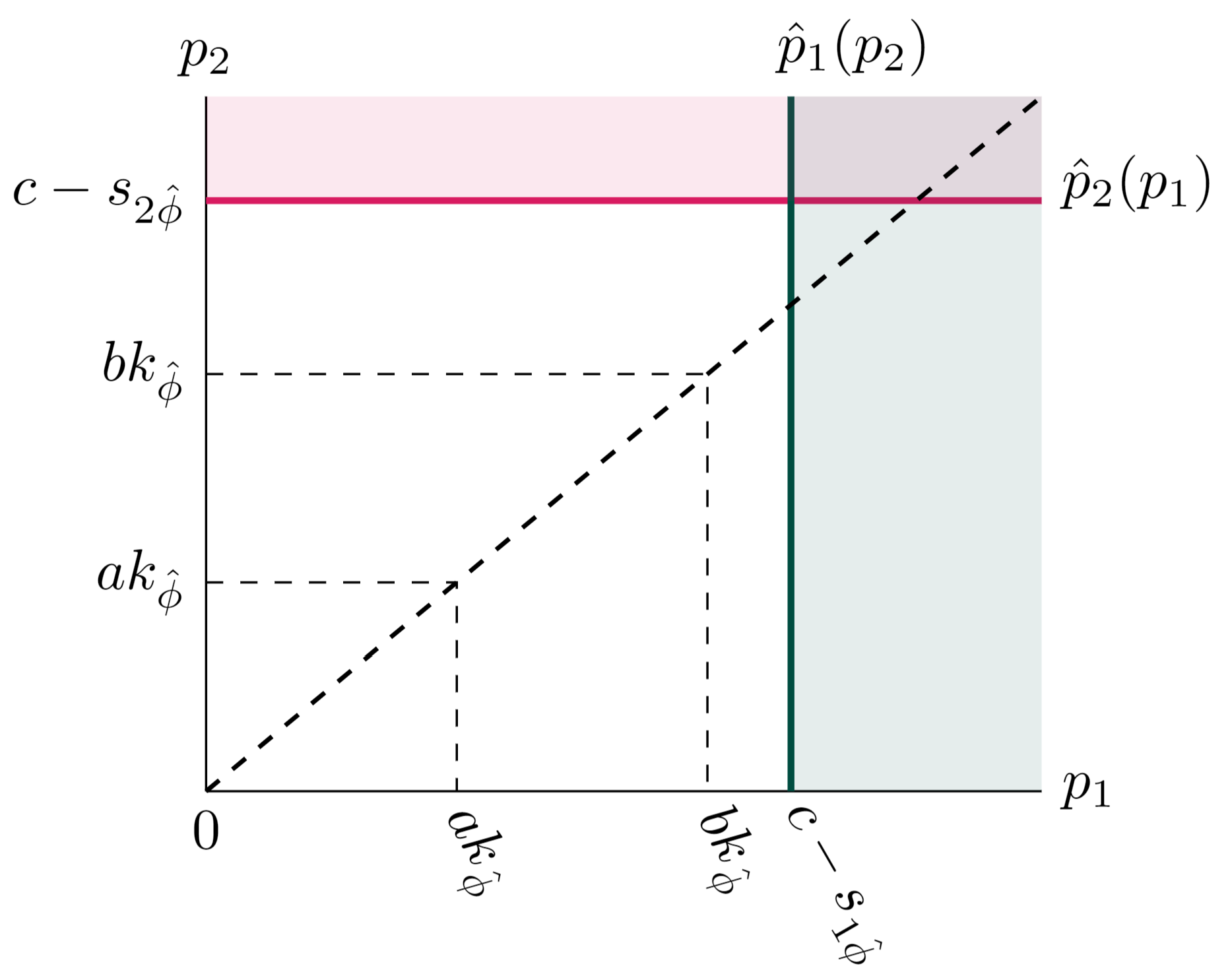}
         \caption{A range of equilibrium bids  ($ak_{\hat\phi}=3$, $bk_{\hat\phi}=6$ $c=10$, $s_{1\hat\phi}=3$, $s_{2\hat\phi}=1.5$)}
         \label{fig:2d}
     \end{subfigure}
        \caption{When the municipality sets unequal subsidy levels for the two firms, a much broader range of equilibrium bid pairs and outcomes can arise. Three such cases are illustrated here. First, we show a case where firm 2 receives a higher subsidy than firm 1 and bids a price that marginally undercuts the lowest bid that firm 1 can make without losing money on treating a tree. The second case corresponds to an outcome where the subsidy level of firm 2 is significantly lower than firm 1. In this case, firm 1 can bid as if they are a monopolist. Finally, we illustrate a case where neither firm receives a subsidy high enough to win a bid and make a profit. In this case the firms are largely indifferent about their bidding, and the tree is never treated. While the outcomes under unequal subsidies are more varied than those that arise under equal subsidy levels, we show that it is never in a municipality's interest to choose unequal subsidy levels for the two firms.}
        \label{fig:unequalSubs}
\end{figure}

Nash equilibrium bids occur when pairs of strategies are mutual best responses. In other words, at intersections of the two firms best response correspondences. There are many possible Nash equilibria in cases where the subsidy levels provided to the two firms by the municipality differ. Some of these are illustrated in Figure~\ref{fig:unequalSubs}, where sometimes the Nash equilibrium is unique and other times there is a continuum of Nash equilibria. However, the Nash equilibrium bids that the firms will select when subsidies are equal, $s_{\hat\phi}=s_{1\hat\phi}=s_{2\hat\phi}$, follow a simple pattern and are
\begin{equation}
 \left(p_{1\hat\phi}^*, p_{2\hat\phi}^*\right) = \begin{cases}
 (c-s_{\hat\phi}, c-s_{\hat\phi})
    & s_{\hat\phi}\geq c-bk_{\hat\phi}\\
    
     (p_{1\hat\phi}\geq c-s_{\hat\phi}, p_{2\hat\phi}\geq c-s_{\hat\phi}) 
     & s_{\hat\phi}< c-bk_{\hat\phi}
\end{cases}
\label{priceEQ}
\end{equation}
with the details of the second case being irrelevant to outcomes because these bids will never lead to treatment of the tree. We decline to compute the conditions that for all possible Nash equilibria under unequal subsidies because we later show that these cases are of little importance; it is never in a municipality's best interest to select unequal subsidies for the firms. 

\subsection{Municipality behavior}\label{suppMuni}
When the municipality sets subsidy levels for the two firms, we assume that it does so to maximize its expected utility. First, we will show that the municipality will never find it advantageous to set different subsidies for each firm. After we show this, we will present the analysis of optimal subsidy choice assuming ex-ante that equal subsidies will be chosen. 

\paragraph{Unequal subsidies}
When subsidies are unequal, we cannot rely on the equilibrium bids derived in the previous section, because these bids corresponded to the case of equal subsidies. First, we consider the case where $ s_{1\hat\phi}<c-bk_{\hat\phi}$ and $ s_{2\hat\phi}< c-bk_{\hat\phi}$. 
From the best responses of the firms we know that subsidy levels this low will lead to bids of $p_{1\hat\phi}\geq c-s_{\hat\phi}$ and , $p_{2\hat\phi}\geq c-s_{\hat\phi}$, which will not lead to treatment of the tree. Therefore, from the perspective of the municipality, in this case their expected utility is $E[U_{m}] = \Pi_{u\hat\phi}$, which does not depend on the subsidy levels. Therefore for low subsidy levels that do not lead to treatment, there is no incentive to set subsidy levels that are unequal. The municipality might as well choose $s_{1\hat\phi}=s_{2\hat\phi}$. 

Therefore, if there is a situation where unequal subsidies are preferable, it will have to occur when either $ s_{1\hat\phi}\geq c-bk_{\hat\phi}$ or $ s_{2\hat\phi}\geq c-bk_{\hat\phi}$. We will proceed by way of contradiction and assume without loss of generality that there exists a setting in which the municipality finds it optimal to choose $s_{1\hat\phi}\geq s_{2\hat\phi}$. 

Given that $s_{1\hat\phi}\geq s_{2\hat\phi}$ and $ s_{1\hat\phi}\geq c-bk_{\hat\phi}$, we can write the best responses of the firms as 
\begin{equation}
\hat p_{1\hat\phi}\left(p_{2\hat\phi}\right) = \begin{cases}
c-s_{1\hat\phi}     &   p_{2\hat\phi}<c-s_{1\hat\phi}\\
p_{2\hat\phi}-\epsilon & c-s_{1\hat\phi}<p_{2\hat\phi}<p_{1\hat\phi}'\\
p_{1\hat\phi}' &p_{1\hat\phi}'<p_{2\hat\phi}
\end{cases}
\end{equation}
 with 
\begin{equation}
    p_{1\hat\phi}'=\max\left\{\frac{1}{2}\left(bk_{\hat\phi}+c-s_{1\hat\phi}\right),ak_{\hat\phi} \right\}
\end{equation}
and
\begin{equation}
\hat p_{2\hat\phi}\left(p_{1\hat\phi}\right) = \begin{cases}
c-s_{2\hat\phi} 
     &  s_{2\hat\phi}< c-bk_{\hat\phi} \text{    or    } p_{1\hat\phi}<c-s_{2\hat\phi}\\  
 p_{1\hat\phi}-\epsilon 
    &s_{2\hat\phi} \geq c-bk_{\hat\phi} \text{  and  }  c-s_{2\hat\phi}\leq p_{1\hat\phi} \leq p_{2\hat\phi}'\\
     p_{2\hat\phi}'
     & s_{2\hat\phi} \geq c-bk_{\hat\phi} \text{  and  } p_{2\hat\phi}'<p_{1\hat\phi} \\
\end{cases}
\end{equation}
where we have 
\begin{equation}
    p_{2\hat\phi}'=\max\left\{\frac{1}{2}\left(bk_{\hat\phi}+c-s_{2\hat\phi}\right),ak_{\hat\phi} \right\}
\end{equation}
as the price the firm 2 would bid if they have a chance to profitably treat the tree and expect the other firm not to submit a competitive bid. 
This can lead to two types of equilibrium bid pairs of, 
\begin{equation}
 \left(p_{1\hat\phi}^*, p_{2\hat\phi}^*\right) = \begin{cases}
 (p_{1\hat\phi}', c-s_{2\hat\phi})
    & c-s_{2\hat\phi}> p_{1\hat\phi}' \\
    
     (c-s_{2\hat\phi}-\epsilon , c-s_{2\hat\phi}) 
     & c-s_{2\hat\phi}\leq p_{1\hat\phi}'
\end{cases}
\end{equation}
given our assumptions.
This makes clear that $p_{1\hat\phi}^*< p_{2\hat\phi}^*$ so if the tree is treated it will be by firm 1. Therefore we can write the expected utility of the municipality as
\begin{equation}
E[U_{m}] = \Pi_{u\hat\phi} +\left(\Delta_m \left(k_{\hat\phi}+l_{\hat\phi}\right)-s_{1\hat\phi}\right)P\left(p_{1\hat\phi}^*\leq \Delta_o k_{\hat\phi}\right)
\end{equation}
which breaks into four cases based on the two types of equilibrium bid pairs and the two different values of $p_{1\hat\phi}'$ that can arise. Writing expected utility in terms of these four cases yields
\begin{equation}
E[U_{m}] = \begin{cases}
 
\Pi_{u\hat\phi} +\frac{\left(\Delta_m \left(k_{\hat\phi}+l_{\hat\phi}\right)-s_{1\hat\phi}\right)\left(bk_{\hat\phi}-c+s_{1\hat\phi}\right)}{2k_{\hat\phi}(b-a)}
    
    & c-p_{1\hat\phi}'>s_{2\hat\phi}  \text{ and } c+bk_{\hat\phi}-2ak_{\hat\phi}\geq s_{1\hat\phi} \\
    
\Pi_{u\hat\phi} +\frac{\left(\Delta_m \left(k_{\hat\phi}+l_{\hat\phi}\right)-s_{1\hat\phi}\right)\left(bk_{\hat\phi}-c+\epsilon+s_{2\hat\phi}\right)}{k_{\hat\phi}(b-a)}
     
     & c-p_{1\hat\phi}'\leq s_{2\hat\phi} \text{ and } c+bk_{\hat\phi}-2ak_{\hat\phi}\geq s_{1\hat\phi}\\
     
\Pi_{u\hat\phi} +\Delta_m \left(k_{\hat\phi}+l_{\hat\phi}\right)-s_{1\hat\phi}
    
    & c-p_{1\hat\phi}'>s_{2\hat\phi}  \text{ and } c+bk_{\hat\phi}-2ak_{\hat\phi}<s_{1\hat\phi} \\
    
\Pi_{u\hat\phi} +\Delta_m \left(k_{\hat\phi}+l_{\hat\phi}\right)-s_{1\hat\phi}
     
     & c-p_{1\hat\phi}'\leq s_{2\hat\phi} \text{ and } c+bk_{\hat\phi}-2ak_{\hat\phi}< s_{1\hat\phi}
\end{cases}
\label{eq:MuniUtil}
\end{equation}
where the first case corresponds to equilibrium where $p_{1\hat\phi}^*=\frac{1}{2}\left(bk_{\hat\phi}+c-s_{1\hat\phi}\right)$, the second case corresponds to the case where the subsidy level for firm 2 is high enough to induce some competition and thus $p_{1\hat\phi}^*=c-s_{2\hat\phi}-\epsilon$. The final two case are where the subsidy level of firm 1 is high enough that the tree will be treated with certainty. In case three,  $p_{1\hat\phi}^*=ak_{\hat\phi}$ and in the final case the equilibrium bid of firm 1 is again, $p_{1\hat\phi}^*=c-s_{2\hat\phi}-\epsilon$. But for these final two cases, this difference in price does not impact the municipality. Both cases lead to equivalent expected utility. 

The municipality chooses subsidy levels that maximize expected utility. Subsidy pairs that lead to either of the final two cases will never be optimal, because in both these cases, expected utility is decreasing in $s_{1\hat\phi}$. Lower values of $s_{1\hat\phi}$ will be chosen and this will move the bid pair out of the parameter space of these cases as they are defined by having large values of $s_{1\hat\phi}$. Therefore, we know that an optimum with $s_{1\hat\phi}>s_{2\hat\phi}$ must fall within the first two cases of equation~\ref{eq:MuniUtil}. 

If we look at the first case in equation~\ref{eq:MuniUtil}, the expected utility of the municipality does not depend on the subsidy level that firm 2 would receive if they treat the tree. Furthermore, on the boundary between case 1 and case 2, when $ s_{2\hat\phi}=c-p_{1\hat\phi}'$, the utility in both of the first two cases is equivalent. Therefore any level of utility that can be attained within case 1 can also arise in case 2. Therefore and optimum with $s_{1\hat\phi}>s_{2\hat\phi}$ must arise in the region of subsidy space where $c-p_{1\hat\phi}'\leq s_{2\hat\phi} \text{ and } c+bk_{\hat\phi}-2ak_{\hat\phi}\geq s_{1\hat\phi}$. Across this region, the expected utility of the municipality is decreasing in $s_{1\hat\phi}$ and increasing in $s_{2\hat\phi}$. Decreasing the value of $s_{1\hat\phi}$ and increasing the value of  $s_{2\hat\phi}$ will always stay within the parameter region that defines case 2. However, this implies that there cannot be an optimum where $s_{1\hat\phi}>s_{2\hat\phi}$ because we showed that such an optimum would have to occur within case 2 of the expected utility function and that within this case the municipality will always want to decrease $s_{1\hat\phi}$ and $s_{2\hat\phi}$ given $s_{1\hat\phi}>s_{2\hat\phi}$. We conclude that a municipality will never find it optimal to set unequal subsidies for the two firms.

\begin{figure}[htbp]
     \centering
     \begin{subfigure}[ht]{0.44\textwidth}
         \centering
         \includegraphics[width=.9\textwidth]{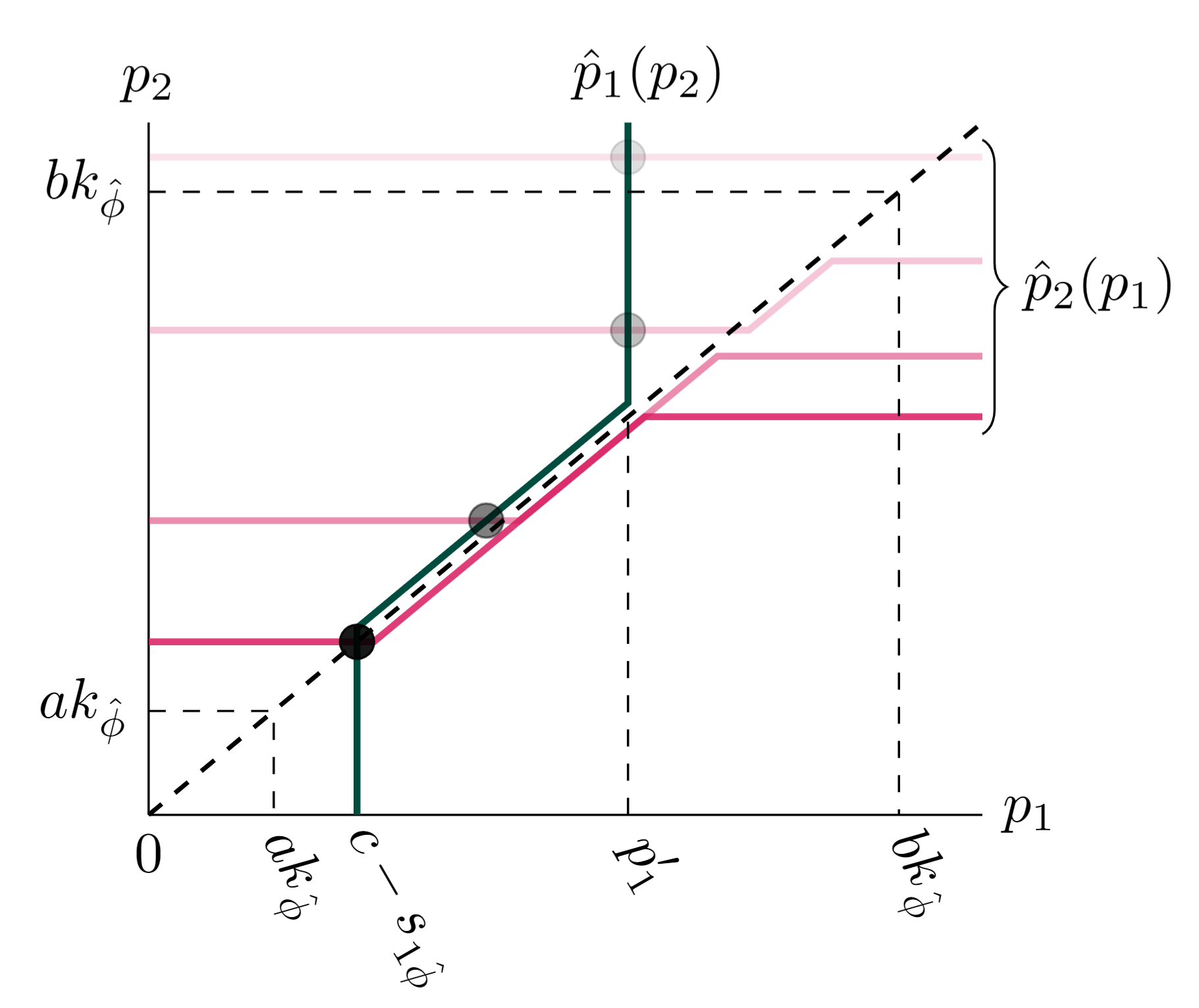}
         \caption{The best response of firm 1 for $s_{1\hat\phi}=7.5$ plotted against four best responses of firm 2 with $s_{2\hat\phi}\in\{0.5,3,5.75,7.5\}$. Progressively darker curves correspond to higher subsidy levels for firm 2. Equilibrium bid pairs are marked with circles. Notice that for $s_{2\hat\phi}<s_{1\hat\phi}$ the circles are in the region of the plane where $p_1<p_2$ and therefore firm 2 will never win the bid. This means that whenever the bid is awarded the municipality pays firm 1 $s_{1\hat\phi}=7.5$. Increasing $s_{2\hat\phi}$ toward $s_{1\hat\phi}$ leads to lower equilibrium prices, increased probability of treatment and no change to the per-tree subsidy paid. Therefore municipalities have an incentive to select equal subsidies for the two firms.  }
         \label{suppfig:1a}
     \end{subfigure}
     \begin{subfigure}[ht]{0.54\textwidth}
         \centering
         \captionsetup{width=.9\linewidth}
         \includegraphics[width=.95\textwidth]{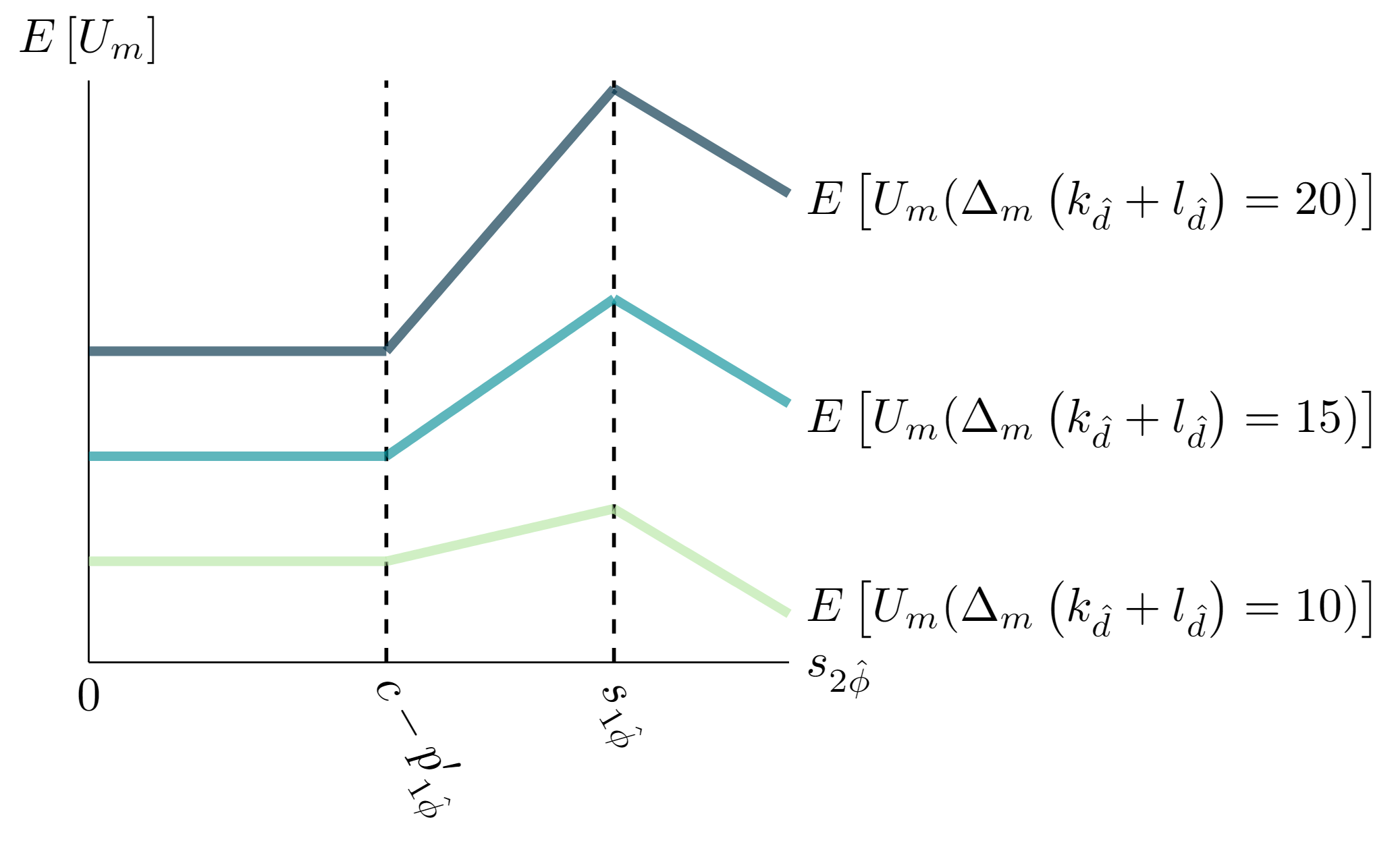}
         \caption{The expected utility of the municipality is shown as a function of the subsidy level of firm 2, $s_{2\hat\phi}$, given three different values of $\Delta_m \left(k_{\hat\phi}+l_{\hat\phi}\right)$. Each of these curves illustrate that for a fixed value of $s_{1\hat\phi}$, expected utility peaks at $s_{2\hat\phi}=s_{1\hat\phi}$. In this case, for $s_{2\hat\phi}<c-p'_{1\hat\phi}$ firm 1 can behave like a monopolist and set their bid price at $p'_{1\hat\phi}$. For values of $s_{2\hat\phi}>s_{1\hat\phi}$ the municipality pays increasing subsidies to firm 2 but this doesn't result in lower prices. This reduces the expected utility of the municipality. For intermediate subsidy levels for firm 2, increasing $s_{2\hat\phi}$, increases expected utility of the municipality because it leads to increased treatment likelihoods without impacting marginal costs to the municipality.}
         \label{suppfig:1b}
     \end{subfigure}
        \caption{($s_{1\hat\phi}=7.5$, $ak_{\hat\phi}=1.5$, $bk_{\hat\phi}=9.5$, $c=10$, $\Pi_{u\phi}=1$)}
        \label{suppfig:uneq}
\end{figure}
Intuition for the result can be understood by looking at the best response relations for gradually increasing values of $s_{2\hat\phi}$, illustrated in Figure~\ref{suppfig:uneq}. At first, increasing $s_{2\hat\phi}$ has no effect on the lowest price bid, but eventually, this increase leads firm 1 to have to lower their price. This increases the likelihood that the tree will be treated but has no impact on the subsidy that the municipality will have to pay to firm 1. Therefore, the municipality will want to keep increasing $s_{2\hat\phi}$ up to the point that the two subsidy levels are equal, because when they are both equal, the likelihood of treatment can be maximized at the least cost, due to competition among the firms. We now turn to the analysis of the optimal equal subsidy level. 

\paragraph{Analysis given equal subsidies}
We will begin by constructing the expected utility function of the municipality when $s_{1\hat\phi}=s_{2\hat\phi}=s_{\hat\phi}$, which is
\begin{equation}
E[U_{m}] = \Pi_{u\hat\phi} +\left(\Delta_m \left(k_{\hat\phi}+l_{\hat\phi}\right)-s_{\hat\phi}\right)P_{t\hat\phi o}
\end{equation}
where $P_{t\hat\phi o}$ is the probability that the focal privately owned tree will be treated, given the subsidies the municipality offers. This probability of treatment depends on the valuations of tree's by their owners, the bids of the firms, as well and how a firm's bidding is impacted by the subsidy level that the municipality chooses. This probability can be broken down as
\begin{equation}
P_{t\hat\phi o}=P\left(\min\left\{p_{1\hat\phi}^*, p_{2\hat\phi}^*\right\}\leq \Delta_o k_{\hat\phi}\right)
\end{equation}
which, given equal subsidies $s_{\hat\phi}\geq0$, is equivalent to 
\begin{equation}
P_{t\hat\phi o}=P\left(c-s_{\hat\phi}\leq \Delta_o k_{\hat\phi}\right)=
\begin{cases}
    1 &  c-ak_{\hat\phi}\leq s_{\hat\phi}\\
    \frac{bk_{\hat\phi}-c+s_{\hat\phi}}{k_{\hat\phi}(b-a)} & c-bk_{\hat\phi} < s_{\hat\phi}< c-ak_{\hat\phi}\\
     0 & s_{\hat\phi}\leq c-bk_{\hat\phi}
\end{cases}
\end{equation}
becuase the firms will offer equal prices equal to $c-s_{\hat\phi}$.
With this, we can rewrite the expected utility function of the municipality as 
\begin{equation}
E[U_{m}]=
\begin{cases}
     \Pi_{u\hat\phi} + \Delta_m \left(k_{\hat\phi}+l_{\hat\phi}\right) - s_{\hat\phi} &  c-ak_{\hat\phi}\leq s_{\hat\phi}\\
    \Pi_{u\hat\phi} +\left(\Delta_m \left(k_{\hat\phi}+l_{\hat\phi}\right)-s_{\hat\phi}\right)\frac{bk_{\hat\phi}-c+s_{\hat\phi}}{k_{\hat\phi}(b-a)} & c-bk_{\hat\phi} < s_{\hat\phi}< c-ak_{\hat\phi}\\
     \Pi_{u\hat\phi} & s_{\hat\phi}\leq c-bk_{\hat\phi}
\end{cases}
\end{equation}
where the first case corresponds to a subsidy level which is high enough to induce all tree owners to treat their tree. The second corresponds to the case where there is a positive probability that the tree will be treated and the final case is when subsidies are set to a level which is too low to induce treatment by any tree owners. We will analyze the optimal behavior of the municipality across three cases and subsequently synthesize the results.

\paragraph{Case 1: $c<ak_{\hat\phi}$}
In this case, the cost of administering treatment is low enough that firms could offer prices and profitably treat trees in the absence of subsidies even for tree owners with the lowest value of avoiding mortality of a tree. Competition among firms will result in bids that will lead to the tree being treated with certainty for any subsidy level. This corresponds to the first case from the expected utility function of the municipality. In this case, utility declines in the level of the subsidy, and therefore the municipality will set their subsidy, $s^*_{\hat\phi}=0$. 

\paragraph{Case 2: $c\in \left[ak_{\hat\phi},bk_{\hat\phi}\right]$}
In this case, the cost of treatment is such that in the absence of any subsidy, there will be a chance that a tree will be treated, given the optimal bidding policies of the firms. Since we restrict our analysis to non-negative subsidy levels, only the first two cases of the expected utility function can occur. The optimal subsidy can be determined from the partial derivative of the utility function with respect to the subsidy level, which is 
\begin{equation}
\frac{\partial E[U_{m}]}{\partial s_{\hat\phi}}=
\begin{cases}
     -1 &  c-ak_{\hat\phi}\leq s_{\hat\phi}\\
    \frac{\Delta_m \left(k_{\hat\phi}+l_{\hat\phi}\right)-bk_{\hat\phi}+c-2s_{\hat\phi}}{k_{\hat\phi}(b-a)} & 0 \leq s_{\hat\phi}< c-ak_{\hat\phi}.
\end{cases}
\end{equation}
This sets an upper bound on the optimal subsidy level of $c-ak_{\hat\phi}$. The lower case has an optimum at $s_{\hat\phi}=1/2\left(\Delta_m \left(k_{\hat\phi}+l_{\hat\phi}\right)-bk_{\hat\phi}+c\right)$
and this falls within the range of the lower case for intermediate values of $\Delta_m \left(k_{\hat\phi}+l_{\hat\phi}\right)$. In sum, the optimal subsidy policy for the municipality is
\begin{equation}
s^*_{\hat\phi}=
\begin{cases}
     c-ak_{\hat\phi} &  c+bk_{\hat\phi}-2ak_{\hat\phi}\leq \Delta_m \left(k_{\hat\phi}+l_{\hat\phi}\right)\\
    1/2\left(\Delta_m \left(k_{\hat\phi}+l_{\hat\phi}\right)-bk_{\hat\phi}+c\right) & bk_{\hat\phi}-c < \Delta_m \left(k_{\hat\phi}+l_{\hat\phi}\right)< c+bk_{\hat\phi}-2ak_{\hat\phi}\\
    0 &  \Delta_m \left(k_{\hat\phi}+l_{\hat\phi}\right)\leq bk_{\hat\phi}-c 
\end{cases}
\end{equation}
which shows how the optimal municipal subsidy depends on the relationship between social and private benefits of treatment and the cost of treatment. 

\paragraph{Case 3: $c> bk_{\hat\phi}$}
This corresponds to a situation where firms cannot profitably treat a private tree in the absence of subsidies. Now, all three cases from the expected utility function can result. The relevant partial derivative is 
\begin{equation}
\frac{\partial E[U_{m}]}{\partial s_{\hat\phi}}=
\begin{cases}
     -1 
        &  c-ak_{\hat\phi}\leq s_{\hat\phi}\\
    \frac{\Delta_m \left(k_{\hat\phi}+l_{\hat\phi}\right)+c-bk_{\hat\phi}-2s_{\hat\phi}}{k_{\hat\phi}(b-a)} 
        & c-bk_{\hat\phi} < s_{\hat\phi}< c-ak_{\hat\phi}\\
    0 
        &   0\leq s_{\hat\phi}  \leq c-bk_{\hat\phi}.
\end{cases}
\end{equation}
The optimum at the middle case is once again $s_{\hat\phi}=1/2\left(\Delta_m \left(k_{\hat\phi}+l_{\hat\phi}\right)+c-bk_{\hat\phi}\right)$, and this will fall within the middle case for intermediate values of $\Delta_m \left(k_{\hat\phi}+l_{\hat\phi}\right)$, following
\begin{equation}
s^*_{\hat\phi}=
\begin{cases}
    c-ak_{\hat\phi} 
        &  c+bk_{\hat\phi}-2ak_{\hat\phi}\leq \Delta_m \left(k_{\hat\phi}+l_{\hat\phi}\right)\\
    1/2\left(\Delta_m \left(k_{\hat\phi}+l_{\hat\phi}\right)+c-bk_{\hat\phi}\right) 
        & c-bk_{\hat\phi} < \Delta_m \left(k_{\hat\phi}+l_{\hat\phi}\right)< c+bk_{\hat\phi}-2ak_{\hat\phi}\\
    0 
        &  \Delta_m \left(k_{\hat\phi}+l_{\hat\phi}\right)\leq c-bk_{\hat\phi}
\end{cases}
\end{equation}
where the final case corresponds to the setting where a subsidy program could result in treatment of trees that would not be in the collective best interest of society. In this bottom case, any subsidy $s_{\hat\phi}\in\left[0,c-bk_{\hat\phi}\right]$ will lead to the same result. However, for simplicity we assume a $0$ subsidy because the municipality is not aiming to increase the likelihood of treatment. The middle case leads to a chance of treatment and the top case is when the social benefits are high enough that a municipality should choose a subsidy that leads everyone to treat their trees. 

\paragraph{Synthesis}
Combining the results of the three possible cases leads to the following optimal subsidy policy across all conditions. We merge the above three cases and derive an optimal subsidy policy for the municipality given by 
\begin{equation}
s^*_{\hat\phi}=
\begin{cases}
    0 
        & c\leq ak_{\hat\phi}\\
    c-ak_{\hat\phi} 
        &  c> ak_{\hat\phi} \;\text{ and }\;c+bk_{\hat\phi}-2ak_{\hat\phi}\leq \Delta_m \left(k_{\hat\phi}+l_{\hat\phi}\right) \\
    1/2\left(\Delta_m \left(k_{\hat\phi}+l_{\hat\phi}\right) +c-bk_{\hat\phi}\right) 
        & c> ak_{\hat\phi} \;\text{ and }\;\mid c-bk_{\hat\phi}\mid < \Delta_m \left(k_{\hat\phi}+l_{\hat\phi}\right) < c+b k_{\hat\phi}-2a k_{\hat\phi}\\
    0 
        &  c> a k_{\hat\phi} \;\text{ and }\; \Delta_m \left(k_{\hat\phi}+l_{\hat\phi}\right) \leq \mid c-b k_{\hat\phi}\mid
\end{cases}
\end{equation}
so that positive subsidies are only used when they increase treatment of trees, and only when the resulting treatments increase the welfare of society. This will lead to different subsidy policies for each assessed tree state. The exact policies for each tree state will depend on the prevalence of the disease, the accuracy of assessment and the effectiveness of treatment at reducing mortality risk. We will explore several case studies in a later section. 
\begin{figure}
    \centering
    \includegraphics[width=.45\textwidth]{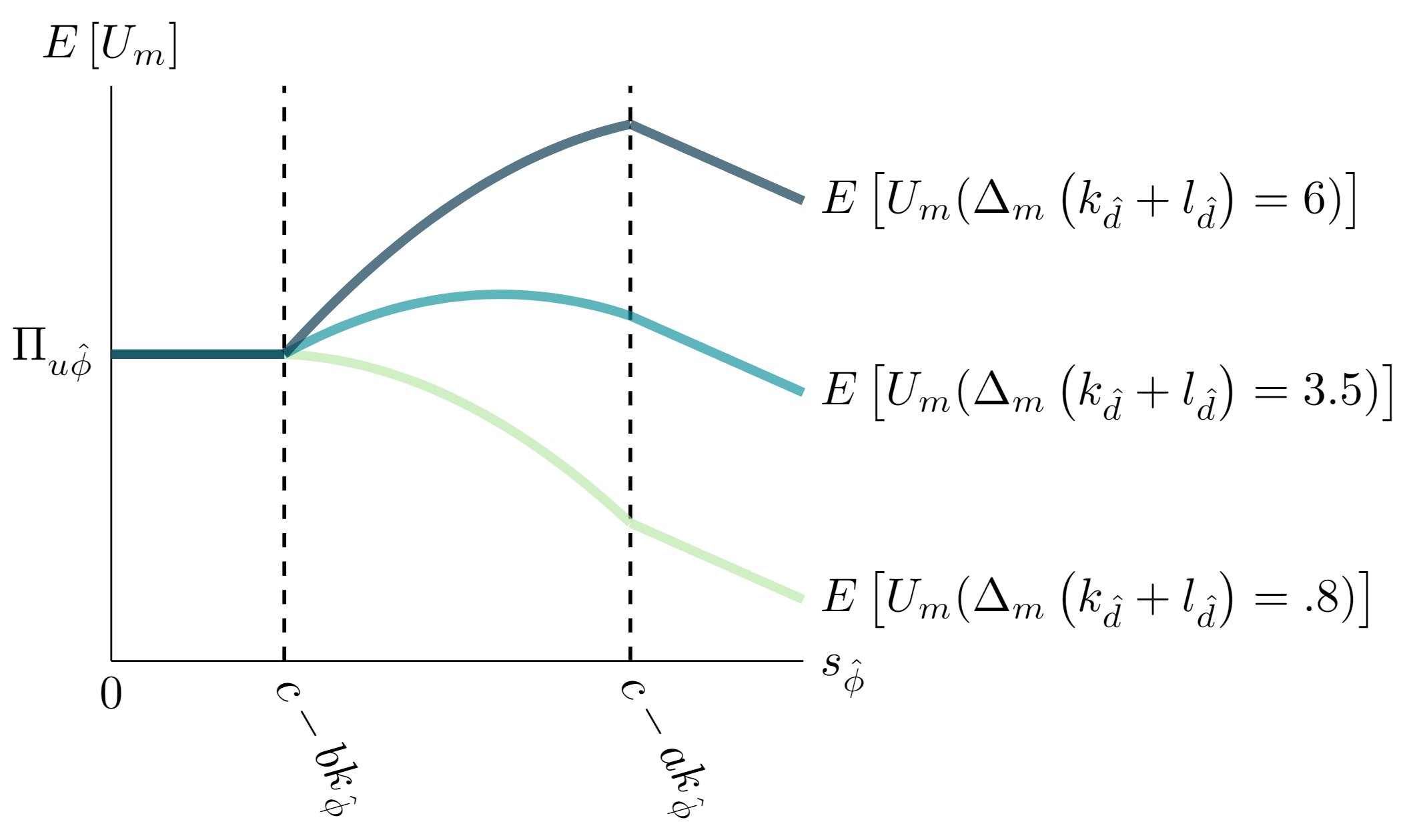}
    \caption{The expected utility of the municipality is shown as a function of the subsidy level, $s_{\hat\phi}$, that they set for a tree assessed in health state $\hat\phi$ for both firms. When the subsidy level chosen by the municipality $s_{\hat\phi}<c-bk_{\hat\phi}$, then there is no chance that a tree in assessed in state $\hat\phi$ will be treated. When $s_{\hat\phi}\geq c-ak_{\hat\phi}$ then a focal tree will be treated with probability 1. Here utility curves are shown for three different social benefit values, $\Delta_m \left(k_{\hat\phi}+l_{\hat\phi}\right)$, of treating a tree assessed in state $\hat\phi$. For low levels of social benefit from treating a tree, the municipality has its highest expected utility when subsidies do not lead to treatment of a tree. For intermediate social value of treatment, expected municipal utility is maximized at an intermediate subsidy level that leads to a positive probability of treatment. For high social values of treatment, expected utility is maximized when the subsidy is set just high enough to guarantee that a tree assessed in state $\hat\phi$ will be treated by its owner. Subsidy levels in excess of $c-ak_{\hat\phi}$ lead to declining expected utility in all cases. ($c=10$, $ak_{\hat\phi}=7$, $bk_{\hat\phi}=9$, $\Pi_{u\hat\phi}=4$) }
    \label{fig:muniUtil}
\end{figure}
\subsection{Outcomes under optimal behavior}\label{suppOutcomes}
Here we explore the outcomes that results from owner, firms, and municipalities adhering to the optimal behaviors identified in preceding sections. Given the subsidy policy above, the firms Nash equilibrium bids will be 
\begin{equation}
p_{1\hat\phi}^*=p_{2\hat\phi}^*=
\begin{cases}
    c 
        & c \leq ak_{\hat\phi}\\
    ak_{\hat\phi} 
        &  c > ak_{\hat\phi} \;\text{ and }\;c+bk_{\hat\phi}-2ak_{\hat\phi}\leq \Delta_m \left(k_{\hat\phi}+l_{\hat\phi}\right)\\
    1/2\left(c+bk_{\hat\phi}-\Delta_m\left(k_{\hat\phi}+l_{\hat\phi}\right)\right)  
        & c > ak_{\hat\phi} \;\text{ and }\;\mid c-bk_{\hat\phi}\mid < \Delta_m \left(k_{\hat\phi}+l_{\hat\phi}\right)< c+bk_{\hat\phi}-2ak_{\hat\phi}\\
    c
        & bk_{\hat\phi}>c> ak_{\hat\phi} \;\text{ and }\; \Delta_m \left(k_{\hat\phi}+l_{\hat\phi}\right) \leq  bk_{\hat\phi}-c \\
    p\geq c 
        & c\geq bk_{\hat\phi}\;\text{ and }\; \Delta_m \left(k_{\hat\phi}+l_{\hat\phi}\right) \leq  c-bk_{\hat\phi}.
\end{cases}
\end{equation}
The likelihood of treatment of a focal tree under the optimal behaviors defined above is 
\begin{equation}
P^*_{t\hat\phi o}=
\begin{cases}
    1 
        & c \leq ak_{\hat\phi}\\
    1 
        &  c> ak_{\hat\phi} \;\text{ and }\;c+bk_{\hat\phi}-2ak_{\hat\phi}\leq \Delta_m \left(k_{\hat\phi}+l_{\hat\phi}\right) \\
    \frac{\Delta_m \left(k_{\hat\phi}+l_{\hat\phi}\right)-c+bk_{\hat\phi}}{2(b-a)k_{\hat\phi}}  
        & c> ak_{\hat\phi} \;\text{ and }\;\mid c-bk_{\hat\phi}\mid < \Delta_m \left(k_{\hat\phi}+l_{\hat\phi}\right)< c+bk_{\hat\phi}-2ak_{\hat\phi}\vspace{3.5pt}\\
    \frac{bk_{\hat\phi}-c}{(b-a)k_{\hat\phi}} 
        & bk_{\hat\phi}>c> ak_{\hat\phi} \;\text{ and }\; \Delta_m\left(k_{\hat\phi}+l_{\hat\phi}\right) \leq  bk_{\hat\phi}-c\\
     0 
        & c\geq bk_{\hat\phi}\;\text{ and }\; \Delta_m\left(k_{\hat\phi}+l_{\hat\phi}\right) \leq  c-bk_{\hat\phi}
\end{cases}
\end{equation}
and this leads to expected utility for the municipality of 
\begin{equation}
E[U_m^*]=
\begin{cases}
    \Pi_{u\hat\phi} + \Delta_m \left(k_{\hat\phi}+l_{\hat\phi}\right) 
        &  c\leq ak_{\hat\phi}\\
    \Pi_{u\hat\phi} + \Delta_m \left(k_{\hat\phi}+l_{\hat\phi}\right) -c+ak_{\hat\phi}
        &  c> ak_{\hat\phi} \;\text{ and }\;c+bk_{\hat\phi}-2ak_{\hat\phi}\leq \Delta_m \left(k_{\hat\phi}+l_{\hat\phi}\right) \\
    \Pi_{u\hat\phi} + \frac{\left(\Delta_m \left(k_{\hat\phi}+l_{\hat\phi}\right)-c+b k_{\hat\phi}\right)^2}{4(b-a)k_{\hat\phi}} 
        & c> ak_{\hat\phi} \;\text{ and }\;\mid c-bk_{\hat\phi}\mid < \Delta_m \left(k_{\hat\phi}+l_{\hat\phi}\right) < c+bk_{\hat\phi}-2ak_{\hat\phi}\vspace{3.5pt}\\
    \Pi_{u\hat\phi} + \Delta_m \frac{bk_{\hat\phi}-c}{b-a} 
        &  bk_{\hat\phi}>c> ak_{\hat\phi} \;\text{ and }\; \Delta_m \left(k_{\hat\phi}+l_{\hat\phi}\right) \leq  bk_{\hat\phi}-c\\
    \Pi_{u\hat\phi} 
        &  c\geq bk_{\hat\phi}\;\text{ and }\; \Delta_m \left(k_{\hat\phi}+l_{\hat\phi}\right) \leq  c-bk_{\hat\phi}
\end{cases}
\end{equation}
and an expected utility for the firms of 
\begin{equation}
E\left[U_{f_1}^*\right]=E\left[U_{f_2}^*\right]=0
\end{equation}
because under all cases where treatment can happen, the firms bid competitively and receive 0 profit. 

A representative tree owner knows their own valuation of their tree. Therefore rather than consider their expected utility, we can instead calculate their utility as a function of their type (i.e. their true underlying valuation of the tree), $\Delta_o$, that other players in the game know is drawn from a uniform distribution. This results in
\footnotesize
\begin{equation}
U_o^*[\Delta_o]=
\begin{cases}
    \pi_{u\hat\phi} + \Delta_o k_{\hat\phi}-c
        &  c\leq ak_{\hat\phi}\\
    \pi_{u\hat\phi} + \Delta_o k_{\hat\phi}-ak_{\hat\phi} 
        &  c> ak_{\hat\phi} \;\text{ and }\;c+bk_{\hat\phi}-2ak_{\hat\phi}\leq \Delta_m \left(k_{\hat\phi}+l_{\hat\phi}\right) \\
    \pi_{u\hat\phi} + \frac{c+bk_{\hat\phi}+2\Delta_ok_{\hat\phi}-\Delta_m\left(k_{\hat\phi}+l_{\hat\phi}\right)}{2}
        & c> ak_{\hat\phi} \text{ and }\mid c-bk_{\hat\phi}\mid < \Delta_m \left(k_{\hat\phi}+l_{\hat\phi}\right)< c+bk_{\hat\phi}-2ak_{\hat\phi} \text{ and } \Delta_o\geq\frac{c+bk_{\hat\phi}-\Delta_m\left(k_{\hat\phi}+l_{\hat\phi}\right)}{2k_{\hat\phi}} \\
    \pi_{u\hat\phi} 
        & c> ak_{\hat\phi} \text{ and }\mid c-bk_{\hat\phi}\mid < \Delta_m \left(k_{\hat\phi}+l_{\hat\phi}\right)< c+bk_{\hat\phi}-2ak_{\hat\phi} \text{ and } \Delta_o<\frac{c+bk_{\hat\phi}-\Delta_m \left(k_{\hat\phi}+l_{\hat\phi}\right)}{2k_{\hat\phi}}\\
    \pi_{u\hat\phi}+ \Delta_o k_{\hat\phi}-c 
        &  bk_{\hat\phi}>c> ak_{\hat\phi} \;\text{ and }\; \Delta_m \left(k_{\hat\phi}+l_{\hat\phi}\right) \leq  bk_{\hat\phi}-c \;\text{ and }\; \Delta_o\geq c / k_{\hat\phi}\\
    \pi_{u\hat\phi}  
        &  bk_{\hat\phi}>c> ak_{\hat\phi} \;\text{ and }\; \Delta_m\left(k_{\hat\phi}+l_{\hat\phi}\right) \leq  bk_{\hat\phi}-c \;\text{ and }\; \Delta_o< c / k_{\hat\phi}\\
    \pi_{u\hat\phi} 
        &  c\geq b k_{\hat\phi}\;\text{ and }\; \Delta_m \left(k_{\hat\phi}+l_{\hat\phi}\right) \leq  c-b k_{\hat\phi}
\end{cases}
\end{equation}
\normalsize
where the top case corresponds to the setting where the cost of treatment is low enough that all types of tree owners will treat their trees in the absence of subsidies, the second case arises when the social benefit of treatment is great enough that the municipality chooses a subsidy level to assure that all types of tree owners will treat their trees. The next two cases occur when the municipalities best option is an intermediate subsidy which leads tree owners with high valuations of their tree to treat and low valuations not to treat. This cutoff is calculated in terms of the tree owners true valuation of avoiding the mortality of a tree, $\Delta_o$. The next two cases arise when the municipality chooses a subsidy of zero, but there are still some tree owners that will treat their tree. The final case occurs when the social benefit of treatment does not justify a positive subsidy level, and the private benefit is small enough that no types of tree owners opt for treatment.

\section{Special case where only firm 1 is subsidized}\label{suppS2=0}
Here we assume that $s_{2\hat\phi}=0$, and $bk_{\hat\phi}<c$. These restrictions imply that firm 2 does not receive a subsidy, that no private owners value treatment enough to offset its cost in the absence of a subsidy. We use this setting as a simplified setting within which to compare outcomes under the optimal policy of equal subsidies with outcomes that arise when only one firm receives a subsidy. 

\subsection{Tree owner best response}
As before, given a pair of bids, $(p_{1\hat\phi}, p_{2\hat\phi})$, the tree owner maximizes their utility by following the strategy defined by 
\begin{equation}
g_{o\hat\phi}^*\left[\Delta_o\right] = \begin{cases}
     \text{treat w/}f_1 & \text{if } p_{1\hat\phi}<p_{2\hat\phi} \text{  and  } p_{1\hat\phi}\leq \Delta_o k_{\hat\phi}\\
    \text{treat w/}f_2 &\text{if } p_{2\hat\phi}<p_{1\hat\phi} \text{  and  } p_{2\hat\phi}\leq \Delta_o k_{\hat\phi}\\
    \text{flip coin} &\text{if } p_{1\hat\phi}=p_{2\hat\phi} \text{  and  }  p_{2\hat\phi}\leq \Delta_o k_{\hat\phi}\\
    \text{don't treat} &\text{if } \Delta_o k_{\hat\phi}<\text{min}\{p_{1\hat\phi},p_{2\hat\phi}\}
\end{cases}
\end{equation}
Which has not changed from earlier cases because subsidies have no direct impact on the tree owner. 

\subsection{Behavior of the firms}
In this case, firms will maximize their expected utility, given the bid of the opposing firm and their beliefs about the value of treatment to the tree owner. 

Previously, we derived best response correspondences for unequal subsidy levels. We will apply these to the special case where $s_{2\hat\phi}=0$, and $bk_{\hat\phi}<c$. 

Firm 1 has the same best response as in the general case, which given 
$$p_{1\hat\phi}'=\max\left\{\frac{1}{2}\left(bk_{\hat\phi}+c-s_{1\hat\phi}\right),ak_{\hat\phi} \right\}
$$
is
\begin{equation}
\hat p_{1\hat\phi}\left(p_{2\hat\phi}\right) = \begin{cases}
p_{1\hat\phi}\geq c-s_{1\hat\phi} 
     &  s_{1\hat\phi}< c-bk_{\hat\phi} \text{    or    } s_{1\hat\phi}<c-p_{2\hat\phi}\\  
 p_{2\hat\phi}-\epsilon 
    &s_{1\hat\phi} \geq c-bk_{\hat\phi} \text{  and  }  c-s_{1\hat\phi}\leq p_{2\hat\phi} \leq p_{1\hat\phi}'\\
     p_{1\hat\phi}'
     & s_{1\hat\phi} \geq c-bk_{\hat\phi} \text{  and  } p_{1\hat\phi}'<p_{2\hat\phi} .
\end{cases}
\end{equation}
However, due to the cost of treatment and the lack of subsidy, the best response of firm 2 becomes 
\begin{equation}
\hat p_{2\hat\phi}\left(p_{1\hat\phi}\right) = \begin{cases}
p_{2\hat\phi}\geq c
\end{cases}
\end{equation}
because firm 2 is unsubsidized and the cost of treatment is high enough that no owners will accept the minimum profitable bid that they could submit. To further simplify our analysis, we assume that $$\hat p_{2\hat\phi} \left( p_{1\hat\phi} \right) = c.$$
These best responses can lead to three types of equilibrium bid pairs, given our assumptions.   We can aggregate the Nash equilibria across the three sub-cases and use this to determine the best subsidy level from the perspective of the municipality. We can define the cases in terms of the subsidy level set by the municipality, $s_{1\hat\phi}$ as
\begin{equation}
(p_{1\hat\phi}^*, p_{2\hat\phi}^*) = \begin{cases}
   ( c-s_{1\hat\phi},c)
        & s_{1\hat\phi}< c-bk_{\hat\phi}\\
   \left(\frac{1}{2}\left(bk_{\hat\phi}+c-s_{1\hat\phi}\right),c\right) 
        & c-bk_{\hat\phi} < s_{1\hat\phi}< c+bk_{\hat\phi}-2ak_{\hat\phi}\\
   (ak_{\hat\phi},c) 
        &  c+bk_{\hat\phi}-2ak_{\hat\phi}\leq s_{1\hat\phi}
\end{cases}
\end{equation}
where the first case implies that the subsidy for firm 1 is not high enough to lead to a positive probability of treatment, so the details here are not important. The second case arises when the subsidy level is high enough to allow for a chance of firm 1 profitably treating the tree. The final case arises when the subsidy level is high enough for firm 1 to bid a price that will lead to treatment with certainty. Across all cases, firm 2 never has a chance at winning the bid. 

\subsection{Optimal municipal subsidy}
Taking into account the above responses to their subsidy levels, the municipality will seek to choose a subsidy level, $s_{1\hat\phi}\geq0$ that maximizes their expected utility. The expected utility function of the municipality in this case is
\begin{equation}
E\left[U_m(s_{1\hat\phi})\right] = \begin{cases}
    \Pi_{u\hat\phi}, 
        & s_{1\hat\phi}<c-bk_{\hat\phi}\\
    \Pi_{u\hat\phi}+\left(\Delta_m \left(k_{\hat\phi}+l_{\hat\phi}\right)-s_{1\hat\phi}\right)\frac{bk_{\hat\phi}-c+s_{1\hat\phi}}{2k_{\hat\phi}(b-a)}, 
        & c-bk_{\hat\phi}<s_{1\hat\phi}<c+bk_{\hat\phi}-2ak_{\hat\phi}\\
    \Pi_{u\hat\phi}+\Delta_m \left(k_{\hat\phi}+l_{\hat\phi}\right)-s_{1\hat\phi}, 
        &c+bk_{\hat\phi}-2ak_{\hat\phi}<s_{1\hat\phi}
\end{cases}
\end{equation}
where the first case implies that subsidies are too low for trees to get treated, the middle case corresponds to subsidy levels which lead a positive probability of the tree getting treated and the final case corresponds to the expected utility that the municipality receives when subsidies are set high enough to assure that the tree will be treated. 

We can find the optimal subsidy by setting the partial derivative
\begin{equation}
\frac{\partial}{\partial s_{1\hat\phi}}E\left[U_m(s_{1\hat\phi})\right] = \begin{cases}
    0, 
        & s_{1\hat\phi}<c-bk_{\hat\phi}\\
    \frac{\Delta_m \left(k_{\hat\phi}+l_{\hat\phi}\right)-bk_{\hat\phi}+c-2s_{1\hat\phi}}{2k_{\hat\phi}(b-a)}, 
        & c-bk_{\hat\phi}<s_{1\hat\phi}<c+bk_{\hat\phi}-2ak_{\hat\phi}\\
    -1, 
        &c+bk_{\hat\phi}-2ak_{\hat\phi}<s_{1\hat\phi}
\end{cases}
\end{equation}
equal to zero. For the middle case this implies an optimal subsidy level of 
\begin{equation}
    s^\prime_{1\hat\phi}=\frac{\Delta_m \left(k_{\hat\phi}+l_{\hat\phi}\right)+c-bk_{\hat\phi}}{2}
\end{equation}
which will lead to a positive probability of a tree being treated. 

Recalling that we assumed at the outset of this section that $bk_{\hat\phi}<c$, we can show that an optimal subsidy that leads to treatment, $s^\prime_{1\hat\phi}>c-bk_{\hat\phi}$ will arise when $\Delta_m \left(k_{\hat\phi}+l_{\hat\phi}\right)>c-bk_{\hat\phi}>0$. The optimal subsidy policy for the municipality is 
\begin{equation}
    s^*_{1\hat\phi} = \begin{cases}
    0,
        & \Delta_m \left(k_{\hat\phi}+l_{\hat\phi}\right) <c-bk_{\hat\phi}\\
    \frac{1}{2}(\Delta_m \left(k_{\hat\phi}+l_{\hat\phi}\right) +c-bk_{\hat\phi}), 
        & c-bk_{\hat\phi}\leq  \Delta_m \left(k_{\hat\phi}+l_{\hat\phi}\right) <c+3bk_{\hat\phi}-4ak_{\hat\phi}\\
    c+bk_{\hat\phi}-2ak_{\hat\phi}, 
        & c+3bk_{\hat\phi}-4ak_{\hat\phi}\leq \Delta_m \left(k_{\hat\phi}+l_{\hat\phi}\right)
\end{cases}
\end{equation}
with the first implying that a subsidy is not in the municipal best interest, the second case leading to a positive probability that the tree will be treated and the third case corresponding to the minimum subsidy that assures that the tree will be treated.

This subsidy policy leads to equilibrium firm bids of 

\begin{equation}
    (p_{1\hat\phi}^*, p_{2\hat\phi}^*) = \begin{cases}
    (c,c),
        & \Delta_m \left(k_{\hat\phi}+l_{\hat\phi}\right) <c-bk_{\hat\phi}\vspace{4pt}\\ 
    \left(\frac{c+3bk_{\hat\phi}-\Delta_m\left(k_{\hat\phi}+l_{\hat\phi}\right)}{4},c\right), 
        & c-bk_{\hat\phi}\leq  \Delta_m \left(k_{\hat\phi}+l_{\hat\phi}\right) <c+3bk_{\hat\phi}-4ak_{\hat\phi}\vspace{4pt}\\
    \left(ak_{\hat\phi},c\right), 
        & c+3bk_{\hat\phi}-4ak_{\hat\phi}\leq \Delta_m \left(k_{\hat\phi}+l_{\hat\phi}\right) 
\end{cases}
\end{equation}
Which implies that the tree owner will never contract with firm 2 and firm 1 will effectively have a local monopoly. These bids will lead to a probability function that the tree owner will treat their tree with firm 1 given by 
\begin{equation}
     P^*_{t\hat\phi o} = \begin{cases}
    0,
        & \Delta_m\left(k_{\hat\phi}+l_{\hat\phi}\right) <c-bk_{\hat\phi}\\
    \frac{ \Delta_m\left(k_{\hat\phi}+l_{\hat\phi}\right) -c+bk_{\hat\phi}}{4(b-a)k_{\hat\phi}}, 
        & c-bk_{\hat\phi}\leq  \Delta_m \left(k_{\hat\phi}+l_{\hat\phi}\right) <c+3bk_{\hat\phi}-4ak_{\hat\phi}\\
    1, 
        & c+3bk_{\hat\phi}-4ak_{\hat\phi}\leq \Delta_m\left(k_{\hat\phi}+l_{\hat\phi}\right). 
\end{cases}
\end{equation}
With the strategy choices in hand, we can proceed to consider a welfare analysis. 

\subsection{Welfare analysis}
Now we will examine the utility of each player that results when the all players adhere to the Nash equilibrium strategies determined above. 
For firm 1 their Nash equilibrium expected utility is 
\begin{equation}
E\left[U_{f_1}^*\right] = \begin{cases}
    0,
        & \Delta_m \left(k_{\hat\phi}+l_{\hat\phi}\right) <c-bk_{\hat\phi}\\
    \frac{(\Delta_m\left(k_{\hat\phi}+l_{\hat\phi}\right)-c+bk_{\hat\phi})^2}{16(b-a)k_{\hat\phi}}, 
        &c-bk_{\hat\phi}\leq  \Delta_m \left(k_{\hat\phi}+l_{\hat\phi}\right) <c+3bk_{\hat\phi}-4ak_{\hat\phi}\\
    k_{\hat\phi}(b-a), 
        & c+3bk_{\hat\phi}-4ak_{\hat\phi}\leq \Delta_m \left(k_{\hat\phi}+l_{\hat\phi}\right) 
\end{cases}
\end{equation}
which is clearly at least as great as that which occurs under equal subsidies because there is a chance for positive utility for the firm. Firm 2 still has an expected utility of zero in all cases. The municipality has an expected utility of 
\begin{equation}
E\left[U_{m}^*\right] = \begin{cases}
    \Pi_{u\hat\phi},
        & \Delta_m \left(k_{\hat\phi}+l_{\hat\phi}\right) <c-bk_{\hat\phi}\\
\Pi_{u\hat\phi}+\frac{(\Delta_m\left(k_{\hat\phi}+l_{\hat\phi}\right) -c+bk_{\hat\phi})^2}{8(b-a)k_{\hat\phi}}, 
        &c-bk_{\hat\phi}\leq  \Delta_m \left(k_{\hat\phi}+l_{\hat\phi}\right)<c+3bk_{\hat\phi}-4ak_{\hat\phi}\\
\Pi_{u\hat\phi} + \Delta_m \left(k_{\hat\phi}+l_{\hat\phi}\right) - c -bk_{\hat\phi} +2ak_{\hat\phi}, 
        & c+3bk_{\hat\phi}-4ak_{\hat\phi}\leq \Delta_m \left(k_{\hat\phi}+l_{\hat\phi}\right)
\end{cases}
\end{equation}
which is never greater than the utility that results from an optimal equal subsidy policy. When treatment occurs with certainty, the expected utility of the municipality is equivalent to the equal subsidy case only when $a=b$. In other words, when there is no uncertainty about the value of avoiding tree mortality to its owner. As $a$ decreases and/or $b$ increases, which leads to uncertainty about the valuation that people have for their trees, then the utility declines for the municipality below that which would be achieved under equal subsidies. The intermediate case can only arise when there is some uncertainty about the value tree owners place on their trees ($b>a$) and when this intermediate case occurs, it leads to lower expected utility than that which would result under equal subsidies. Further, the range of values of $\Delta_m \left(k_{\hat\phi}+l_{\hat\phi}\right)$ for which this case arises is broader in this case where  $s_{2\hat\phi}=0$ than it is the the case where $s_{1\hat\phi}=s_{2\hat\phi}=s_{\hat\phi}$. 

A tree owner knows their own type, $\Delta_o$, and therefore they can calculate they utility rather than expected utility, which is 
\small
\begin{equation}
U_o^*[\Delta_o] = \begin{cases}
    \pi_{u\hat\phi},
        & \Delta_m\left(k_{\hat\phi}+l_{\hat\phi}\right) <c-bk_{\hat\phi}\\
\pi_{u\hat\phi}+\Delta_o k_{\hat\phi}-\frac{c+3bk_{\hat\phi}-\Delta_m\left(k_{\hat\phi}+l_{\hat\phi}\right) }{4},
        &c-bk_{\hat\phi}\leq \Delta_m \left(k_{\hat\phi}+l_{\hat\phi}\right) <c+3bk_{\hat\phi}-4ak_{\hat\phi} \text{ and } \Delta_o\geq \frac{c+3bk_{\hat\phi}-\Delta_m\left(k_{\hat\phi}+l_{\hat\phi}\right)}{4k_{\hat\phi}} \\
    \pi_{u\hat\phi}, 
        &c-bk_{\hat\phi}\leq  \Delta_m \left(k_{\hat\phi}+l_{\hat\phi}\right) <c+3bk_{\hat\phi}-4ak_{\hat\phi} \text{ and } \Delta_o< \frac{c+3bk_{\hat\phi}-\Delta_m\left(k_{\hat\phi}+l_{\hat\phi}\right)}{4k_{\hat\phi}}\\
    \pi_{u\hat\phi}+\Delta_o k_{\hat\phi}-ak_{\hat\phi}, 
        & c+3bk_{\hat\phi}-4ak_{\hat\phi}\leq \Delta_m \left(k_{\hat\phi}+l_{\hat\phi}\right)
\end{cases}
\end{equation}
\normalsize
and which is never greater than the utility that results when the municipality chooses equal optimal subsidies. 

In sum, the analysis of the case where $s_{2\hat\phi}=0$ and $c>bk_{\hat\phi}$ showed two main results. First, when there is no uncertainty about the value of avoiding tree mortality to its owner, $\Delta_o$, then the outcomes for this case exactly correspond to those that would occur for equal optimal subsidy levels for both firms. However, when the municipality and the firms are uncertain about the value avoiding tree mortality, and $\Delta_o\sim U[a,b]$ for $b>a$, then the outcomes that arise when $s_{2\hat\phi}=0$ diverge from those that occur under equal subsidies. Firm 1 is better off, because the fact that the other firm is unsubsidized means that they can behave like a local monopolist and achieve positive profits. This comes at a cost to the expected utility of the municipality and the tree owner. Firm 1 does not pass all of the subsidy on to the tree owner and therefore for any particular subsidy level, the likelihood of treatment will typically be lower (or equivalent in some cases) to that which would result under equal subsidies. Therefore it is more costly for the municipality to treat the same number of trees. For tree owners, for any given subsidy level, there is a decreased chance that they will find a suitable bid from firm 1 (or firm 2) if they do choose to treat their tree they will tend to have lower utility than they would have had under an equal subsidy policy.  

While these results were framed as applying to a case where only one of the two firms is subsidized, similar outcomes would arise in the case where there is only one tree care firm in a municipality. Should this arise, the results of this section indicate that it would be in the interest of tree owners and the municipality to reduce uncertainty around the value of treatment, because less uncertainty leads to higher (expected) utility for the municipality and the tree owner.

\section{Optimization model for municipally owned trees}\label{supp: muniOpt}
A mosaic of private and public ownership of tree is characteristic of urban forests, where parcels of private and public land intermingle to create a complex management setting where uniform management across these ownership types can be difficult to achieve. Here we assume that public trees can be treated at a cost, $c$, by the municipality. The problem for a forest manager is to identify the subset of trees for which the social benefit of treatment is greater than the public cost. While the social value of a surviving tree, $V_m$, remains unchanged, the social cost of mortality of a municipally owned tree may be higher than the social cost of a private tree's mortality. In the case of a municipally owned tree, we assume this cost is inclusive of removal cost of the tree. Thus we define $W'_m$ as the social cost of tree mortality, including removal costs. This can be used to construct the social payoffs from a focal public tree of 
\begin{align}
    \Pi_{t\hat \phi} &=
     P( h \mid \hat \phi)\left[V_m-\Delta'_m\left(\mu_{th}+\lambda_{th}\right)\right]
    +P( i \mid \hat \phi)\left[V_m-\Delta'_m\left(\mu_{ti}+\lambda_{ti}\right)\right]
    -P(d \mid \hat \phi)W'_m\\
    \Pi_{u\hat \phi} &=
     P( h \mid \hat \phi)\left[V_m-\Delta'_m\left(\mu_{uh}+\lambda_{uh}\right)\right]
    +P( i \mid \hat \phi)\left[V_m-\Delta'_m\left(\mu_{ui}+\lambda_{ui}\right)\right]
    -P(d \mid \hat \phi)W'_m
\end{align} 
for treatment or non-treatment, where the net social benefit of avoiding tree mortality is $\Delta'_m=V_m+W'_m$. Therefore the objective of the municipality for the treatment of public trees is to choose treatment probabilities to maximize
\begin{equation}
U_{m\hat\phi} = \begin{cases}
    \Pi'_{t\hat \phi}-c, & \text{if tree assessed as $\hat \phi$ is treated}\\
    \Pi'_{u\hat \phi} & \text{if tree assessed as $\hat \phi$ is untreated}.
\end{cases}
\end{equation}
The expected benefit of treatment is 
\begin{equation}
\Pi'_{t\hat \phi} - \Pi'_{u\hat \phi} = \Delta'_m \left(k_{\hat \phi}+l_{\hat\phi}\right)
\end{equation}
where, as before, we denote the change in the survival probability in response to treatment for a focal tree assessed in state $\hat\phi$ as $k_{\hat\phi}$ and the spillover effect of treatment on changes in expected community-level tree survival as $l_{\hat\phi}$.
This can be used to define the cases where a municipal tree should be treated with the optimal probability of treating a public tree given by 
\begin{equation}
P^*_{t\hat\phi m}=
\begin{cases}
    1        &  c\leq \Delta'_m\left(k_{\hat\phi}+l_{\hat\phi}\right)\\
    0        &  c > \Delta'_m\left(k_{\hat\phi}+l_{\hat\phi}\right).
\end{cases}
\end{equation}
This equation provides a simple proscription for treatment of public trees. However, considerable nuance is encoded within the $k_{\hat\phi}+l_{\hat\phi}$ because they depend on treatment effectiveness, assessment accuracy, the state of EAB infestation within the community and the expected direct and spillover mortality risks to trees in the community.

\begin{table}[htbp]
\footnotesize
\begin{center}
    \begin{tabular}{c|c|c|c|c|c|c|c } 
     Parameter & Fig.~\ref{fig:optSub1} & Case Study & Fig.~\ref{supp:fig:Assessment}(a)(d) & Fig.~\ref{supp:fig:Assessment}(c)(f) & Fig.~\ref{supp:fig:spillovers}(a)(c) & Fig.~\ref{supp:fig:no uncertaintyin Delta_o} & Fig.~\ref{supp:fig:highly uncertain}\\ 
     \hline
     $a$ &  675 & 675 & 675 &675 &675 & 887.5&0  \\
     $b$ & 1100 &1100 & 1100&1100 &1100 & 887.5& 1775 \\
     $c$ & 200 & 250 &250 & 250& 250& 250&  250\\
     $\Delta_m$ & 0-1000 & 1150&1150 & 1150& 1150&1150 & 1150  \\
     $\Delta'_m$ & N/A  & 1850&1850 &1850 & 1850& 1850&  1850 \\
     $\Delta_o$ & 675-1100 &675-1100& 675-1100&675-1100 &675-1100 & 887.5& 0-1775   \\
     $\epsilon_h$ & 0.99 & 0.97&0.97 & .097&0.97 &0.97 & 0.97 \\
     $\epsilon_i$ & 0.8 & 0.5 &0.5 & .05& 0.5& 0.5& 0.5 \\
     $\beta$  &1 & 1&1 &1 & 1&1 &1  \\
     $\alpha$  &1 & 1&1 &1 & 1& 1& 1 \\
     $\gamma$  & 0.3&0.3&0.3 & 0.3&0.3 &0.3 & 0.3 \\
     $\tau^*$ & 3 & 3& 3&3 &3 &3 &  3 \\
     $P(h)$ & 0.8& &   &   &   &   &    \\
     $P(i)$ & 0.15&   &   &   &   &   &    \\
     $P(d) $ & 0.05&   &   &   &   &   &    \\
     $P(\hat h \mid h) $ &0.89 & 0.89&1/3 &1 &0.89 &0.89 &0.89  \\
     $P(\hat i \mid h) $ &0.1 & 0.1 &1/3 &0 & 0.1& 0.1& 0.1 \\
     $P(\hat d \mid h) $ & 0.01&0.01 &1/3 &0 &0.01 & 0.01&  0.01 \\
     $P(\hat h \mid i) $ & 0.49&0.49 &1/3 &0 &0.49 &0.49 & 0.49  \\
     $P(\hat i \mid i) $ &0.5 & 0.5 &1/3 & 1&0.5 & 0.5& 0.5 \\
     $P(\hat d \mid i) $ & 0.01&0.01 &1/3 &0 &0.01 &0.01 & 0.01  \\
     $P(\hat h \mid d) $ &0.01 &0.01 & 1/3&0 &0.01 &0.01 & 0.01  \\
     $P(\hat i \mid d) $ & 0.19&0.19 & 1/3&0 &0.19 & 0.19&  0.19 \\
     $P(\hat d \mid d) $ & 0.8 & 0.8 &1/3 &1 &0.8 &0.8 & 0.8 \\
     $l_{\hat\phi}$ &   &  & & &0 & &  \\
    \end{tabular}
     \caption {Parameter values, which apply to main text Figures~\ref{fig:optSub1}-\ref{fig:simplexDynam} as well as Supplementary Material Figures, as noted. The Parameters for the EAB case study are applied in Figures~\ref{fig:simplex1}-\ref{fig:simplexDynam}, Figure~\ref{supp:fig:dynamics}, Figure~\ref{supp:fig:Assessment}(b)(e), Figure~\ref{supp:fig:spillovers}(b)(d), and Figure~\ref{supp:fig:costsbenefits}. Blank entries correspond to values that vary dynamically in the corresponding figure.}
     \label{param_table}
\end{center}
\end{table}
\normalsize
\section{Epidemiological model}\label{suppCompartmentalModel}
Here, we derive a dynamic model of an urban tree population that is subject to forest pest infestation and management. We assume that municipally and privately owned trees are separate classes of trees with distinct management practices, yet the spread of the forest pest does not distinguish these differences. We will use the model to illustrate the performance of a range of treatment scenarios, including the optimal subsidy and treatment policies for public (municipally owned) and private trees.  

We project the dynamics of an urban forest subject to a pest using a compartmental model of disease dynamics~\citep{kermack1927contribution}. In these models, populations are categorized into classes with distinct properties that influence the spread of a pathogen. We study the dynamics of pest spread across a population of municipal and private trees by subdividing the population into healthy, infested, and dying classes for municipally owned and privately owned trees. Rates of transfer between compartments describe disease spread, and tree death or recovery. We model a population of municipal trees divided into three classes described by the fraction of healthy trees, $H_m$, infested trees, $I_m$, and dead or dying trees, $D_m$. The municipally owned trees interact with three classes of trees with private owners, $H_o$, $I_o$, and $D_o$. Since we measure the fraction of trees in each category, we have $H_m+I_m+D_m+H_o+I_o+D_o=1$. 

\subsection{Infestation rate}
We assume that a infested tree emits pests at a rate of $\beta_1$ regardless of whether or not it has been treated and regardless of whether it is publicly or privately owned (however, treatment can result in the tree becoming healthy which would stop the spread of a pest). The probability that a healthy municipal tree (for example) is exposed to the pest will therefore be $\beta_1 H_m(I_m+I_o)$ following the mass action kinetics assumptions of compartmental models. However, not all exposures result in infestation. We assume the probability of the establishment of an infestation given exposure is $\beta_2$ for untreated trees and $\beta_2 (1-\epsilon_h)$ for treated trees, so that treatment lessens the likelihood of the establishment of an infestation in proportion to the effectiveness of treatment of healthy trees, $\epsilon_h$. This results in an infestation rate for healthy  municipal trees of 
\begin{equation*}
\left(P_{thm}\beta_2 (1-\epsilon_h)+(1-P_{thm})\beta_2\right)\beta_1 H_m(I_m+I_o),
\end{equation*}
where $P_{thm}$ is the probability of treatment for healthy municipally owned trees. Combining terms and defining the transmission rate of the pest as $\beta=\beta_1\beta_2$, which combines spread and establishment under baseline conditions, results in a simplified expression for the infestation rate of healthy privately owned trees, with
\begin{equation}
\left(1-\epsilon_hP_{thm}\right)\beta H_m(I_m+I_o).
\end{equation}
Following the same approach, we can write the infestation rate for privately owned trees as 
\begin{equation}
\left(1-\epsilon_hP_{tho}\right)\beta H_o(I_m+I_o)
\end{equation}
where $P_{tho}$ is the probability of treatment for healthy privately owned trees. These expressions show that the rate at which public and private trees become infested varies mainly through differences in treatment probabilities, but these differences can lead to feedback effects that change the prevalence of healthy  and infested trees, which shapes the rate of infestation in the future.

\subsection{Recovery Rate}
In our model, we allow for recovery of infested trees that are treated. We assume that the effectiveness of treatment at leading to tree recovery is $\epsilon_i\in[0,1]$ and is equal across municipal and private trees. The recovery rate for treated trees is $\alpha$ when the treatment is effective, and therefore we can write the recovery rate of infested privately owned trees as
\begin{equation}
\alpha \epsilon_i P_{tio}  I_o     
\end{equation} 
for private trees, and 
\begin{equation}
    \alpha\epsilon_i P_{tim} I_m 
\end{equation}
for municipally owned trees, where $P_{tio}$ is the probability of treatment for infested trees with private owners and $P_{tim}$ is the probability of treatment for infested municipally owned trees. 

\subsection{Death rate}
While some treated infested trees can recover, others will eventually experience mortality. We assume that the pest-induced death rate for infested trees is $\gamma$, however this only applies to trees that are untreated or for which treatment was ineffective (a $1-\epsilon_i$ probability). Therefore the mortality rate for infested municipal trees is 
\begin{equation*}
    \left(\gamma(1-P_{tim})+\gamma (1-\epsilon_i)P_{tim}\right)I_m
\end{equation*}
which can be simplified to 
\begin{equation}
    \left(1-\epsilon_i P_{tim}\right)\gamma I_m.
\end{equation}
Following the same approach for privately owned trees results in a mortality rate of 
\begin{equation}
    \left(1-\epsilon_i P_{tio}\right)\gamma I_o.
\end{equation}

\subsection{Dynamical system}
Given these descriptions of the infestation, recovery and death rates for municipal and private trees, we can define a system of equations that describes the dynamics of each health class of tree given its ownership status. We denote the rate of change of the fraction of healthy municipal trees, for example, as $$\frac{dH_m}{d\tau}=\dot H_m$$ for simplicity of notation. Using this notation can write a complete description of the dynamical system as 
\begin{align}
    &\dot H_m = -\beta\left(1-\epsilon_h P_{thm}\right) H_m(I_m+I_o) + \alpha\epsilon_iP_{tim} I_m \\
    &\dot H_o = -\beta\left(1-\epsilon_h P_{tho}\right) H_o(I_m+I_o) + \alpha\epsilon_iP_{tio} I_o \\
    &\dot I_m = \beta\left(1-\epsilon_h P_{thm}\right) H_m(I_m+I_o) - \gamma\left(1-\epsilon_i P_{tim}\right) I_m - \alpha\epsilon_i P_{tim} I_m\\
    &\dot I_o = \beta\left(1-\epsilon_h P_{tho}\right) H_o(I_m+I_o) - \gamma\left(1-\epsilon_i P_{tio}\right) I_o - \alpha\epsilon_i P_{tio} I_o\\
      &\dot D_m = \gamma\left(1-\epsilon_i P_{tim}\right) I_m \\
      &\dot D_o = \gamma\left(1-\epsilon_i P_{tio}\right) I_o
\end{align}
and this system of equations can be applied to arbitrary treatment scenarios defined by 6 treatment probabilities. These probabilities can be fixed throughout the simulation or be dependent on the system's state. We proceed to explore treatment scenarios reflective of both these possibilities including scenarios where treatment probabilities are the result of the optimal subsidy policies and treatment choices analyzed in the game theoretic model examined in previous sections. 

\begin{figure}[htbp]
     \centering
    \includegraphics[width=.85\textwidth]{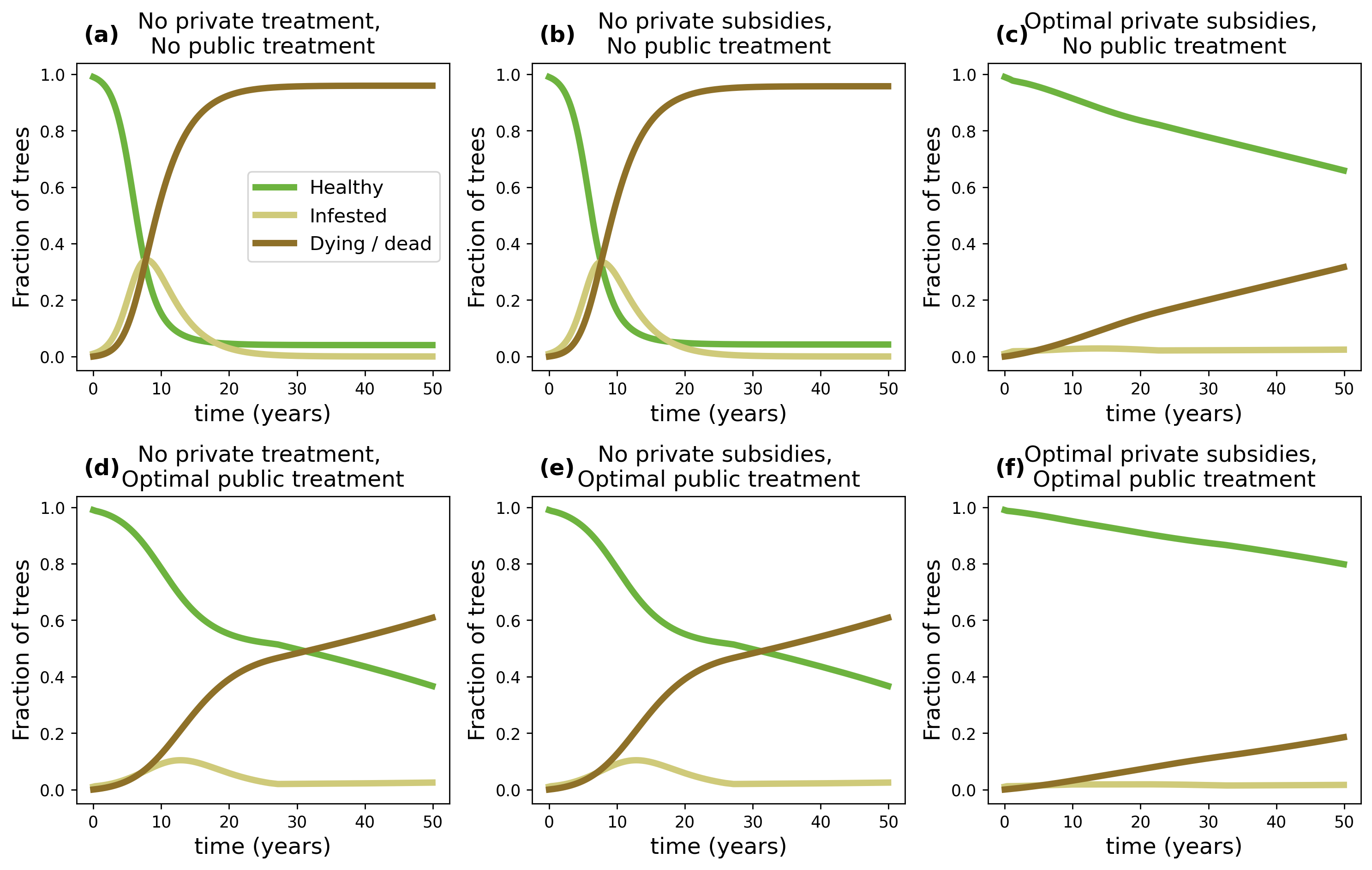}
     \caption{Using parameter values for spread rates drawn from the literature we illustrate the expected dynamics of EAB in a mixed landscape of public and private tree ownership over time across six treatment scenarios. These plots show aggregate dynamics of EAB spread averaged across public and private ownership. The first row shows cases where publicly owned trees are not treated and the second row shows cases where public trees are treated optimally. The first column shows simulations where privately owned trees are untreated, the second column shows the dynamics of EAB spread when there are no subsidies for the treatment of privately owned trees, and the final column shows cases where optimal subsidy levels are set for the treatment of private trees. If no treatments are administered, this model predicts that over 80\% of trees will be dying within 15 years, a finding that is consistent with EAB literature. When subsidies are not implemented and public trees are untreated, ash mortality is still greater than 80\%, however, reaching this level mortality takes almost 50 years. For the parameter values we consider, optimal subsidy policy combined with optimal treatment of public trees is required to prevent the establishment and spread of EAB within the community.}
    \label{supp:fig:dynamics}
     \end{figure}
     
\section{Treatment scenarios}\label{TreatScenarios}
the dynamics of the epidemiological model of pest spread depend on treatment probabilities that remain characterized. In this section, we construct a range of treatment scenarios based on the results of our optimization and game theoretic models. These scenarios construct equations that describe how treatment probabilities, for example $P_{thm}$ for healthy municipal trees, depend on the parameters of the model and the state of the system.

\subsection{Municipal treatment scenarios}\label{epiMuniTreat}
We have described the dynamics of EAB infestation on a mosaic of interacting public and private land given the probabilities of treatment across these categories of trees. Here we construct treatment scenarios based on the game analysis described earlier in this manuscript. For municipally owned trees, we consider two scenarios. First, we explore the case where no trees are treated. Next, we consider the case where trees are optimally treated, given by
\begin{equation}
P^*_{t\hat\phi m}=
\begin{cases}
    1        &  c\leq \Delta'_m\left(k_{\hat\phi}+l_{\hat\phi}\right)\\
    0        &  c\geq  \Delta'_m\left(k_{\hat\phi}+l_{\hat\phi}\right)
\end{cases}
\end{equation}
which was derived in Section~\ref{supp: muniOpt}.
The assessment of trees is imperfect, so there is not perfect alignment between assessment and the actual state of a trees health. We can write the probability that a healthy public tree is treated as 
\begin{equation}
    P^*_{thm} = P(\hat h\mid h) P^*_{t\hat h m}+ P(\hat i\mid h) P^*_{t\hat i m} + P(\hat d\mid h) P^*_{t\hat d m}
\end{equation}
and an infested public tree's treatment probability is
\begin{equation}
    P^*_{tim} = P(\hat h\mid i) P^*_{t\hat h m} + P(\hat i\mid i) P^*_{t\hat i m} + P(\hat d\mid i) P^*_{t\hat d m}.
\end{equation}
The treatment rates for dead trees have no impact of the dynamics of infestation in our model.

\subsection{Private treatment scenarios}\label{epiPrivTreat}
 For privately owned trees, we consider three scenarios. First, we examine the case where there is no treatment of any type of privately owned trees and $P^*_{tho}=P^*_{tio}=P^*_{tdo}=0$. The second case is where treatment is unsubsidized, i.e. $s_{\hat h}=s_{\hat i}=s_{\hat d}=0$. This results in bids from the firms equal to $c$ and treatment probabilities of 
 \begin{equation}
     P_{t\hat\phi o}^*=
\begin{cases}
1 
        &  c\leq ak_{\hat\phi} \\
\frac{bk_{\hat\phi}-c}{(b-a)k_{\hat\phi}} 
        & ak_{\hat\phi} < c < bk_{\hat\phi} \\
0
        & bk_{\hat\phi} \leq c
\end{cases}
 \end{equation}
 for trees in each assessed state.
 Finally, we consider the case of optimal subsidies for treatment of privately owned trees. Optimal subsidies result in treatment probabilities of trees assessed in state $\hat\phi\in\{\hat h,\hat i,\hat d\}$ of  
\begin{equation}
P_{t\hat\phi o}^*=
\begin{cases}
    1 
        &  c\leq ak_{\hat\phi} \;\text{ or }\;c+bk_{\hat\phi}-2ak_{\hat\phi}\leq \Delta_m \left(k_{\hat\phi}+l_{\hat\phi}\right) \\
    \frac{\Delta_m \left(k_{\hat\phi}+l_{\hat\phi}\right)-c+bk_{\hat\phi}}{2(b-a)k_{\hat\phi}} 
        & c> ak_{\hat\phi} \;\text{ and }\;\mid c-bk_{\hat\phi}\mid < \Delta_m \left(k_{\hat\phi}+l_{\hat\phi}\right)< c+bk_{\hat\phi}-2ak_{\hat\phi}\vspace{3.5pt}\\
    \frac{bk_{\hat\phi}-c}{(b-a)k_{\hat\phi}} 
        & bk_{\hat\phi}> c > ak_{\hat\phi} \;\text{ and }\; \Delta_m \left(k_{\hat\phi}+l_{\hat\phi}\right) \leq  bk_{\hat\phi}-c\\
     0 
        & c\geq bk_{\hat\phi}\;\text{ and }\; \Delta_m \left(k_{\hat\phi}+l_{\hat\phi}\right) \leq  c-bk_{\hat\phi}.
\end{cases}
\end{equation}
Recalling that assessments are imperfect we can use this expression to write the probability of a tree that is healthy being treated as 
\begin{equation}
    P^*_{tho} = P(\hat h\mid h) P_{t\hat h o}^*+ P(\hat i\mid h) P_{t\hat i o}^* + P(\hat d\mid h) P_{t\hat d o}^*
\end{equation}
and an infested tree's treatment probability is
\begin{equation}
    P^*_{tio} = P(\hat h\mid i) P_{t\hat h o}^*+ P(\hat i\mid i) P_{t\hat i o}^* + P(\hat d\mid i) P_{t\hat d o}^*.
\end{equation}

Combining the municipal and private treatment scenarios yields six distinct cases with aggregated public and private forest community dynamics shown in Figure~\ref{supp:fig:dynamics}. In our case study, we project the dynamics of an ash population under a baseline scenario where there is no treatment of public or private trees or other actions to slow the spread of EAB. Then, we project the ash population under a more realistic baseline where there is no municipal subsidy program for treatment of private trees but some tree owners may nonetheless treat their trees.  Finally, we compare outcomes of these baseline cases with the trajectory of the ash population under the optimal treatment policies for municipal and private trees. For these scenarios, we use model parameters (see Table~\ref{param_table} that reflect the management costs of EAB and the social and private benefits of urban trees.
\section{Additional results}
\begin{figure}[htbp]
     \centering
    \includegraphics[width=.85\textwidth]{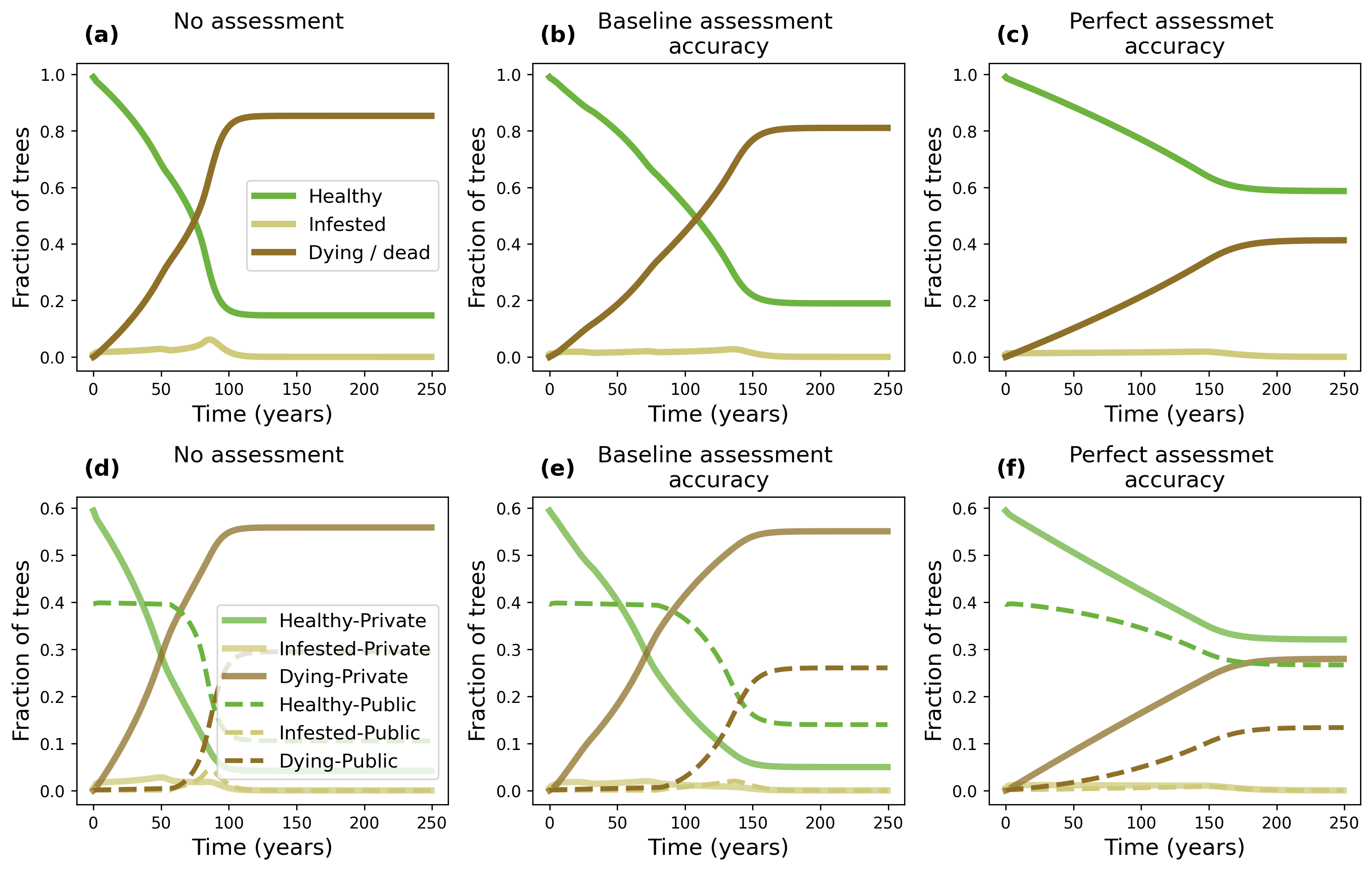}
     \caption{Dynamics of infestation and mortality over a 250-year time span given optimal subsidy and treatment policies under 3 different assessment scenarios. Panels (a) and (d) show a case where there is no assessment of tree health.  Panels (b) and (e) reproduce the results form the main text. Panels (c) and (f) show the case where  assessment is a prefect predictor of a trees underlying health state.}
    \label{supp:fig:Assessment}
     \end{figure}

Using parameter values equivalent to those considered in the EAB case study, Figure~\ref{supp:fig:Assessment} shows the impact of assessment on long term tree mortality. Panels (a) and (d) show a case where subsidy policies and treatment decisions are made in the absence of tree health assessments. Panels (b) and (e) reproduce the results form the main text but over a much longer 250 year time horizon, which shows that assessment with estimated accuracy levels delays mortality significantly but only moderately reduces long term mortality of ash trees. Panels (c) and (f) show that when assessments are prefect predictors of a trees underlying health state, optimal policies are far more effective at reducing long term tree mortality.
 
\begin{figure}[htbp]
     \centering
    \includegraphics[width=.55\textwidth]{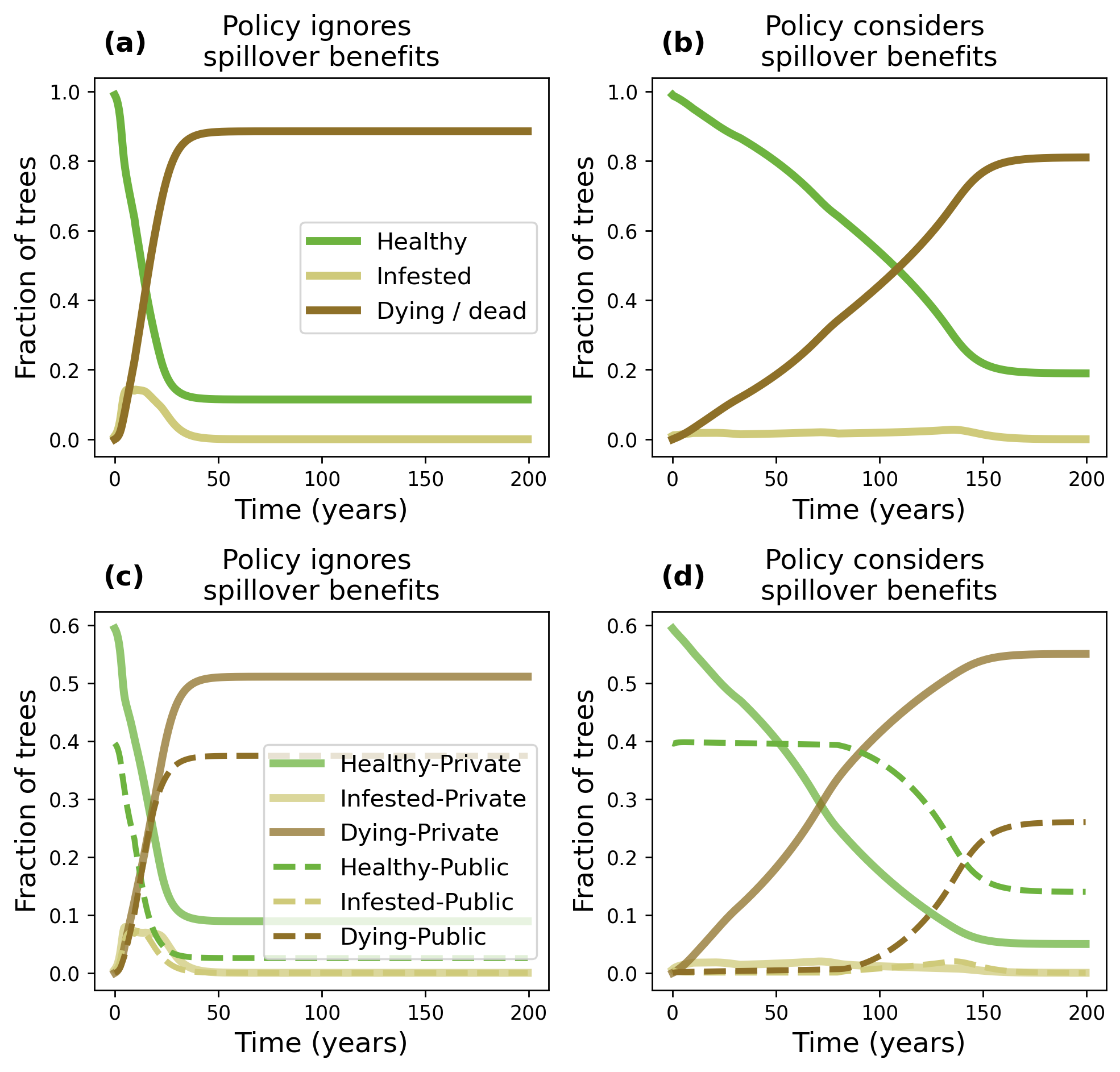}
     \caption{Dynamics of infestation and mortality over a 200-year time span given optimal subsidy and treatment policies under 2 different scenarios. Panels (a) and (c) show a case where the municipality does not consider the spillover effects of treatment on survival of other trees.  Panels (b) and (d) reproduce the results form the main text.}
    \label{supp:fig:spillovers}
     \end{figure}

We introduce a model to assess the spillover effect of treatment on the survival of other trees in the community. When a municipality considers spillover effects, it expands the range of infestation states in which treatment occurs. Panels (a) and (c) in Figure\ref{supp:fig:spillovers} show dynamics of a pest in the absence of such considerations of spillover benefits of treatment in public policy design.  This results in more rapid tree mortality, whereas consideration of spillover effects delays the incidence of mortality by decades.

\begin{figure}[htbp]
 \centering
\includegraphics[width=.85\textwidth]{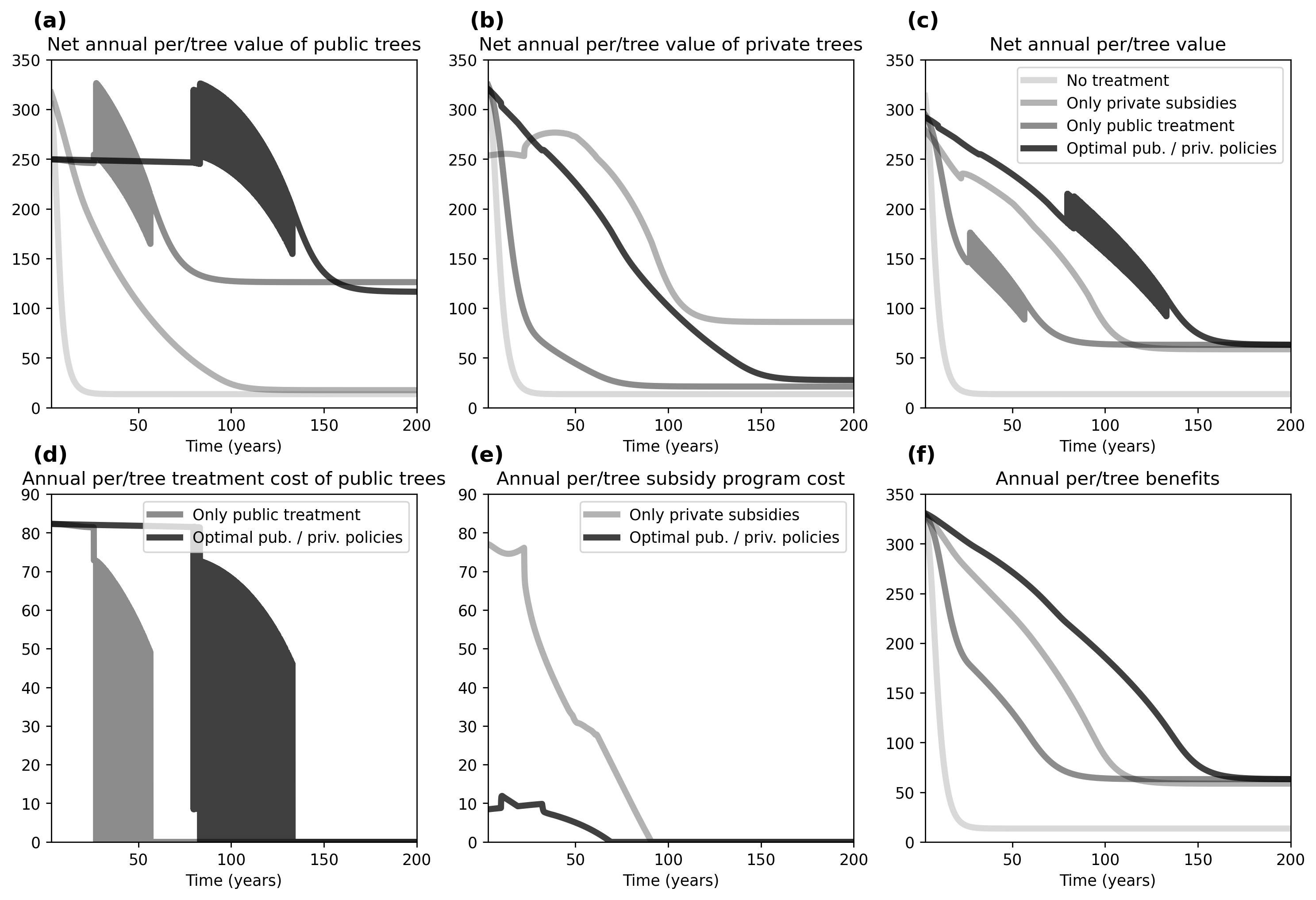}
 \caption{Annual net social value from a representative focal public tree (a), private tree (b), or arbitrary tree (c). The benefits of surviving trees is shown in panel (f). Some of these benefits are offset by the cost of treatment of public trees (d) and subsidies for private trees(e). Removal costs for public trees and other social costs of tree mortality for all trees are also considered in the net annual values reported in panels (a-c). Regions in panels (a), (c), and (d) where the lines appear to fill the space are the result of a dynamic feedback between treatment choices and infestation dynamics which results in repeated alternation of treatment rates for public trees.}
\label{supp:fig:costsbenefits}
 \end{figure}

We study the optimization of treatment subsidies and treatment choices, for private and public trees (respectively). These treatment choices are based on expectation of tree mortality risks that are consistent with our dynamical model, but based on several simplifying assumptions. As such, it may be possible for the `optimal' policies to perform poorly if they are based on risk perceptions that prove to be inaccurate. To test for this, we compute outcomes of these policies on the payoffs from the perspective of the municipality. Because our dynamic model is for a population of trees (where we only track the fraction of trees in each state) we model the costs and benefits from the perspective of the municipality and applied to an individual representative tree.  We consider annualized benefits from tree survival, which is  
\begin{equation}
    \frac{V_m}{\tau^*} \left(H_m+I_m+H_o+I_o\right)
\end{equation}
for municipal and private trees combined, annualized expected costs of treatment of a random municipal tree, which is
\begin{equation}
    \frac{c}{\tau^*} \left(P_{thm}H_m+P_{tim}I_m+P_{tdm}D_m\right).
\end{equation}
We also consider the expected annualized cost of the subsidy program when applied to a random private tree, which is 
\begin{equation}
    \frac{1}{\tau^*} \left(s^*_{\hat h}P_{t\hat ho}\hat H_o+s^*_{\hat i}P_{t\hat io}\hat I_o+s^*_{\hat d}P_{t\hat do}\hat D_o\right)
\end{equation}
where $\hat H$, $\hat I$, and $\hat D$ are the fractions of trees assessed in each health state. We also incorporate the one-time costs associated with tree mortality, $W_m$. These costs are 
\begin{equation}
     \dot D_m W'_m
\end{equation}
for a public tree and 
\begin{equation}
     \dot D_o W_m
\end{equation}
for a private tree at any point in time, and we annualize these values. In Figure~\ref{supp:fig:costsbenefits}, we show how the annual net values of a representative public tree (panel (a)), a representative private tree (panel(b)), are predicted to change over time. We average across ownership types (panel (c)), showing that our optimal policies scenario outperforms other scenarios throughout the course of our 200 year simulation.

 \begin{figure}[htbp]
 \centering
\includegraphics[width=.95\textwidth]{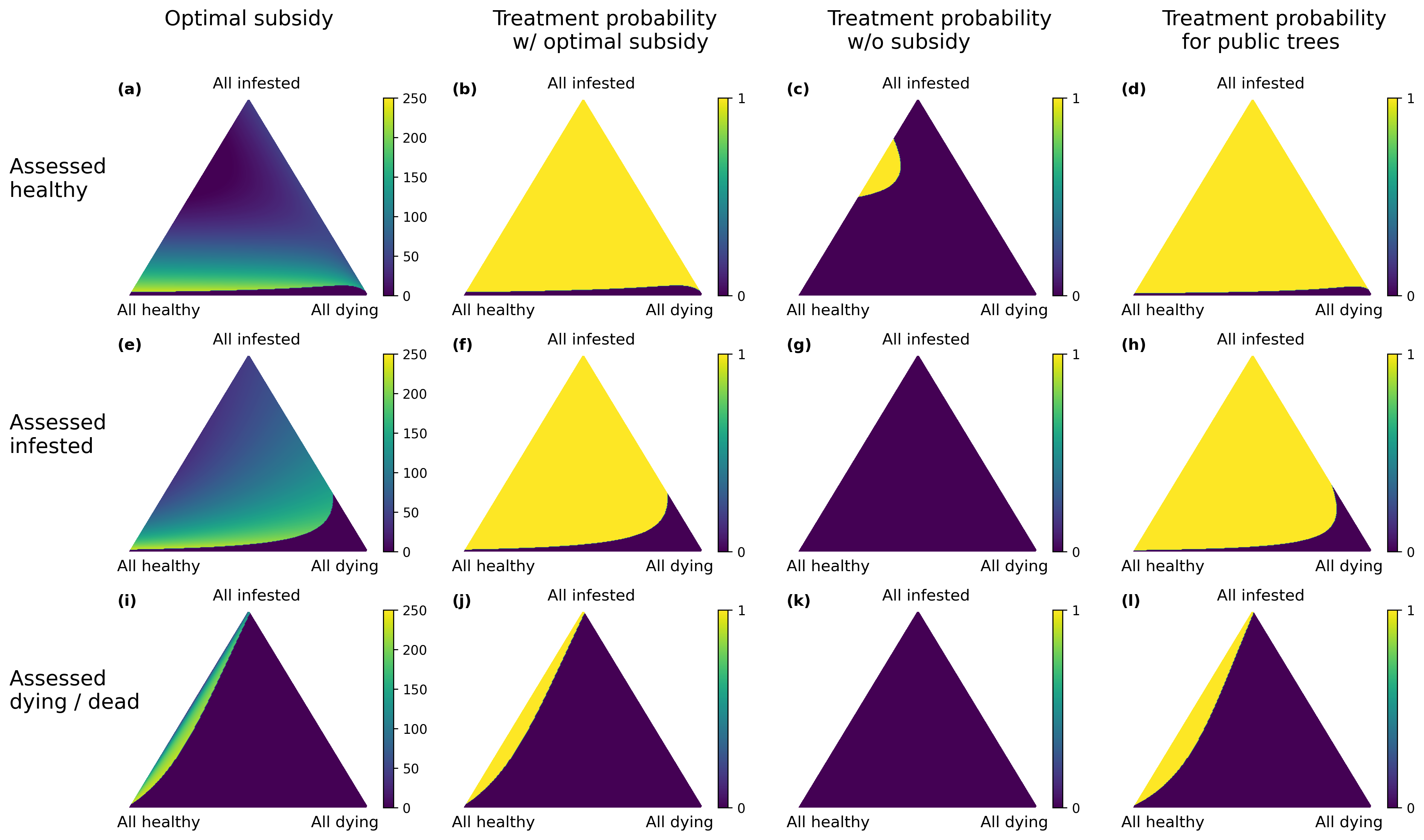}
 \caption{Subsidy and treatment outcomes under no uncertainty about tree owner values for avoiding tree mortality, $\Delta_o$.}
\label{supp:fig:no uncertaintyin Delta_o}
 \end{figure}

  \begin{figure}[htbp]
 \centering
\includegraphics[width=.95\textwidth]{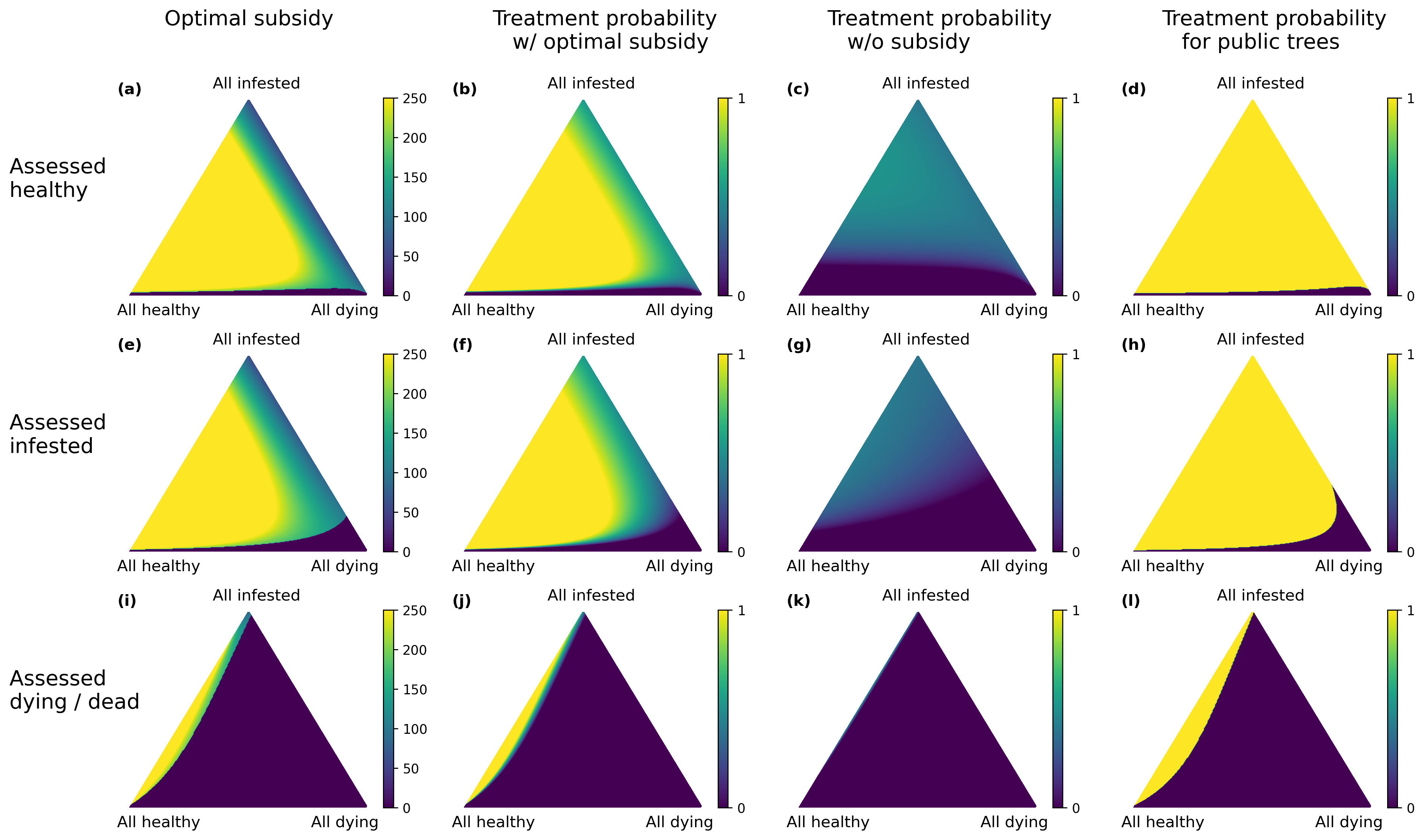}
 \caption{Subsidy and treatment outcomes under high uncertainty about tree owner values for avoiding tree mortality with the same expected value for $\Delta_o$ as in Figure~\ref{supp:fig:no uncertaintyin Delta_o}.}
\label{supp:fig:highly uncertain}
 \end{figure}

\section{EAB case study parameter selection}\label{suppEABpars}
 We let our pest spread and mortality rate parameters be $\beta=1$ and $\gamma=.3$ because this results in approximately 80\% of a community's ash trees dying within 15 years (Figure~\ref{fig:dynamics}, panel(a)), a result that is consistent with the EAB literature \citep{mccullough2012evaluation,knight2013factors,burr2014condition,sadof2017tools,kappler2019evaluating}. We let the recovery rate of treated trees be $\alpha=1$ because treatment can quickly kill EAB larvae~\citep{mccullough2011evaluation}, so we expect the recovery rate to be faster than the death rate. A leading treatment option for EAB is a trunk injection with emamectin benzoate, which studies have shown to be effective for 2-4 years~\citep{sadof2022factors}. Therefore we choose a three-year planning horizon, $\tau^*=3$. We assume a treatment cost of $c=250$ which is consistent with a 20 inch DBH tree being charged \$12.50 per inch DBH for treatment, numbers broadly consistent with mature ash sizes and treatment costs. 

Studies on the effectiveness of insecticide treatments for different health classes of ash trees~\citep{mccullough2011evaluation,flower2015treat}, find that treatment is highly effective for healthy trees, but diminishes for highly infested trees. We assume there is a $97\%$ effectiveness of treatment of EAB-free trees at preventing infestation, but that the likelihood of recovery of an infested ash tree in response to treatment is only $50\%$, and thus we set $\epsilon_h=.97$ and $\epsilon_i=.5$. We use the three year planning horizon, epidemiological model parameters, and these treatment effectiveness parameters to calculate the perceived direct ($\mu$) and spillover (($\lambda$) mortality risks for treated and untreated trees in each assessed health class.    

We assume that the value to tree owners of avoiding tree mortality over this planning horizon is uniformly distributed between \$675 and \$1100 (i.e. $\Delta_o\sim U(675,1100)$).  This range reflects annual tree values of \$25 to \$100 estimated from hedonic property value studies  \citep{sander2010value,kovacs2022tree} (multiplied by the three year planning horizon) combined with a one-time tree removal cost range of \$600 to \$800.  We consider a case where the municipality's value of avoiding private tree mortality over the planning horizon is $\Delta_m=V_m+W_m=1150$, which excludes consideration of removal costs experienced by a private tree owner. The municipality's value of avoiding public tree mortality over the planning horizon includes removal costs because these will be paid by the municipality and is $\Delta'_m=1850$. These values were chosen because studies on the economics of tree cover have shown that the social value of a percentage change in tree cover in a neighborhood can exceed the tree owner's value by a factor of five or more \citep{kovacs2022tree}, thus we expect $V_m$ to fall within $375-1500$. Because we assume that the municipality does not consider a tree owners' removal costs in their assessment of private trees we assume that $W_m$ will be relatively small and $w'_m$ (which does include removal costs) will be similar to the removal costs of a private tree. The values of $\Delta_m$ and $\Delta'_m$ we consider fall within these ranges.  

We rely on studies of ash health assessment under EAB risk to come to reasonable values for assessment accuracy probabilities~\citep{flower2013relationship,knight2014monitoring}. For ash trees with no EAB present, we assume there is a very small chance that the tree will be classified as dying (1\%) with a higher chance of being misclassified as infested (10\%) and most EAB-free trees accurately assessed as healthy (89\%). For trees with an infestation of EAB, the visual characteristics of early stages of crown dieback can be seen but only once a sufficiently large population size of EAB larva are present. Therefore there is a chance that such a tree will be categorized as healthy (49\%) or infested (50\%) with only a small possibility of an infested tree being assessed as dying (1\%). For trees with late-stage untreatable EAB infestations, advanced crown dieback can signal that it may be too late for treatments to be effective. We assume that in these cases there is only a 1\% chance that the tree will be assessed as healthy, a 19\% chance it will be assessed as infested, and a 80\% chance it will accurately be assessed as dying.

\end{document}